\documentclass[letterpaper,twocolumn,8pt]{article}

\usepackage{usenix2019_v3}

\usepackage{booktabs}

\usepackage{amsmath}
\usepackage{url}

\usepackage{xcolor}
\usepackage{cleveref}
\usepackage{graphicx}
\usepackage{subcaption}

\newcommand{\eg}{\emph{e.g.,} }
\newcommand{\ie}{\emph{i.e.,} }
\newcommand{\etal}{\emph{et al.} }

\begin{document}
%
\title{Siren Song: Manipulating Pose Estimation in XR Headsets Using Acoustic Attacks}









\author{
{\rm Zijian Huang$^1$, Yicheng Zhang$^2$, Sophie Chen$^1$,  Nael Abu-Ghazaleh$^2$ and Jiasi Chen$^1$}\\
\emph{$^1$University of Michigan, Ann Arbor\thanks{\{zijianh, sophiecc, jiasi\}@umich.edu}}\\
\emph{$^2$University of California, Riverside\thanks{\{yzhan846, nael.abughazaleh\}@ucr.edu}}\\
}

\thispagestyle{plain}
\pagestyle{plain}

\maketitle

\begin{abstract}
Extended Reality (XR) experiences involve interactions between users, the real world, and virtual content.
A key step to enable these experiences is the XR headset sensing and estimating the user's pose in order to accurately place and render virtual content in the real world.
XR headsets use multiple sensors (\eg cameras, inertial measurement unit) to perform pose estimation and improve its robustness, but this provides an attack surface for adversaries to interfere with the pose estimation process.
In this paper, we create and study the effects of acoustic attacks that create false signals in the inertial measurement unit (IMU) on XR headsets, leading to adverse downstream effects on XR applications.
We generate resonant acoustic signals on a HoloLens 2 and measure the resulting perturbations in the IMU readings, and also demonstrate both fine-grained and coarse attacks on the popular ORB-SLAM3 and an open-source XR system (ILLIXR).
With the knowledge gleaned from attacking these open-source frameworks, we demonstrate four end-to-end proof-of-concept attacks on a HoloLens 2: manipulating user input, clickjacking, zone invasion, and denial of user interaction.
Our experiments show that current commercial XR headsets are susceptible to acoustic attacks, raising concerns for their security.
\end{abstract}

\section{Introduction}
\label{sec:intro}

Extended reality (XR) is growing in popularity, with recent or soon-to-be-released commercial headset prototypes from Apple, Meta, and Google.
XR seamlessly blends real and virtual content on the user's display, enabling a range of exciting applications in entertainment, healthcare, public safety, etc.
In light of these new devices and their sensing capabilities, securing XR devices is of increasing concern and has attracted recent attention from researchers and industry~\cite{lebeck2017securing,ruth2019secure,corbett2023bystandar,rajaram2023eliciting,rajaram2023reframe}.
Of particular interest are attacks that can cause issues with the user interface (UI), preventing users from interacting with UI elements or interfering with their perception of the physical world, for example by blocking the view of important real-world objects or generating auditory sounds to interfere with user attention ~\cite{cheng2024user,cheng2023exploring,lebeck2017securing}.

Implementing UI attacks is not easy, as often an in-band threat model is assumed (\eg a software library that has been compromised~\cite{cheng2023exploring}, or malware that has been accidentally installed from the app store~\cite{slocum2023going}).
Out-of-band attacks, including acoustic~\cite{trippel2017walnut,tu2018injected} or wireless signals~\cite{davidson2016controlling,gluck2020spoofing},
can interfere with sensor readings and cause adverse downstream effects on smartphones and autonomous vehicles.
Acoustic attacks work because sound waves interact with springs in micro-electro-mechanical systems (MEMS) in the inertial measurement unit (IMU), causing incorrect readings of the accelerometer and gyroscope.
The adverse effects of acoustic attacks have been demonstrated on smartphones, drones, and other devices~\cite{son2015rocking,tu2018injected}.

\begin{figure}
    \centering
    \includegraphics[width=0.25\textwidth]{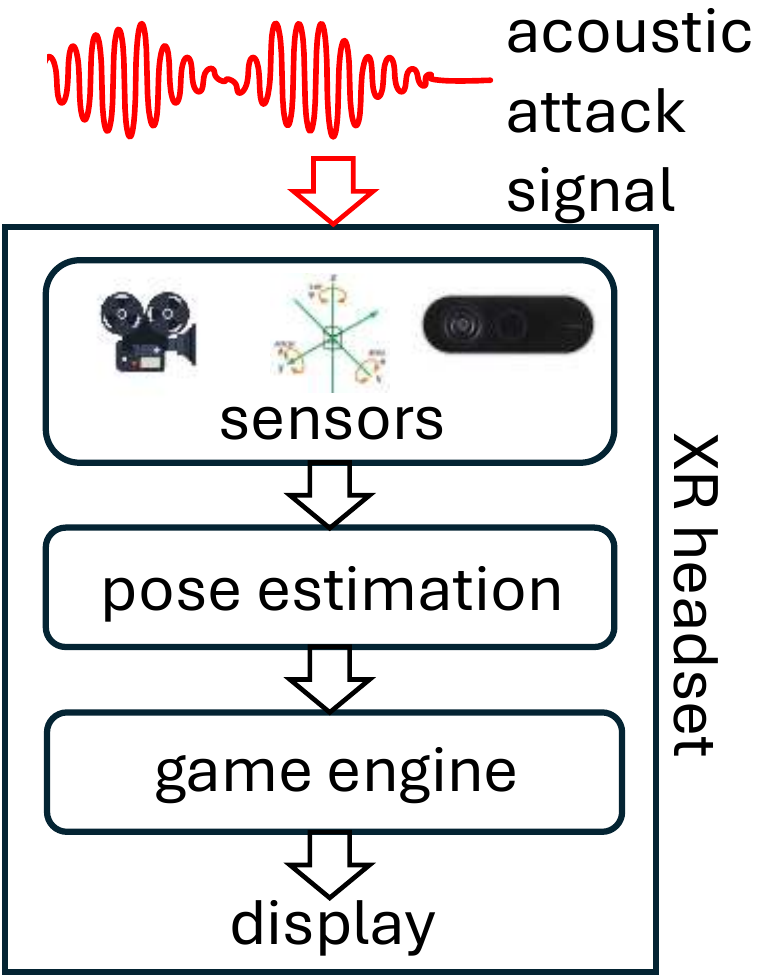} 
    \caption{Scenario overview. An XR headset is subjected to acoustic signals, which affects the pose estimation and final visual outputs.}
    \label{fig:overview}
\end{figure}

The hypothesis in this paper is that XR headsets rely on IMUs to display virtual content and therefore may also be susceptible to MEMS-driven acoustic attacks.
However, the full effects of attacks on the XR visualization are depend on the XR processing pipeline (shown in \Cref{fig:overview}), which ingests the perturbed sensor readings, processes them through a series of software algorithms, and finally outputs to the display.
These data processing steps include pose estimation and XR game engines world generation and rendering (\eg Unity, Unreal).
The key question is whether the low-level sensor inputs can be  appropriately perturbed by the acoustic signals so that they impact the XR application layer, displaying visual outputs that hurt the user experience.

To answer this question and demonstrate real-world attacks, we had to overcome several challenges.
(1) \emph{Attack setting.}
We work with commercial-off-the-shelf (COTS) XR headsets and a non-invasive setup.
We did not assume access to the internals of the device, as in other works~\cite{trippel2017walnut}.
(2) \emph{XR processing pipeline.} 
Pose estimation using simultaneous localization and mapping (SLAM) methods is a key step in the XR processing pipeline. 
We characterize how the perturbed sensor inputs affect pose estimation under the relatively weak assumption of only one attacked sensor modality (IMU) alongside benign visual inputs, and under different implementations of SLAM tracking failure recovery methods.
(3) \emph{Impact on user experience.}
Naive application of acoustic injection attacks may have little impact on user experience if the attack occurs when there is no relevant virtual content.
We crafted four proof-of-concept application scenarios to showcase how acoustic injection could either help or harm user experience.


The main findings of this work are that open-source pose estimation methods are susceptible to acoustic injection attacks, resulting in a variety of effects: ``misleading'' (the virtual content has a constant position offset), ``snapback'' (the virtual content resets its position) and ``drift away'' (the virtual content flies out of the field-of-view).
In closed-source commercial headsets, the ``snapback'' effect is reproducible in end-to-end attacks, as well as an additional ``small drift'' effect, and can be used to demonstrate beneficial or harmful effects to users in four scenarios: VR gaming, clickjacking, denial of user interaction, and destroying privacy zones.
We find that key factors impacting attack success rate are the volume and movement of the headset.
Taken together, these attacks illustrate that modern XR devices and their pose estimation methods are vulnerable to acoustic injection attacks.

In summary, this work explores how UI-level security attacks can be performed via a novel threat model: out-of-band acoustic injection.
The contributions of this work are:
\begin{itemize}
\item We evaluated COTS XR headsets (Microsoft Hololens 2 and Meta Quest 3) to determine their susceptibility to acoustic injection attacks.
\item To understand the impacts of acoustic injection attacks on individual stages of the XR processing pipeline, we performed controlled trace-driven simulations on open-source XR components (ILLIXR~\cite{huzaifa2022illixr} and ORB-SLAM3~\cite{campos2021orb}).
\item We perform end-to-end attacks on Hololens 2 and show how they can benefit or harm the user experience in four proof-of-concept scenarios. We also evaluate the effect of the physical configuration, such as acoustic volume, direction, and headset mobility on attack success.
\end{itemize}
The paper is organized as follows. \Cref{sec:related} discusses related work and \Cref{sec:prelim} describes the threat model and background on acoustic attacks. \Cref{sec:exp} describes our experiments with open-source XR frameworks and \Cref{sec:exp-hololens2} describes our end-to-end attacks on COTS headsets. \Cref{sec:discussion} discusses limitations and \Cref{sec:conclusions} concludes.

\section{Related Work} \label{sec:related}

\textbf{Attacks on SLAM.} 
The robustness of computer vision systems is being actively investigated. With the emergence of adversarial images in the digital domain by adding optimized noise directly to images~\cite{szegedy2013intriguing,carlini2017towards}, researchers find that such attacks also exist physically in the real world \cite{eykholt2018robust,song2018physical,zhao2019seeing}. To fill the gap between attacks in the digital and physical worlds, recent studies have demonstrated that attacks on real-world computer vision systems are practical \cite{eykholt2018robust,li2019adversarial,man2020ghostimage,sharif2016accessorize,zhao2019seeing,zhou2018invisible}. However, attacks on traditional computer vision methods such as SLAM are relatively less explored. \cite{yoshida2022adversarial} proposes an attack against the scan matching algorithm in LiDAR-based SLAM, while most SLAMs in AR/VR devices rely on different sensors like RGB/depth cameras and IMUs. \cite{ikram2022perceptual} and \cite{chen2024adversary} mislead visual SLAM by poisoning the images with special patterns, and \cite{wang2021can} causes the camera to fail using infrared light. In our work, we demonstrate attacks on Visual-Inertial SLAM (VI-SLAM) by perturbing the IMU readings, rather than cameras, and showing its impact on XR user experience. 

\textbf{Acoustic Injection Attacks.} Among various physical attacks, acoustic injection attacks are attractive due to their low cost. Son~\etal~\cite{son2015rocking} were the first to introduce acoustic attacks on MEMS gyroscopes, demonstrating how these attacks could lead to sensor denial-of-service and result in drone crashes. WALNUT~\cite{trippel2017walnut} expanded on this by developing output biasing and control attacks that enable precise manipulation of MEMS accelerometer outputs using modulated sound waves. Wang et al.~\cite{wang2017sonic} demonstrated a sonic gun, showcasing the vulnerability of various smart devices (\eg drones and self-balancing vehicles) to acoustic attacks. Tu et al. \cite{tu2018injected} designed side-swing and switching attacks to alter the outputs of MEMS gyroscopes and accelerometers. Furthermore, Ji et al. \cite{ji2021poltergeist} fool the object detectors by applying acoustic attack to the image stabilizers commonly used in modern cameras. However, none of the existing works study the relationship between the acoustic injections and SLAM outputs on recent XR devices. 


\textbf{XR Security and Privacy.} 
For single-user XR systems, researchers have demonstrated various side-channel attacks to extract sensitive information (\eg keystrokes) through video feeds~\cite{ling2019know}, head movements~\cite{nair2023unique, slocum2023going}, architectural hints~\cite{zhang2023its,shang2020arspy}, power usage~\cite{li2024dangers}, and EM side-channel leakages~\cite{al2021vr}. In multi-user XR systems, Su et al.~\cite{su2024remote} use avatar motion data to infer keystrokes in shared VR environments. Slocum et al.~\cite{slocum2024doesn} reveal vulnerabilities in the shared state frameworks of multi-user AR. Similarly, Lebeck et al.~\cite{lebeck2017securing} highlight risks like deceptive virtual objects and emphasize access control for managing shared physical and virtual spaces. Ruth et al.~\cite{ruth2019secure} further propose a secure multi-user AR framework focusing on content sharing and permissions.
Chandio et al.~\cite{chandio2024stealthy} 
simultaneously manipulated visual and inertial sensors to disrupt XR pose estimation. However, their study evaluated the attack using offline datasets and assumed the attacker's capability to manipulate IMU data streams through acoustic means, without real experiments. Ours is the first to demonstrate acoustic injection attacks on recent XR devices, like the Hololens 2, in the real world.

\section{Preliminaries} \label{sec:prelim}

\subsection{Threat Model}

\textbf{Attack Scope.} We assume that the attackers can play acoustic sounds nearby to affect the integrity of sensor data, but cannot directly access the digitized sensor readings or physically touch the sensors on the device. This is because commercial products in XR disable write access to inertial sensors both in physical ways and through software APIs, in an attempt to help improve the security and privacy of users, which are widely explored by researchers~\cite{michalevsky2014gyrophone,aviv2012practicality,owusu2012accessory,dey2014accelprint,marquardt2011sp}.
Attackers may use the attack to either benefit themselves as users of the XR headset (\eg to cheat in a game) or to harm victims who are wearing the XR headset (\eg to inhibit their performance in a game).

\textbf{Sensor Access.} Although we assume that attackers cannot physically access internals of the target devices when they conduct attacks, we do allow the adversary to have access to a device in the same model, which means that the attacker can profile the device's behavior under different acoustic frequencies and amplitudes with a sample device. This assumption is reasonable since attackers can purchase XR headsets on their own to study their behavior. The transferability of attacks between different devices of the same model has been confirmed by previous studies~\cite{trippel2017walnut}. 

\textbf{Speaker Access.} We allow the attacker to generate acoustic signals from any direction around the victim device, at frequencies in the range of human audible sound to ultrasonic (2-30 kHz). This can be done by an attacker playing a sound file while following the user or by speakers in the environment.
Although sound in the human audible range may be noticeable for users, we allow it in our threat model because its perceptibility depends on the ambient sound volume or music playing in the environment.
Further, users who seek to benefit from the attack by causing themselves advantages in the XR experience would not care about acoustic perceptibility.


\subsection{Background on Acoustic Attacks on IMU}
\label{sec:background}
Modeling the effects of acoustic signals on MEMS IMUs has been previously studied~\cite{trippel2017walnut}. Here we will briefly describe the model as background for our later simulations and experiments. Because our experimental results later demonstrate successful attacks on the accelerometer, 
here we will use the accelerometer for the purposes of explanation, but similar models apply to gyroscopes.

We denote electrical acceleration signals generated by true acceleration $s(t)$ and those generated
by acoustic interference $s_a(t)$. In general, the measured acceleration can be modeled as a linear combination of the true acceleration and acoustic acceleration, which means the measured acceleration $\hat{s}(t)$ by the sensor can be expressed as
\begin{align}
    \hat{s}(t) = s(t) + A_0 s_a(t)
\end{align}
where $A_0$ represents the attenuation of the acoustic signal while in transit to the target device. Because the acoustic acceleration can be modeled as $s_a(t) = A_1 \cos(2\pi F_a t+\phi)$, with frequency $F_a$, amplitude $A_1$, and phase $\phi$, the measured acceleration can be re-written as
\begin{align}
    \hat{s}(t) = s(t) + A_0 A_1  \cos(2\pi F_a t+\phi).
\end{align}
Note that vibrating these systems at their resonant frequencies achieves maximum displacement of the spring mass, \ie $A_0 = 1$.

According to \cite{trippel2017walnut}, there are two kinds of possible attacks: (1) \textbf{Output Biasing Attack} by utilizing sampling deficiencies of the Analog-to-Digital Converter (ADC); and (2) \textbf{Output Control Attack} due to insecure amplifiers, where accelerometers exhibit constant shifted false measurements at their resonant frequencies. 

\subsubsection{Output Biasing Attack}
The output biasing attack consists of two main steps: \textbf{Stablizing} and \textbf{Reshaping}
\begin{enumerate}
    \item \textbf{Stablizing.} The first step is to utilize a DC alias of the acceleration signal at the ADC to generate constant false measurements. This happens when the analog signal’s frequency is an integer multiple of the sampling frequency $F_{\text{samp}}$. We denote the sampling times at discrete intervals $k$ as $t_k=k\cdot\frac{1}{F_{\text{samp}}}$. Because acoustic signals with frequency near the resonant frequency can achieve nearly the same resonant result, and the sampling frequency of IMUs are generally much lower than the resonant frequencies of accelerometer (and gyroscopes), the attacker can find a frequency $F_a = F_{\text{res}} + f_\epsilon = N\cdot F_{\text{samp}}$ where $F_\text{res}$ is the resonant frequency of the accelerometer or the gyroscope, $f_\epsilon$ is the smallest deviation between $F_\text{res}$ and $N\cdot F_{\text{samp}}$, and $N\in\{1,2,3...\}$. Therefore, we have
    \begin{align}
        \hat{s}(t_k) &= s(t_k) + A_0 A_1 \cos(2\pi F_a t_k+\phi) \\
        &= s(t_k) + A_0 A_1 \cos(2\pi Nk+\phi) \\
        &= s(t_k) + A_0 A_1 \cos(\phi),
    \end{align}
    \item \textbf{Reshaping.} To further shape the output signal, the attacker employs either amplitude or phase modulation techniques to tune the parameters $A_1, \phi$ to get the desired output.
\end{enumerate}

\subsubsection{Output Control Attack} 
When the amplifier or the low-pass filter (LPF) is insecure, the attacker can achieve fine-grained control over a sensor’s output using amplitude modulation, which indefinitely controls an accelerometer’s output. 

In conclusion, the accelerometers and the gyroscopes' readings can be manipulated to either have a relative offset or set to a specific constant, depending on whether the ADC or the LPF is  vulnerable.
We will later characterize real XR headset (Hololens 2) in terms of their vulnerabilities to these types of attacks (\Cref{sec:frequency_sweep}).

\section{Experiments on Open-Source Systems}
\label{sec:exp}

The goal of this section is to understand how acoustic attacks impact the first stage of the pipeline in \Cref{fig:overview}, pose estimation, in isolation, before considering end-to-end effects (\cref{sec:exp-hololens2}).
We will study the impact of perturbations on two  pose estimation frameworks:
ORB-SLAM3 \cite{campos2021orb}, which is an open-source VIO library widely deployed as the basis of many SLAM systems 
(\Cref{subsec:exp-orbslam3}), as well as the ILLIXR runtime~\cite{huzaifa2021illixr}, which is an open testbed developed by academia (\Cref{subsec:exp-illixr}).
Together, these two evaluation platforms enable us to understand how different, popular pose estimation algorithms behave under controlled inputs.

\subsection{Attack on ORB-SLAM3}
\label{subsec:exp-orbslam3}
We simulate the effects of acoustic injection attacks on ORB-SLAM3 \cite{campos2021orb}
by adding noise to input IMU values in one of two ways: (1) in a fine-grained way, creating a constant bias, as studied by \cite{trippel2017walnut}; and (2) in a coarse-grained way, by building a data-driven model based on real headset measurements.

\subsubsection{Constant IMU Perturbations} 
\label{subsubsec:exp-orbslam3-constant}
\begin{figure}[t]
    \centering
    \includegraphics[width=0.45\textwidth]{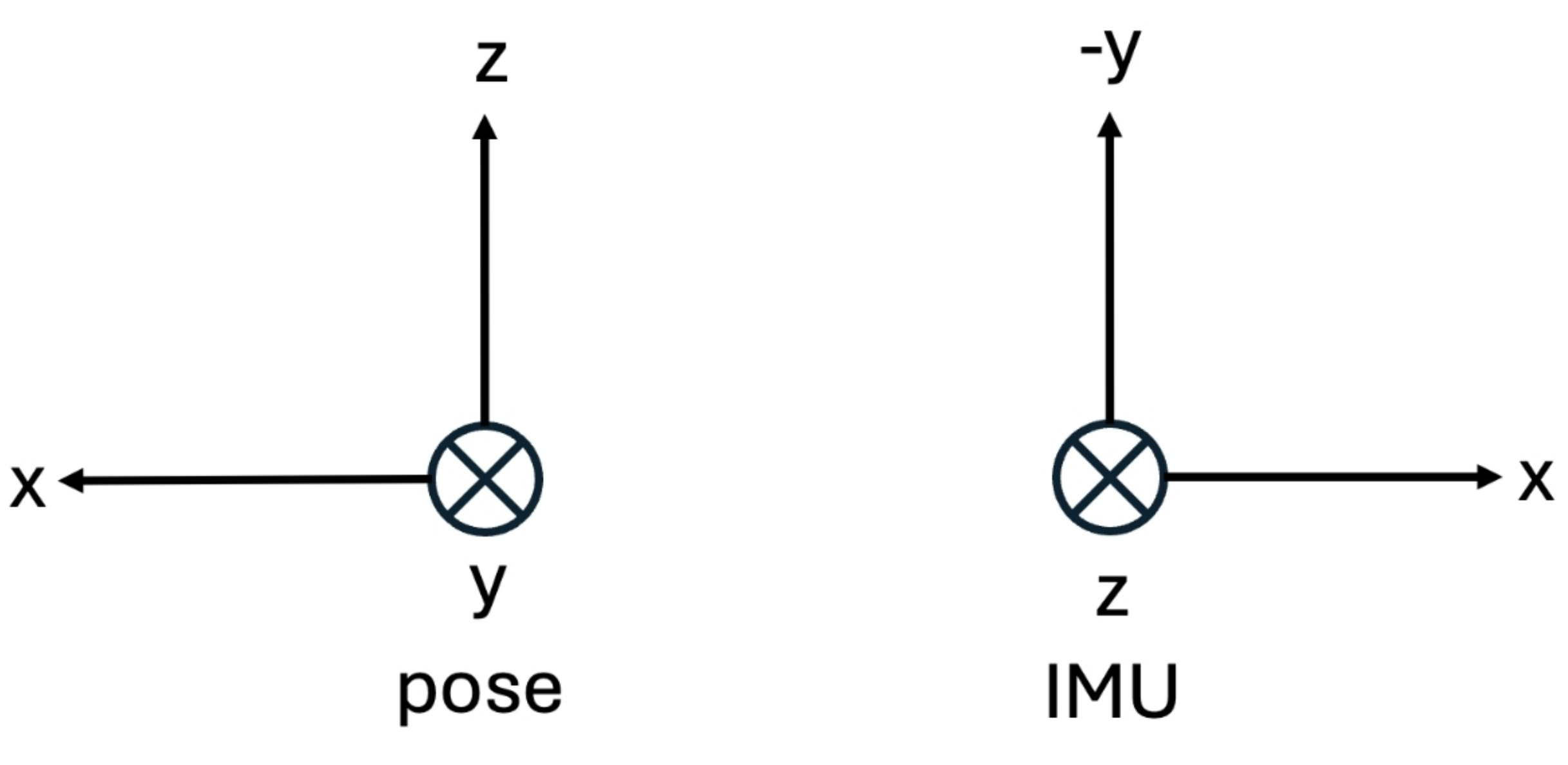} 
    \caption{The coordinate systems of ORB-SLAM3 (left) and IMU on the RealSense D435i camera (right) differ.}
    \label{fig:coordinate_systems}
\end{figure}

\textbf{\emph{Setup.}}
As a first step, we start with the simplest scenario: adding constant bias to the IMU readings, which is possible in practical scenarios~\cite{trippel2017walnut}.
We add or subtract a constant value to the $x, y, z$ axes of accelerometer or gyroscope readings throughout the entire trace.
For realism, we constrain the range of perturbation to $[-g,+g]$ for accelerometer readings, where $g$ is the gravitational acceleration of 9.8 m/s$^2$, and $[-2\;\text{rad/s}, +2\;\text{rad/s}]$ for gyroscope readings, as reported in a previous study~\cite{jeong2023rocking}.
We leave the camera images unmodified as we do not assume an attack vector for the camera.
To understand the fundamental impact of the constant perturbation, we created a trace with a simple but common walking pattern of moving forward, recording camera images and IMU readings using a RealSense D435i~\cite{hu2024apple}.
In this trace, the user moves about 4 meters along the positive $y$ direction of the world frame for about 10 seconds. 
Note that the coordinate system of ORB-SLAM3 differs from that of the RealSense camera, as shown in \Cref{fig:coordinate_systems}, so we have to transform the coordinates of the camera data appropriately before feeding it into ORB-SLAM3.

\begin{figure*}[h]
    \centering
    \begin{tabular}{ccc}
        \begin{subfigure}[b]{0.32\textwidth}
            \centering
            \includegraphics[width=\textwidth]{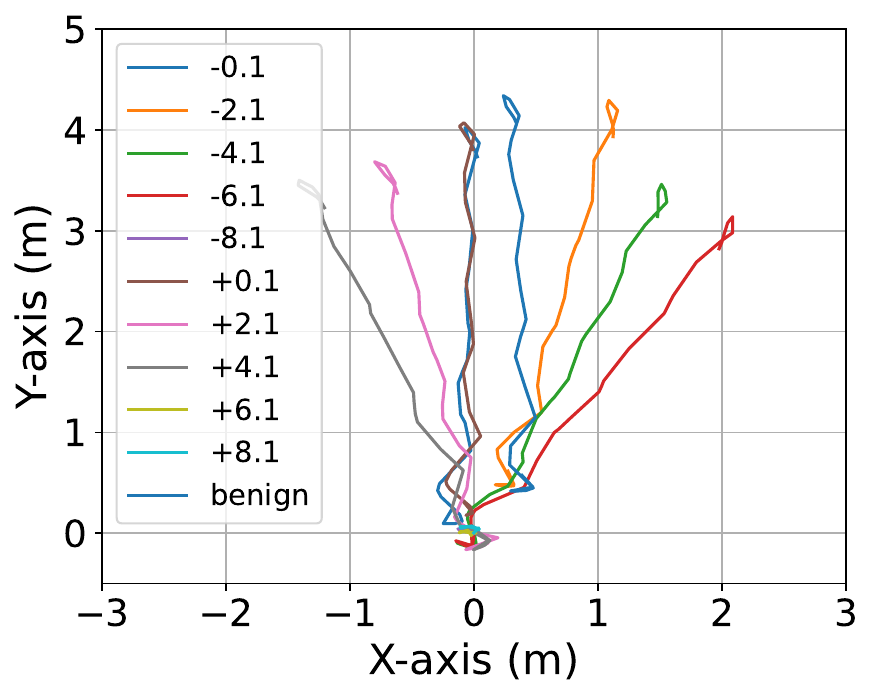}
            \caption{Attack on accelerometer X axis, X-Y plot in ORB-SLAM3 world frame}
            \label{subfig:orbslam_acc_constant_1}
        \end{subfigure}
        ~
        
        \begin{subfigure}[b]{0.32\textwidth}
            \centering
            \includegraphics[width=\textwidth]{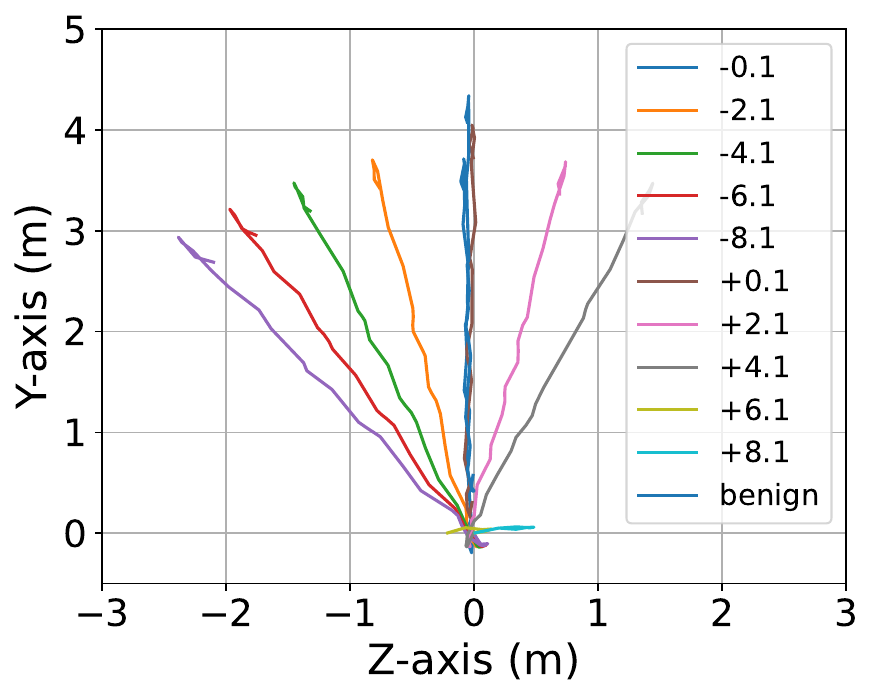}
            \caption{Attack on accelerometer Z axis, Z-Y plot in ORB-SLAM3 world frame}
            \label{subfig:orbslam_acc_constant_6}
        \end{subfigure}
        ~
        \begin{subfigure}[b]{0.32\textwidth}
            \centering
            \includegraphics[width=\textwidth]{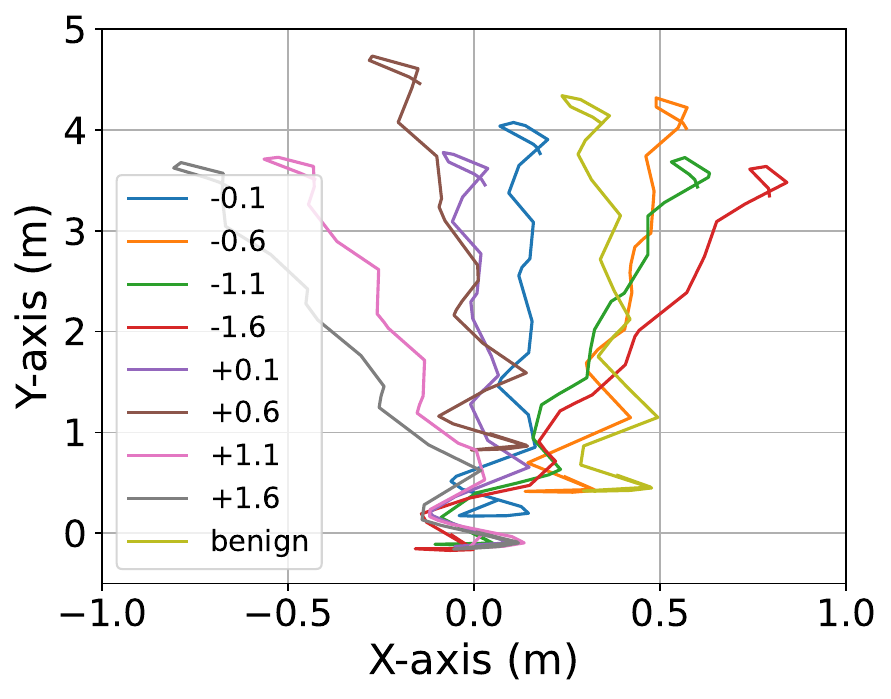}
            \caption{Attack on gyroscope Z axis, X-Y plot in ORB-SLAM3 world frame}
            \label{subfig:orbslam_gyro_constant_3}
        \end{subfigure} \\
    \end{tabular}
    \caption{
    Device pose estimated by ORB-SLAM3 under constant perturbation on the IMU readings.
    Legend denotes perturbation magnitude.
    Increased magnitude of perturbations leads to increased pose error (\textbf{Misleading attack}), and beyond a threshold, devices default to the origin (\textbf{Snapback attack}).
    }
    \label{fig:orbslam_acc_constant}
\end{figure*}

\textbf{\emph{Results of constant IMU perturbations.}}
We plot the trajectory output by ORB-SLAM3 in \Cref{subfig:orbslam_acc_constant_1,subfig:orbslam_acc_constant_6,subfig:orbslam_gyro_constant_3} for different magnitudes and directions of the input IMU data with perturbations.
The benign case is a simple forward and backward movement (blue line).
When the $x$ axis of the accelerometer, $z$ axis of the accelerometer, or the $z$ axis of gyroscope are perturbed,
we can see that the trajectory error increases roughly proportionally with magnitude and duration of the perturbation.
For example, when the perturbation is $x+2.1\,\text{m/s}^2$, the final position is about 1.1 meters off, and when the perturbation is $x+4.1\,\text{m/s}^2$, the final position is about 1.8 meters off. 
We call this effect the \textbf{Misleading attack}.
The impact on the displayed virtual content in an XR headset would be displacement; \eg if the final estimated device pose is 2 meters off to the right, the virtual content would be displayed 2 meters to the left (similar visualizations will be shown later in \Cref{subsec:exp-illixr} and \Cref{sec:exp-hololens2}). As shown in \Cref{subfig:orbslam_gyro_constant_3}, the perturbation on the gyroscope has a similar effect, causing the trajectory to veer off course since the IMU perceives an angular rotation. 

\begin{figure*}[htb]
    \centering
    \begin{tabular}{cccc}
        \begin{subfigure}[t]{0.23\textwidth}
            \centering
            \includegraphics[width=\textwidth]{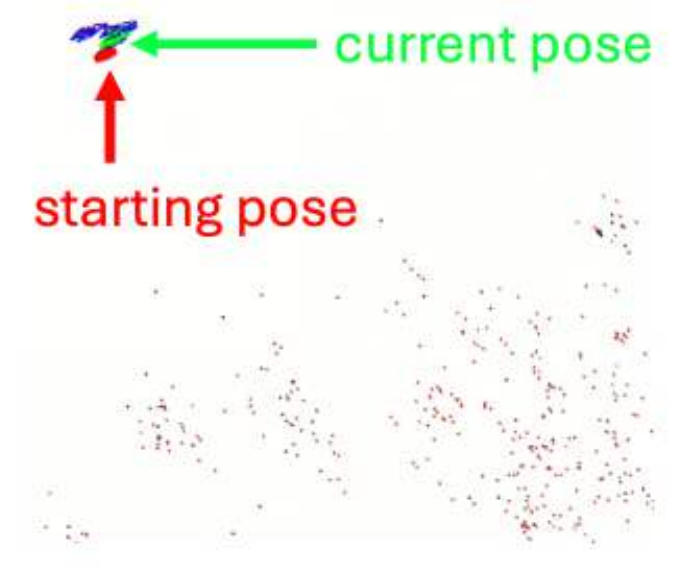}
            \caption{time=0: device at origin}
            \label{subfig:firsthalf_snapback_frame1}
        \end{subfigure} &
        \begin{subfigure}[t]{0.23\textwidth}
            \centering
            \includegraphics[width=\textwidth]{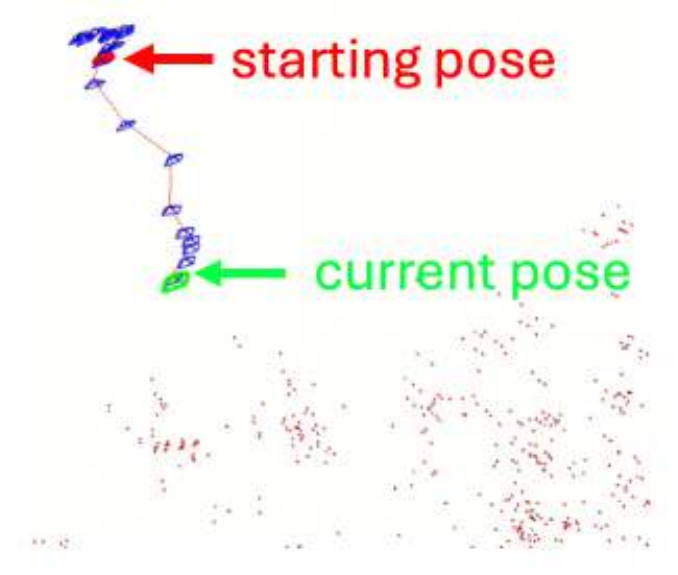}
            \caption{time=1: device moves normally}
            \label{subfig:firsthalf_snapback_frame2}
        \end{subfigure} &
        \begin{subfigure}[t]{0.23\textwidth}
            \centering
            \includegraphics[width=\textwidth]{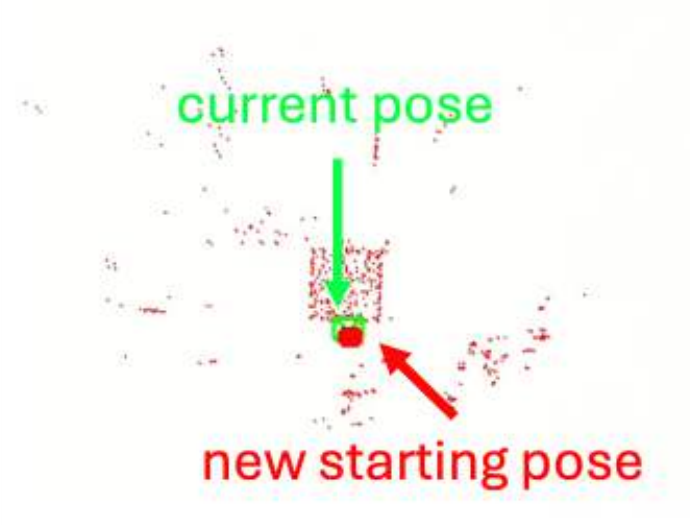}
            \caption{time=2: snapback occurs, device thinks it's at origin}
            \label{subfig:firsthalf_snapback_frame3}
        \end{subfigure} &
        \begin{subfigure}[t]{0.23\textwidth}
            \centering
            \includegraphics[width=\textwidth]{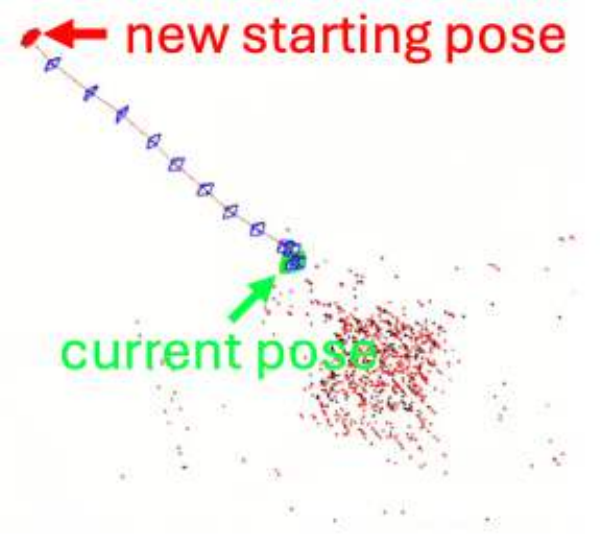}
            \caption{time=3: device moves normally from new origin}
            \label{subfig:firsthalf_snapback_frame4}
        \end{subfigure} 
    \end{tabular}
    \caption{Detailed visualization of snapback attack in ORB-SLAM3. The acoustic attack ends at time=2 and snapback occurs. The scatter points represent visual features found in the real-world environment.
    }
    \label{fig:orbslam_snapback}
\end{figure*}

One key finding is that when the perturbation is large (\eg larger than $6.1\,\text{m/s}^2$ for the x-axis of the accelerometer or larger than $1.6\,\text{rad/s}$ for the gyroscope axis), the device's estimated pose will go back to zero.
As seen in \Cref{subfig:orbslam_acc_constant_1}, the device's estimated pose remains at the origin (0,0,0).
We call this the \textbf{Snapback attack}.
Regarding timing, we find that the snapback effect mainly happens at the onset or termination of the sound.
\Cref{fig:orbslam_snapback} shows a more detailed visualization of the snapback effect, where the snapback happens when the sound ends at time=2.

The snapback effect happens because only the IMU is perturbed and the camera is not; the mis-matched sensor readings confuse ORB-SLAM3 and cause it will lose track of its pose estimate, calling a failure recovery function that re-initializes the map and setting the device pose to its default at the origin~\cite{campos2021orb}.
This kind of failure recovery is common in commercial SLAM frameworks, such as that deployed in the Hololens 2, and can cause the snap back effect as we will show later in \Cref{sec:exp-hololens2}.

We note that even though the change in the estimated pose corresponds to the direction of the perturbation when we apply it on $x$-axis of the accelerometer as shown in \Cref{subfig:orbslam_acc_constant_1}, the result in \Cref{subfig:orbslam_acc_constant_6,subfig:orbslam_gyro_constant_3} shows that this is not always the case.
For example, in \Cref{subfig:orbslam_acc_constant_6}, adding perturbations to the forward-backward direction produces incorrect pose estimates in the up-down direction.
We hypothesize that this is due to the complexity of VIO-SLAM processing, where the final pose estimate depends not only on the IMU readings but also on camera images through a series of non-linear optimizations~\cite{campos2021orb}.
Our real experiments later on (\Cref{sec:exp-hololens2}) also confirm this non-intuitive mapping between the axis of acoustic injection and the axis of pose mis-estimation.
Therefore, we argue that the attacker needs to conduct careful profiling ahead of time if he wants to create wrong pose estimates in certain directions for a misleading attack.

\subsubsection{Vulnerability of XR Headsets to Acoustic Attacks}
\label{sec:frequency_sweep}

While the constant IMU perturbations simulated in \Cref{subsubsec:exp-orbslam3-constant} can experimentally be demonstrated through fine-grained tuning of the acoustic signal's amplitude and phase~\cite{trippel2017walnut,tu2018injected}, such fine control is difficult to achieve in practical scenarios, due to the IMU sensor being embedded in the headset, the difficulty of determining the phase offset, unstable environment factors, etc.
Therefore, we need to create a more realistic model of IMU perturbations by characterizing what perturbations are possible on a real XR headset.
To do this, we subject the Hololens 2 and Quest 3 to acoustic waves at varying frequencies (this subsection), and use this data to create a better model of IMU perturbations (next subsection).

\begin{figure}[h]
    \centering
    \includegraphics[width=0.5\textwidth]{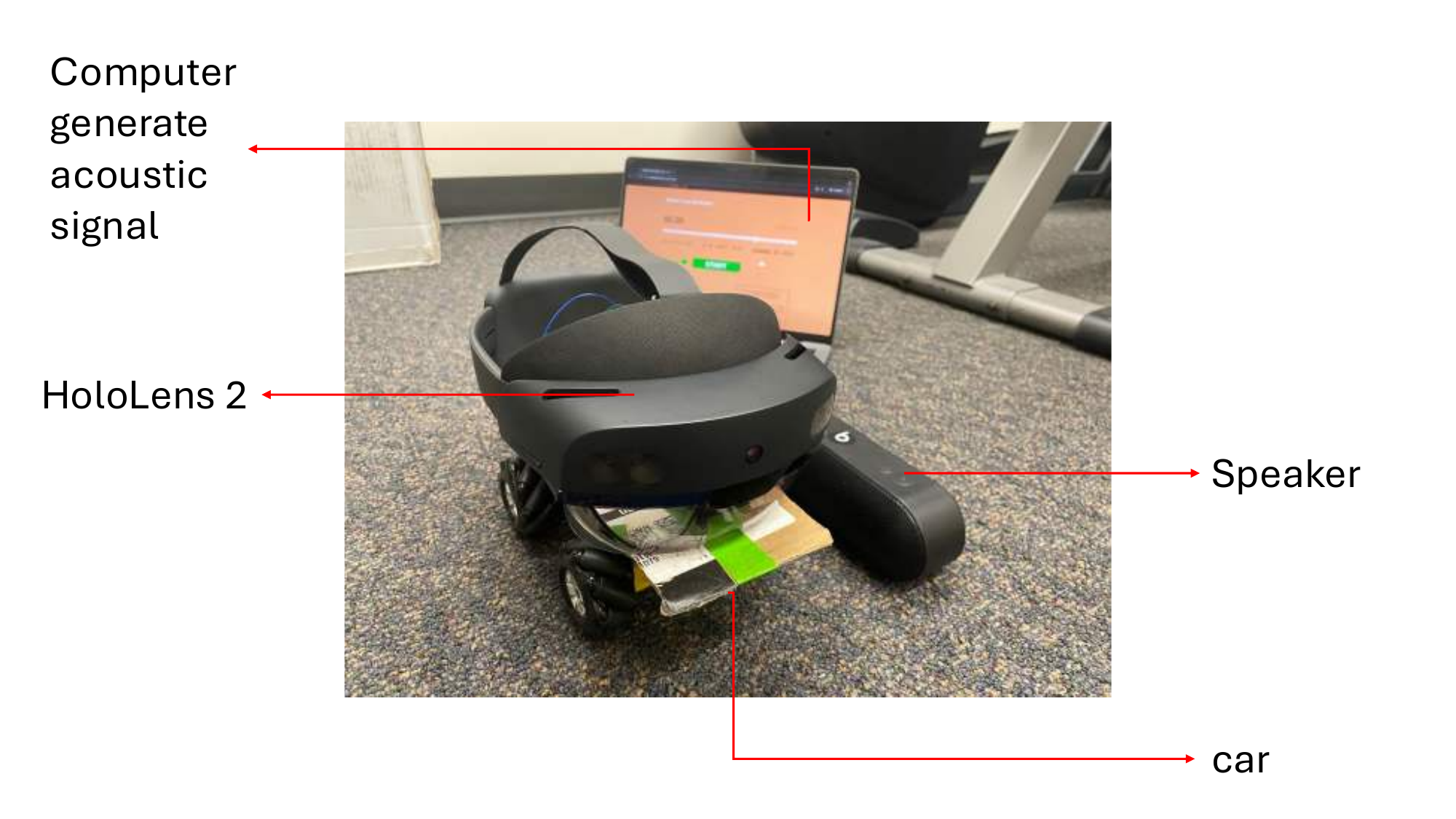} 
    \caption{Experimental setup with XR headset, speaker and sound source, and remote-control car for mobility.}
    \label{fig:exp_setup}
\end{figure}

\textbf{\emph{Setup.}}
We use the experimental setup depicted in \Cref{fig:exp_setup} to test the effects of the acoustic attack on the IMU embedded in HoloLens 2 \cite{ungureanu2020hololens}.
Specifically, a portable speaker (Beats Pill+) plays a pre-generated acoustic signal produced from the laptop, while the Hololens 2 remains stationary.
We verified the output of the portable speaker compared to a function generator (Agilent 33220A) plus amplifier and found little difference, so we used the portable speaker for ease of use.
The speaker is placed to the front or right of the headset at a distance of 5-10 cm.
The pre-generated acoustic signal sweeps across a frequency range from 2-30 kHz, with a step size of 50 Hz, playing for 30 seconds at each step, at a volume of 85 dB.
We log the accelerometer's and gyroscope's readings using the \texttt{hl2ss} library~\cite{dibene2022hololens}.

\begin{figure*}[h]
    \centering
    \begin{tabular}{ccc}
        \begin{subfigure}[b]{0.3\textwidth}
            \centering
            \includegraphics[width=\textwidth]{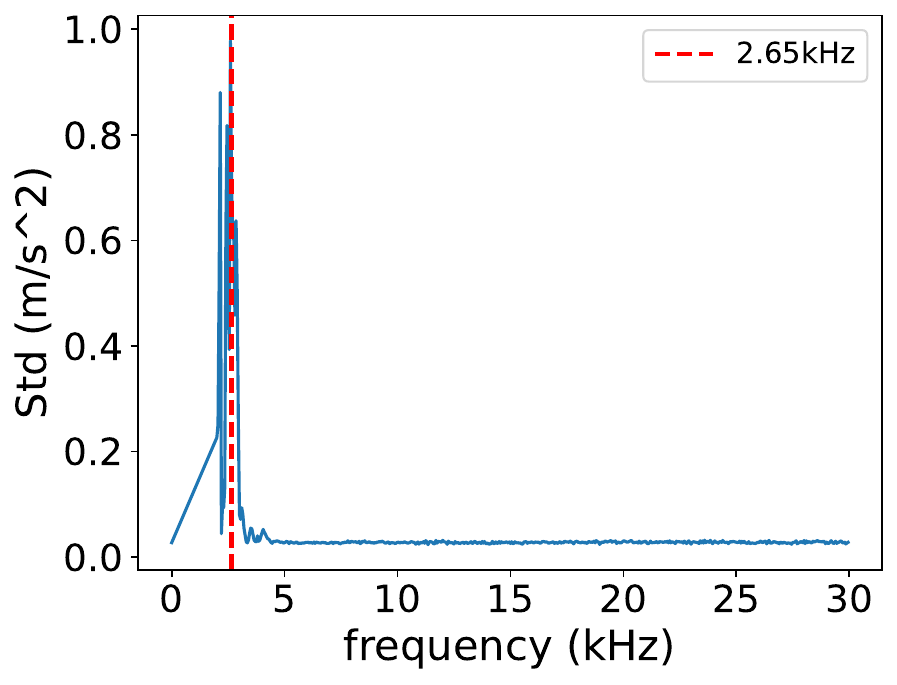}
            \caption{Accelerometer X-axis}
            \label{subfig:right_std_1}
        \end{subfigure} &
        \begin{subfigure}[b]{0.3\textwidth}
            \centering
            \includegraphics[width=\textwidth]{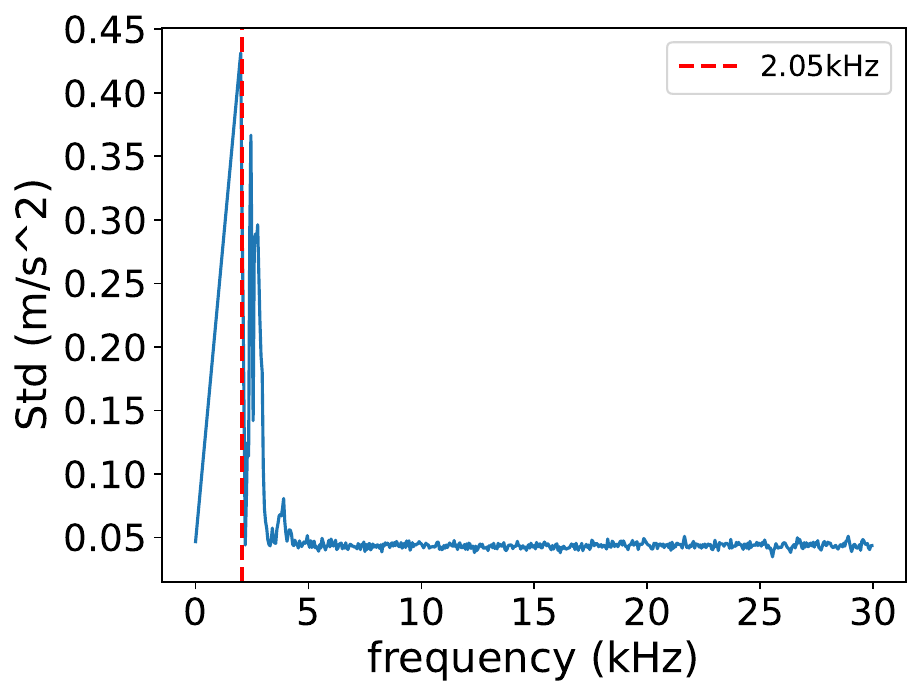}
            \caption{Accelerometer Y-axis}
            \label{subfig:right_std_2}
        \end{subfigure} &
        \begin{subfigure}[b]{0.3\textwidth}
            \centering
            \includegraphics[width=\textwidth]{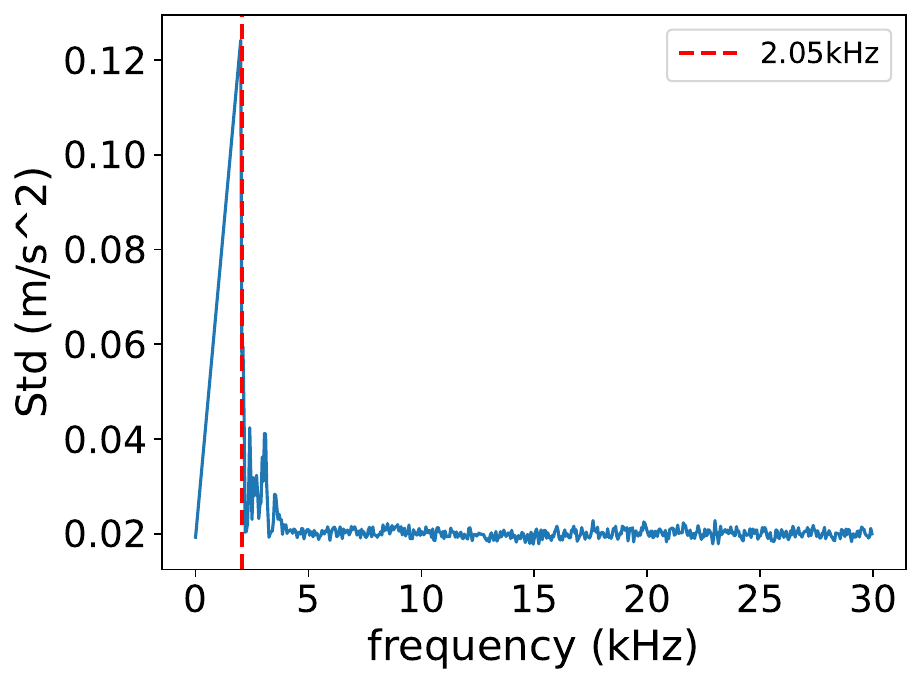}
            \caption{Accelerometer Z-axis}
            \label{subfig:right_std_3}
        \end{subfigure} \\
        
        \begin{subfigure}[b]{0.3\textwidth}
            \centering
            \includegraphics[width=\textwidth]{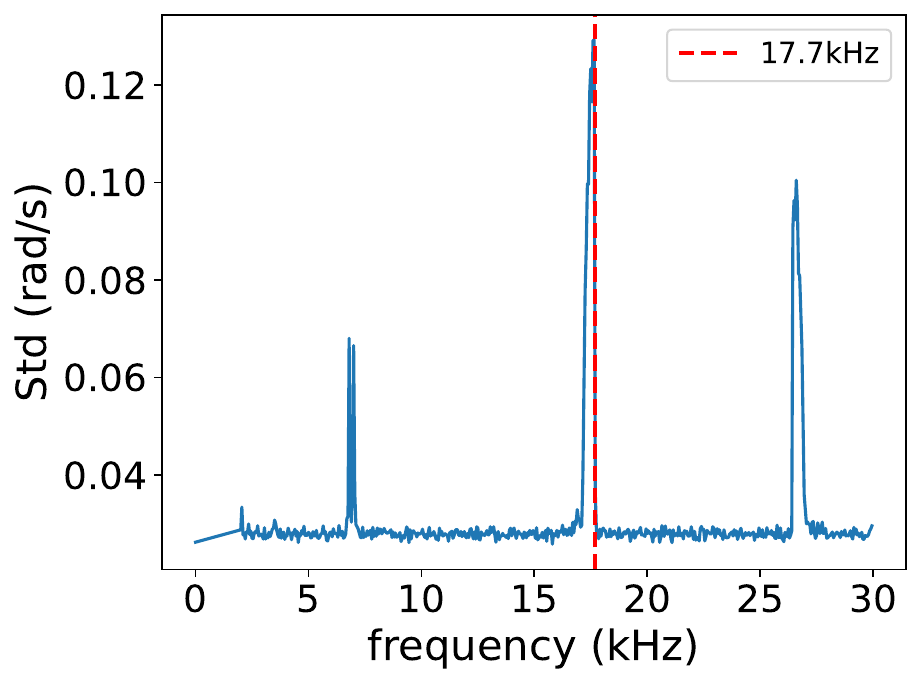}
            \caption{Gyroscope X-axis}
            \label{subfig:right_std_4}
        \end{subfigure} &
        \begin{subfigure}[b]{0.3\textwidth}
            \centering
            \includegraphics[width=\textwidth]{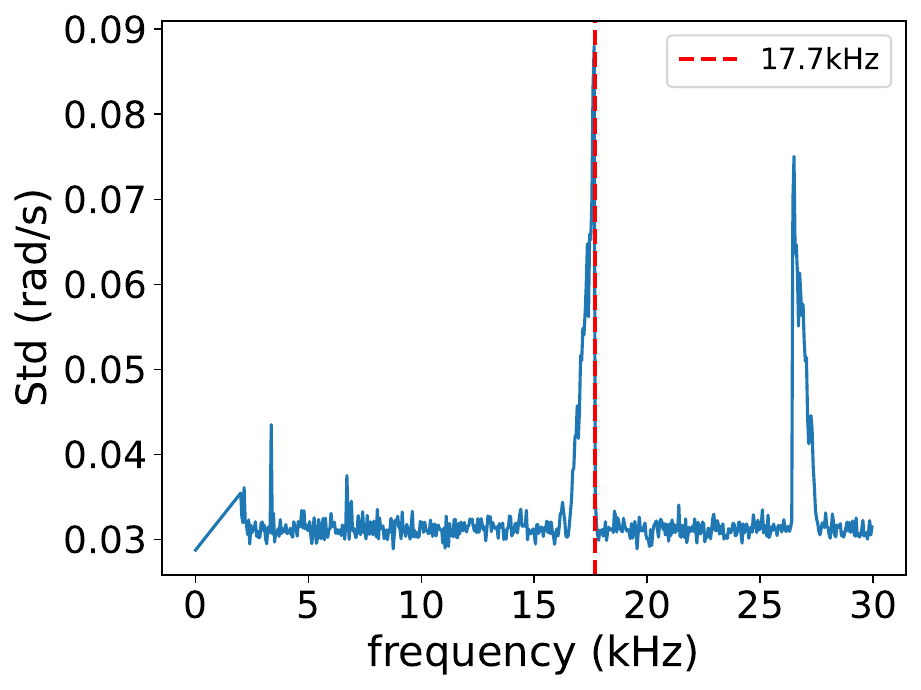}
            \caption{Gyroscope Y-axis}
            \label{subfig:right_std_5}
        \end{subfigure} &
        \begin{subfigure}[b]{0.3\textwidth}
            \centering
            \includegraphics[width=\textwidth]{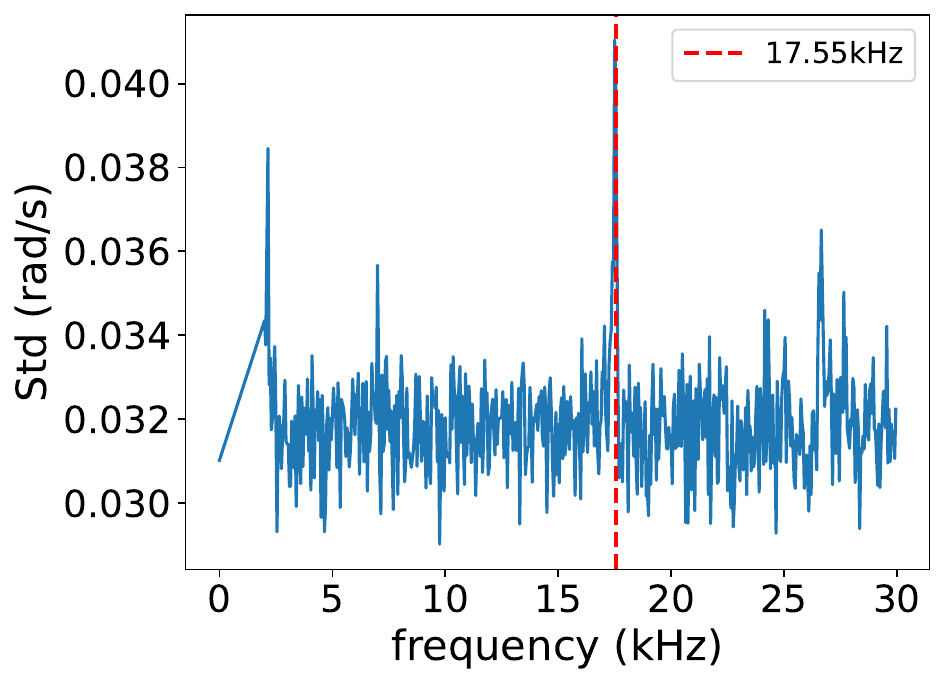}
            \caption{Gyroscope Z-axis}
            \label{subfig:right_std_6}
        \end{subfigure}
    \end{tabular}
    \caption{Frequency response of the Hololens 2 IMU. The red dashed line shows the resonant frequency where large changes in the sensor readings occur. The accelerometer is vulnerable at 2-2.6 kHz and the gyroscope at 17.7 kHz.
    }
    \label{fig:right_std}
\end{figure*}

\textbf{\emph{Results of frequency sweep.}}
\Cref{fig:right_std} show the response of the headset's IMU to acoustic signals played at different frequencies.
We plot the mean and standard deviation of the accelerometer and gyroscope readings, corresponding to potential output bias or output control vulnerabilities~\cite{trippel2017walnut}.
From the results, we observe that there are multiple spikes in the mean or standard deviation values, for different axes, for different sensors. 
We find that the resonant effect is more significant in terms of the standard deviation, rather than the mean, suggesting an output biasing vulnerability (see \Cref{sec:background}.
The resonant frequency is 2.65 kHz, 2.05 kHz, and 2.05 kHz for the accelerometer's $x,y,z$-axis, respectively, and 17.7 kHz, 17.7 kHz, and 17.55 kHz for the gyroscope $x,y,z$-axis.
This aligns with previous findings \cite{jeong2023rocking} that the resonant frequency for the accelerometer is in the lower range of human hearing range and the resonant frequency for the gyroscope is close to or in the ultrasonic range.
We also performed frequency sweeping for the Meta Quest 3,  but it does not have obvious spikes in terms of mean and standard deviation values. We hypothesize that that this might be the result of the physically enclosed case around the sensors, or the internal positioning of the IMU.


\subsubsection{Data-Driven IMU Perturbations} 
\label{subsubsec:orb-slam gmm}

\begin{figure}[h]
    \centering
        \begin{subfigure}[b]{0.22\textwidth}
            \centering
            \includegraphics[width=\textwidth]{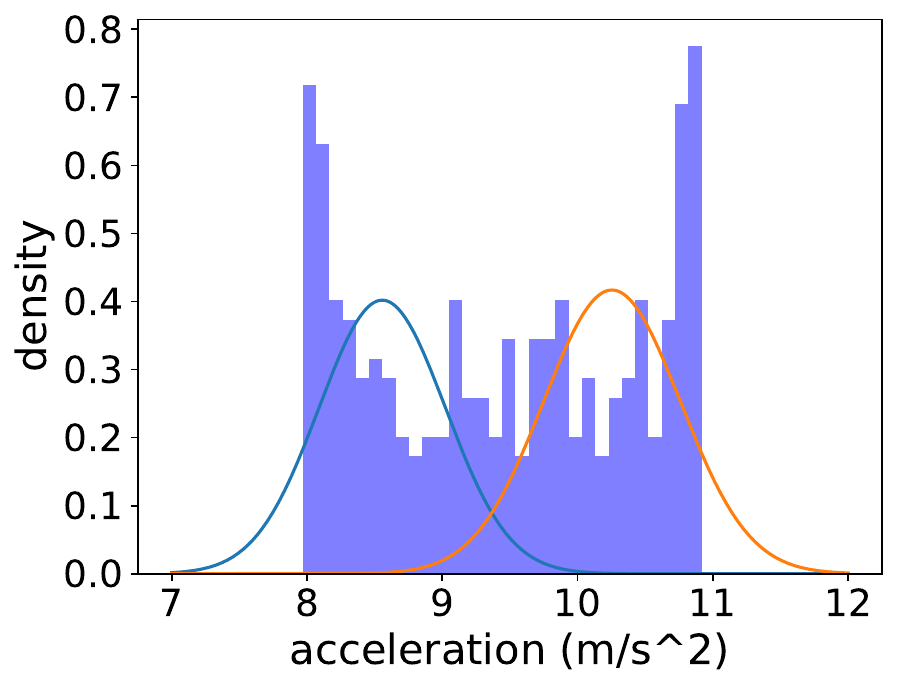}
            \caption{Accelerometer X-axis}
            \label{subfig:right_distribution_1}
        \end{subfigure}
        ~
        \begin{subfigure}[b]{0.22\textwidth}
            \centering
            \includegraphics[width=\textwidth]{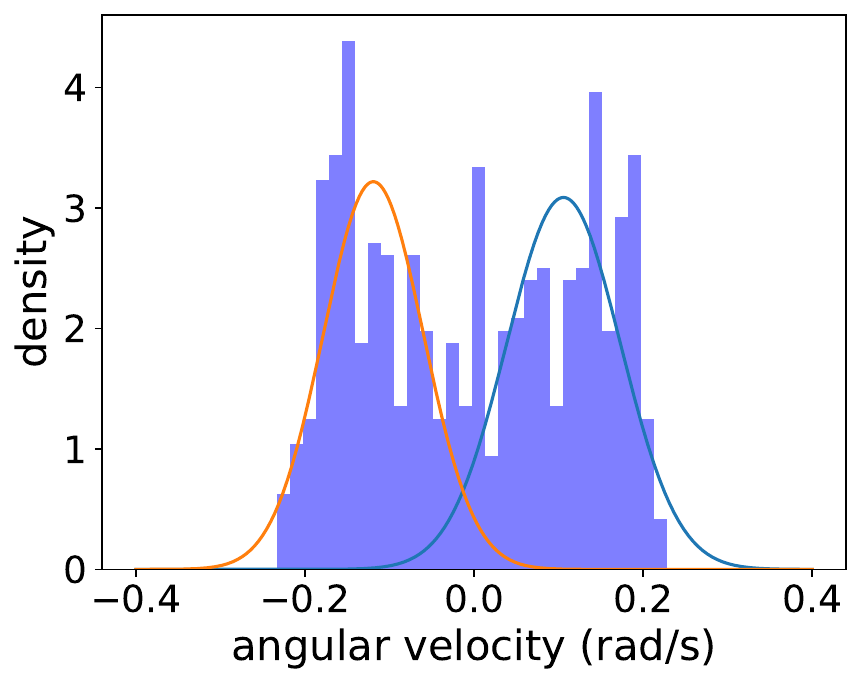}
            \caption{Gyroscope X-axis}
            \label{subfig:right_distribution_4}
        \end{subfigure} 
    \caption{Modeling the distribution of IMU sensor readings of the HoloLens 2 IMU using a Gaussian mixture model, under attack by their resonant frequency.}
    \label{fig:right_distribution}
\end{figure}
With the knowledge gleaned from the experiments on the Hololens 2 in the preceding subsection, we next seek to create a model of the IMU perturbations that can be realized in practice on the headset.
To do this, we plot the distribution of the sensor readings in \Cref{fig:right_distribution}.
Without attack, we would expect the readings to be around 0 (or 9.8 m/s$^2$ for gravity), but with the acoustic attack, the sensor readings exhibit a spread.
We fit the data to a Gaussian Mixture Model (GMM) using the expectation-maximization algorithm.

Based on the fit of the GMMs, we choose a plausible range of mean and standard deviation values for the perturbed IMU readings.
Specifically, for the accelerometer, we set the standard deviation to 0.1 and the mean from 0 to $6.1\,\text{m/s}^2$; for the gyroscope, we set the same standard deviation and the mean from 0 to $1.6 \,\text{rad/s}$.
These ranges of values are based on a combination of our own measurements and prior reported data~\cite{jeong2023rocking}.

\begin{figure}[h]
    \centering
    \includegraphics[width=0.42\textwidth]{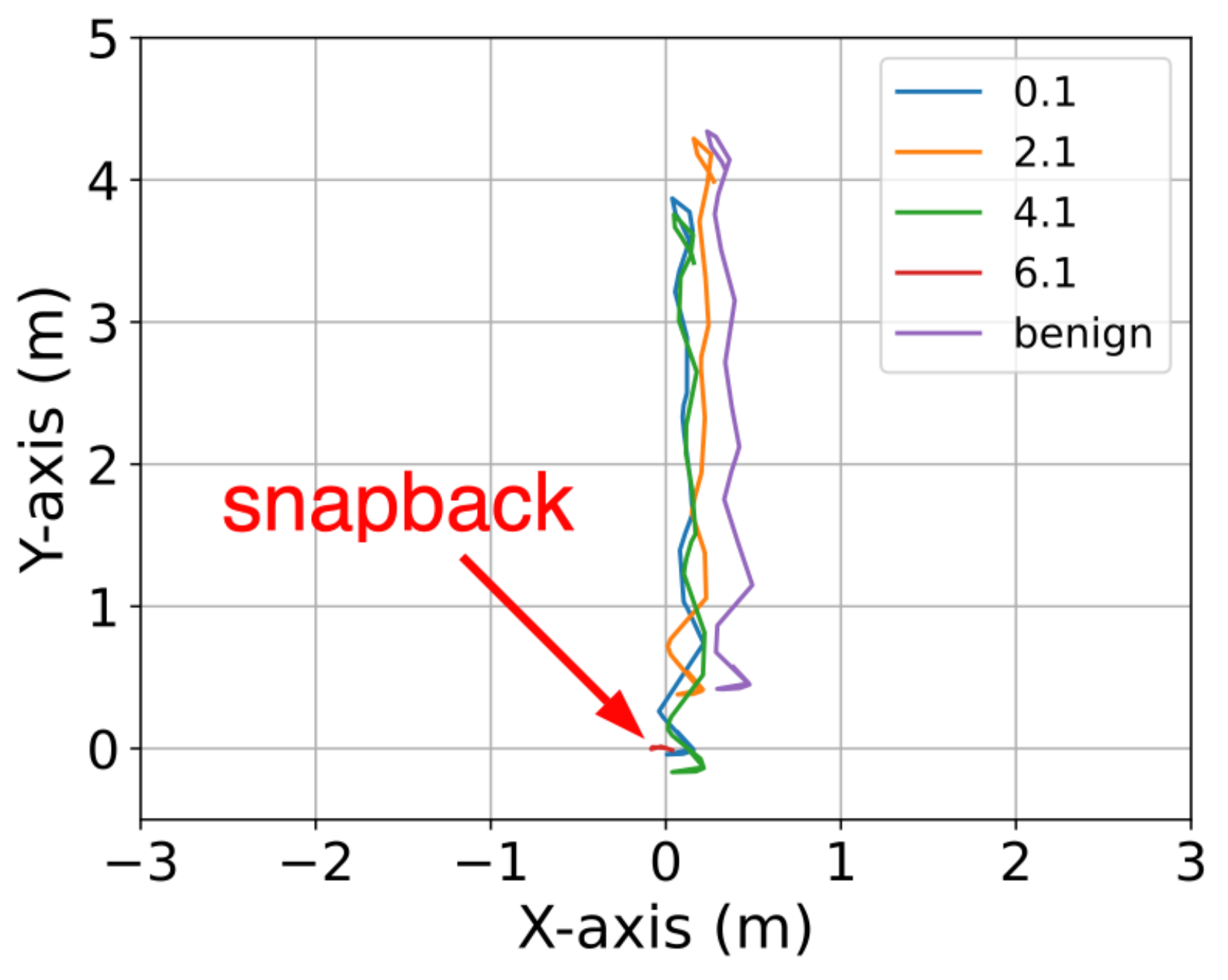}
    \caption{
    Device pose estimated by ORB-SLAM3 under data-driven GMM perturbation on the IMU readings.
    Legend denotes absolute value of GMM mean.
    Beyond a threshold (6.1), the \textbf{snapback} effect occurs.}
    \label{fig:orbslam_acc_gmm}
\end{figure}

To use this data-driven GMM model of IMU perturbations, 
we take sampled values from the GMM and add them to the corresponding IMU readings from the user trace.
We plot the impact of this perturbed IMU on the pose estimates of ORB-SLAM3 in \Cref{fig:orbslam_acc_gmm} for accelerometer inputs (gyroscope results are similar).
This more realistic model also exhibits the snapback effect from \Cref{subsec:exp-orbslam3} and \Cref{fig:orbslam_acc_constant}: namely, when the magnitude of the perturbations exceeds a threshold (in the GMM case, a mean of 6.1 m/s$^2$), the snapback effect occurs and the device re-initializes its pose to the origin.
For smaller mean perturbations, we observe less of a misleading effect, with up to 0.5 meters difference from the ground truth trajectory.
Overall, these results provide further evidence that ORB-SLAM3 based pose estimation will exhibit snapback effects when under acoustic attacks.

\subsection{Attacks on ILLIXR}
\label{subsec:exp-illixr}

The processing pipelines in commercial XR headsets are typically closed-source.
Therefore, to understand the range of possible effects from acoustic attacks, we experiment with another open XR research testbed, ILLIXR ~\cite{huzaifa2022illixr}.
ILLIXR uses a different pose estimation method, OpenVINS~\cite{geneva2020openvins}.
We inject the same two types of perturbation as in \Cref{subsec:exp-illixr}, constant perturbation, and data-driven perturbations.
We apply the perturbations to traces from a standard SLAM dataset, EuRoC~\cite{Burri25012016}. 
These trajectories are more complex than the ones we studied for ORB-SLAM3, enabling us to examine more complex effects of acoustic attacks.

\subsubsection{Constant IMU Perturbations} 

\begin{figure}[h]
    \centering
        \begin{subfigure}[b]{0.22\textwidth}
            \centering
            \includegraphics[width=\textwidth]{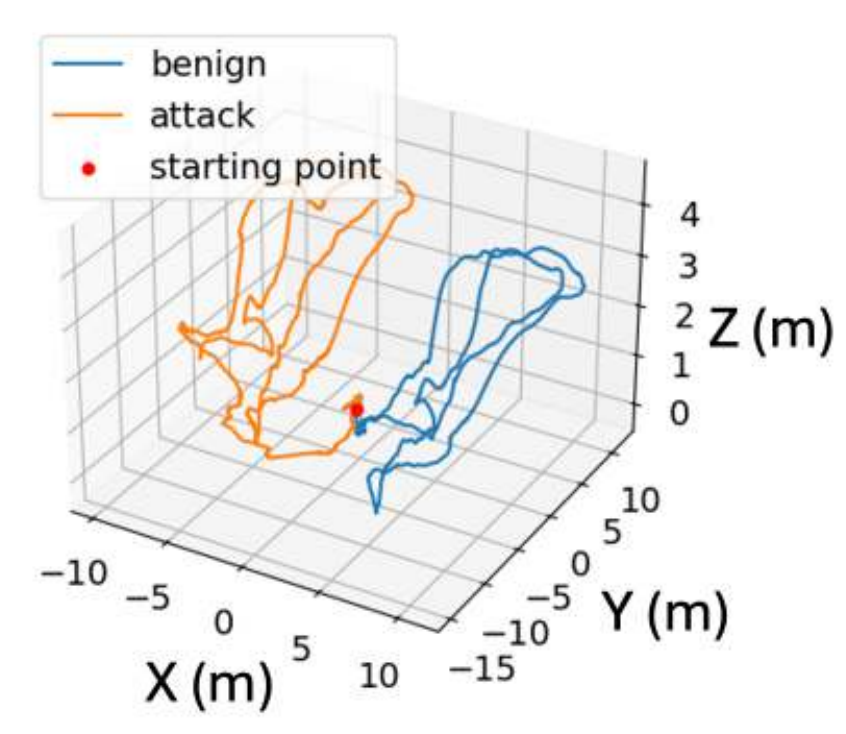}
            \caption{+1 m/s$^2$ on the accelerometer X-axis}
            \label{subfig:illixr_acc_x_constant_3}
        \end{subfigure} ~
        \begin{subfigure}[b]{0.22\textwidth}
            \centering
            \includegraphics[width=\textwidth]{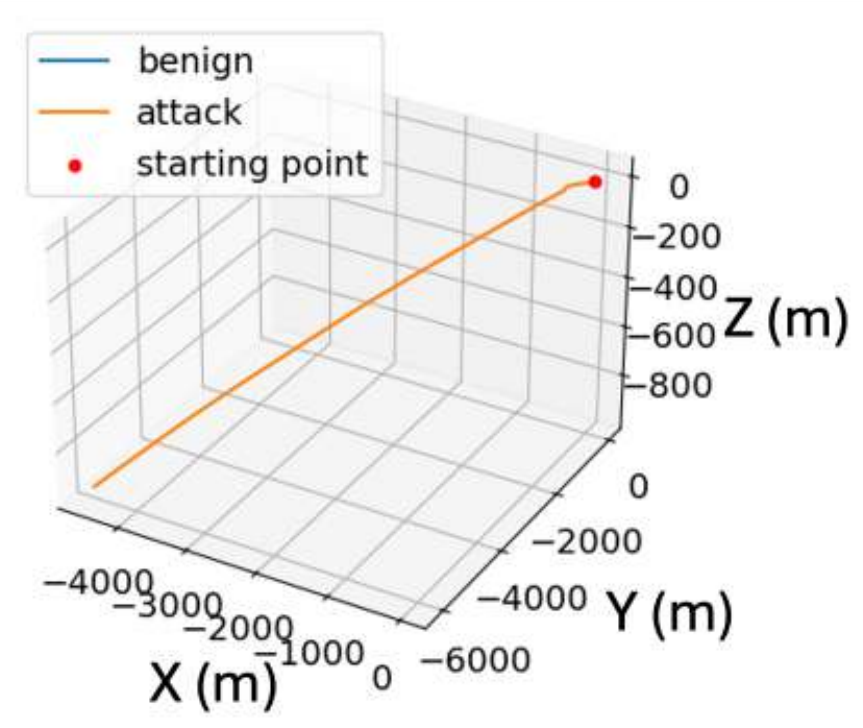}
            \caption{+2 m/s$^2$ on the accelerometer X-axis}
            \label{subfig:illixr_acc_x_constant_4}
        \end{subfigure}
    \caption{
    Device pose estimated by ILLIXR under constant perturbation on the accelerometer.
    For larger perturbation (right), the device loses track of its pose and the \textbf{drift away} effect occurs.
    }
    \label{fig:illixr_acc_constant}
\end{figure}
\begin{figure}[h]
    \centering
        \begin{subfigure}[b]{0.23\textwidth}
            \centering
            \includegraphics[width=\textwidth]{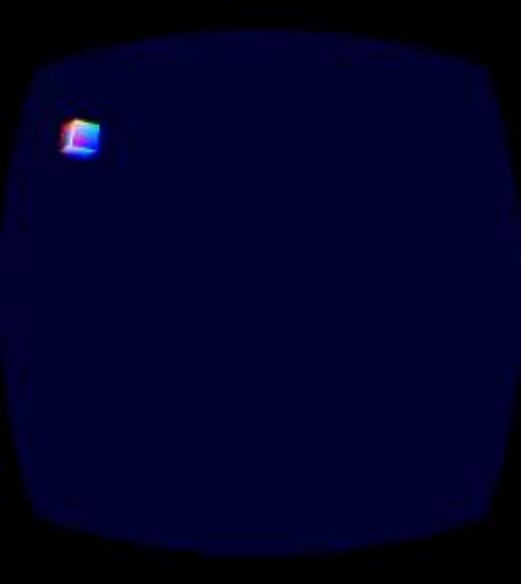}
            \caption{time=0}
            \label{subfig:illixr_driftaway_1}
        \end{subfigure}~
        \begin{subfigure}[b]{0.23\textwidth}
            \centering
            \includegraphics[width=\textwidth]{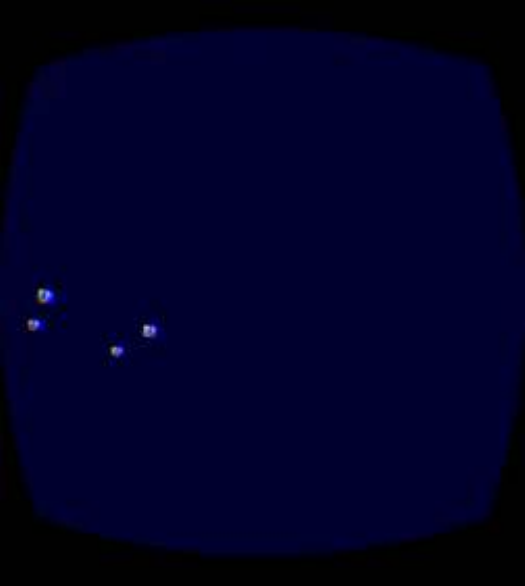}
            \caption{time=1}
            \label{subfig:illixr_driftaway_2}
        \end{subfigure}
    \caption{Visualization of \textbf{drift away} attack in ILLIXR (virtual cube flies away) when the device's pose estimation does not reset to the origin to recover from tracking failure.}
    \label{fig:illixr_driftaway}
\end{figure}

For illustrative purposes, we focus on a particular trace from the EuRoC dataset (MH05), due to its relative simplicity compared to other traces in the dataset.
\Cref{fig:illixr_acc_constant} shows the effect of adding a small constant perturbation (+1 m/s$^2$) to the x axis of the accelerometer.
The results in \Cref{subfig:illixr_acc_x_constant_3} show that the misleading attack is possible in one direction, with the new trajectory being precisely offset from the ground truth.
However, different from the snap back attack observed on ORB-SLAM3, with ILLIXR, when the perturbation is larger (+2 m/s$^2$ in \Cref{subfig:illixr_acc_x_constant_4}), 
pose estimation will fail, causing the \textbf{Drift away attack}.
This means that the device's estimated pose drifts away to infinity.
In terms of the effects on the AR display, screenshots from ILLIXR of the drifting away effect are shown in \Cref{fig:illixr_driftaway}, where the virtual object (colorful cube) flies away out of view.
The differing responses of ORB-SLAM3 and ILLIXR to IMU perturbations are due to how the pose estimation modules handle tracking loss.

ILLIXR uses OpenVINS \cite{geneva2020openvins}, which does not re-initialize the device pose (\ie reset the pose to (0,0,0)) when the estimated pose is deemed unreliable, while ORB-SLAM does.

\subsubsection{Data-Driven IMU Perturbations} 
\begin{figure}[h]
    \centering
    \includegraphics[width=0.45\textwidth]{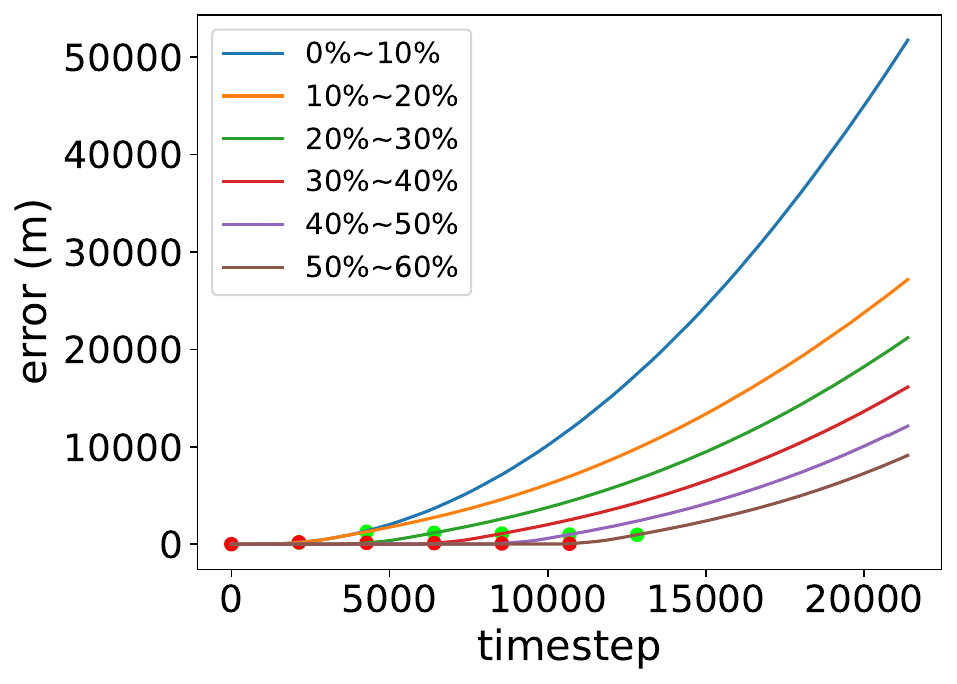}
    \caption{Effect of attack timing on the \textbf{drift away} attack (start=red dot, end=green dot), measured by feeding GMM-perturbed IMU readings to ILLIXR. Position error grows exponentially towards the end of the trace. 
    }
    \label{fig:illixr_acc_gmm}
\end{figure}
Using the same GMM model as in  \Cref{subsubsec:orb-slam gmm}, we apply the perturbations for different time frames in the overall trace (from $0\%$ to $10\%$, $30\%$ to $40\%$, $50\%$ to $60\%$ and $70\%$ to $80\%$). 
The goal is to understand whether the drift away attack occurs nearly instantaneously when the IMU perturbations start/stop (as they do with ORB-SLAM3), or whether there is a time delay to see the effects of the acoustic attack.
The time series of the device's estimate pose are shown in \Cref{fig:illixr_acc_gmm}.
We observe two effects: \emph{\textbf{(1) Time Delay}}: When we apply the perturbation for a short time period, whether it is at the beginning or in the middle (\eg [0\%,10\%] or [30\%,40\%]), the pose estimation is still normal during the attack, but it will cause problems after some time (\eg around timestep 10,000); \emph{\textbf{(2) Exponential Drift Error}}: Even though no acoustic attack is present towards the end of the trace, the pose error will increase exponentially during the drift away attack. 
We believe both of these effects are due to the integrations during state estimation in SLAM, which cause errors to accumulate over time before eventually exploding.

\section{Proof-of-Concept Attacks on HoloLens}
\label{sec:exp-hololens2}

In this subsection, we use the knowledge gained from the attacks on ORB-SLAM3 (\Cref{subsec:exp-orbslam3}) and ILLIXR (\Cref{subsec:exp-illixr}) to demonstrate an end-to-end attack on a commercial XR headset.
We demonstrate several types of attacks on the Microsoft HoloLens 2~\cite{ungureanu2020hololens} using adversarial acoustic signals.
Our main finding is that we are able to replicate the snap back attack on the real device and leverage this effect to carry out four distinct proof-of-concept attacks.




\subsection{Experimental Setup} 

The experimental setup is the same as \Cref{sec:frequency_sweep}, including distance and volume, plus some additional functionality.
Along with the stationary settings discussed earlier, we also wish to experiment with user motion.
To do this in a repeatable and controlled fashion, we place the Hololens 2 on a remote-controlled toy car whose movement can be programmed and synchronized with the acoustic signals. 
The proof-of-concept AR applications are developed using Unity version 2022.3.15f1~\cite{UnitySDK} and the Microsoft Mixed Reality Toolkit (MRTK) version 2.8.3~\cite{mrtk}. During the experiments, we log the traces (including position and rotation) of both the headset and 3D objects in each application using the \texttt{hl2ss} interface~\cite{dibene2022hololens}, while also recording the renderings generated by the headset.

\subsection{Validation of Snapback Attack}
\label{sec:snapback_effect}

\begin{figure}[h]
    \centering
    \includegraphics[width=0.45\textwidth]{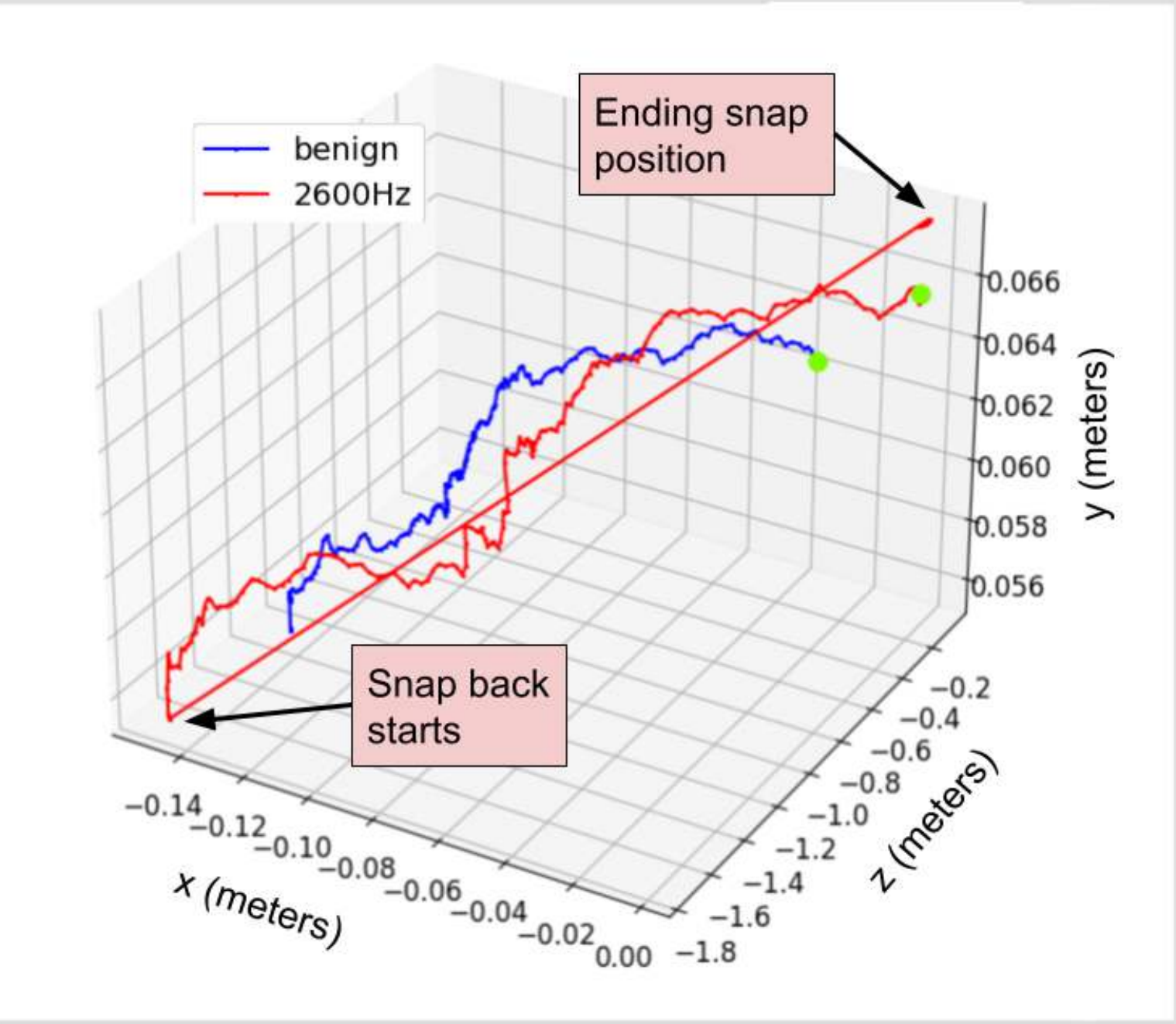}
    \caption{Example of of \textbf{snapback} effect on a Hololens 2 under acoustic injection attack. In the benign case, the headset starts at the green dot, moves forward, and stops. When a 2.6 kHz tone plays, the position of the headset snaps back.}
    \label{fig:holo_acc_basic}
\end{figure}

\textbf{Attack validation.}
Our experiments show that the snapback effect, simulated on ORB-SLAM3  (\Cref{subsec:exp-orbslam3}), can be recreated on the Hololens 2 as follows.
The headset was placed on a stationary remote-controlled car and its pose logged for 10 seconds in silence.
The car with the headset on top of it then drove forward for 10 seconds, while the speaker played a single 10-second tone at 2600 Hz (the resonant frequency of the accelerometer from \ref{sec:frequency_sweep}). Another 10 seconds of silence followed after the car and headset stopped moving.
\Cref{fig:holo_acc_basic} shows a sample trajectory of the headset without the acoustic attack (blue line) and with the acoustic attack (red line).
The headset suddenly estimated its position as (0, 0) after the headset and sound paused.
We repeated this experiment multiple times and achieved success rates $>90\%$.

The key difference between making the snapback attack work in practice, compared to the ORB-SLAM3 simulation, was that the headset should be moving when the acoustic attack occurs.
When the headset was stationary and the acoustic sound was played, we did not observe any effect.
We hypothesize that this is because when stationary, the headset is able to filter out the effects of the acoustic attack as noise. However, when the headset was moving there were legitimate IMU signals mixed in with the noise, the headset was unable to filter out the attack and hence generated inaccurate pose estimates.

\begin{figure}[h]
    \centering
    \includegraphics[width=0.45\textwidth]{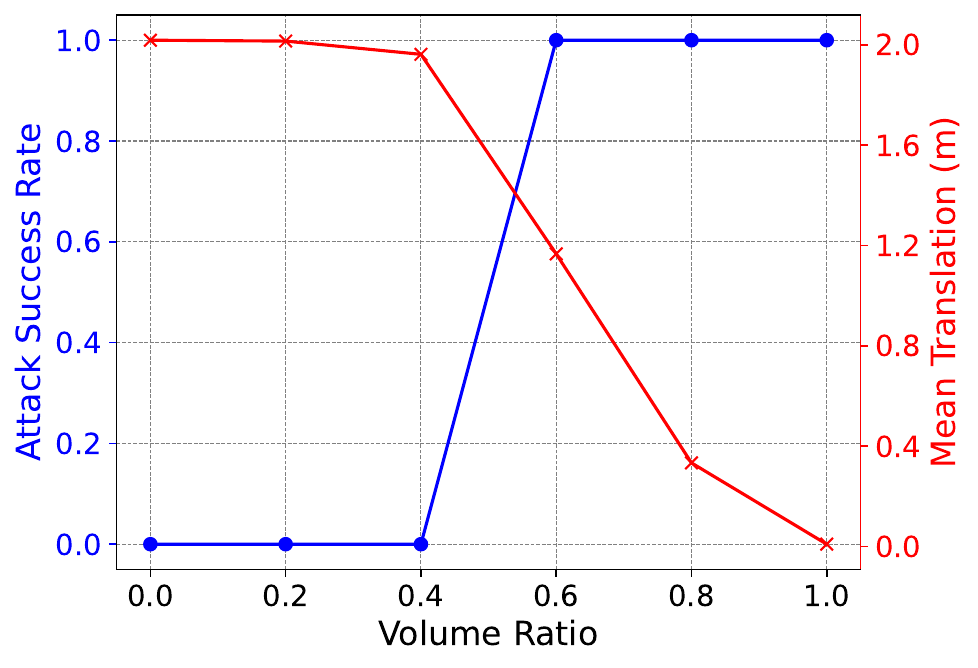}
    \caption{Impact of sound volume on attack success rate and mean translation.
   A successful attack is defined as as \textbf{snapback} happening, and the mean translation is the average headset motion before the snapback occurs.
   A volume ratio of 1.0 corresponds to 85 dB.}
    \label{fig:volume_ASR}
\end{figure}

\textbf{Impact of sound volume.}
We also study the impact of the volume of the acoustic signal and the attack success rate (ASR).
Specifically, we run the same experiment as above, but with each trial at a different volume ratio, relative to the maximum volume of the speaker.
We run 5 trials at each volume and log the fraction of trials that the \textbf{snapback} occurs. In \Cref{fig:volume_ASR}, we observe that there is a threshold of signal power (around 50\% of the maximum speaker volume), below which the attack fails. 
We hypothesize that this is because at low volumes, the resulting small perturbation on the IMU readings can be compensated for by other sensors' observations and corrected by SLAM during sensor fusion, reducing the snapback effect. 

\textbf{Impact of background music.}
Since the resonant frequency of the Hololens 2 is in the audible range, we also conducted experiments to test the effectiveness of 
acoustic attack when masked by background music. 
In detail, we played a fast-paced rock song (``In the End (Instrumental)'' by Linkin Park) alongside the resonant sound on the same speaker at 85 dB, 
which we qualitatively observed can hide the resonant sound to a large extent.
On the Hololens, we observed the same re-initialization and snapback effect.

\begin{figure}
    \centering
    \includegraphics[width=0.45\textwidth]{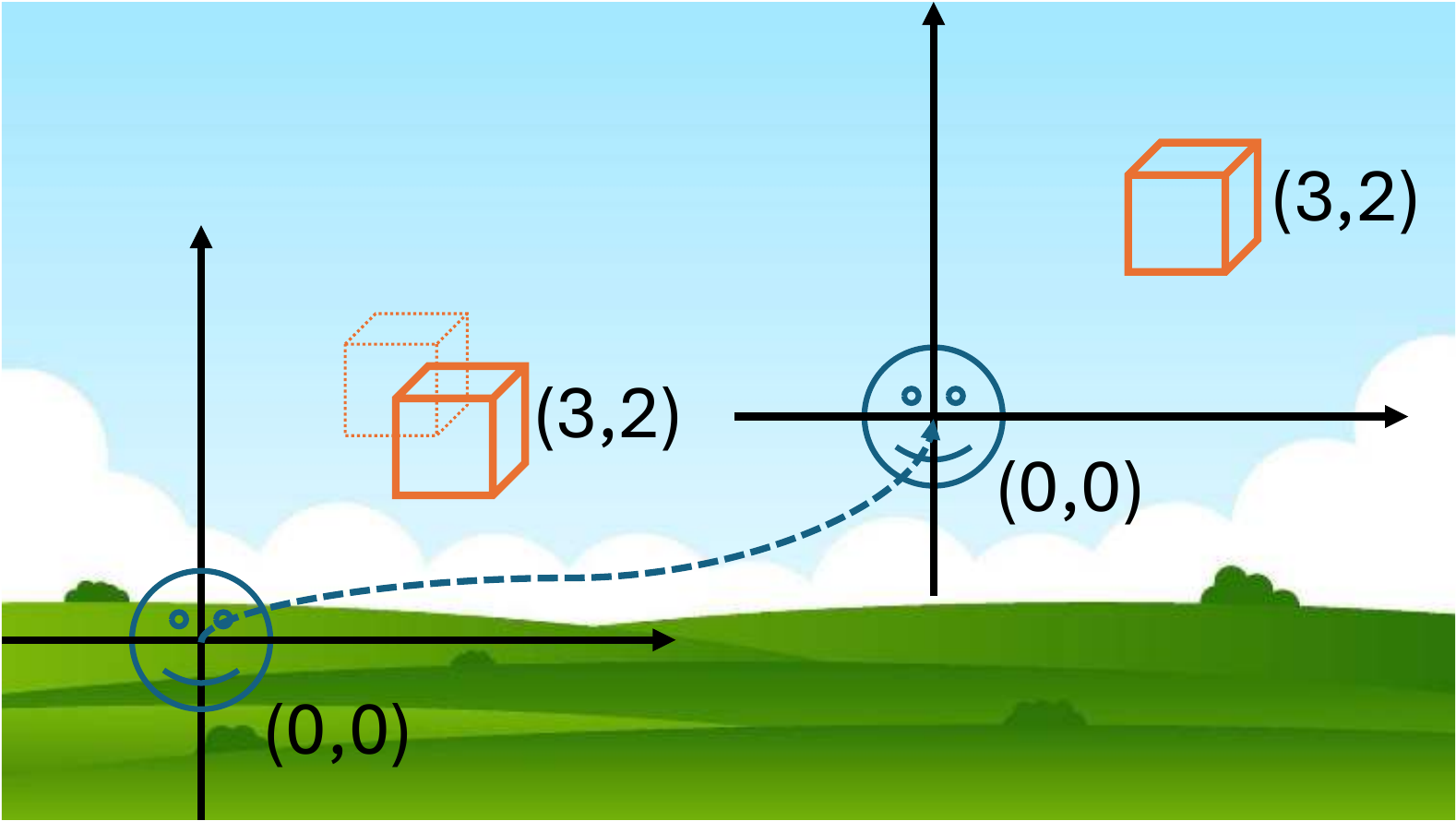}
    \caption{Effect of \textbf{snapback} effect on rendered virtual content. Normally, the world coordinates (black axes) should remain fixed on the grass. Under acoustic attack, when the user (blue smiley) moves across the grass (blue dashed line), the world coordinates reset and cube remains stuck at (3,2) in the user's display.
    }
    \label{fig:snapback_virtual}
\end{figure}

\subsection{Impact on AR Applications} 
While pose re-initialization is commonly used when tracking is lost, preventing exponential drift in modern SLAM systems~\cite{campos2021orb}, it can create adverse visual and UI effects in XR applications.
Below, we present four demonstration XR apps that illustrate two primary impacts of the snapback effect: (1) \textbf{potential harm to the user} and (2) \textbf{potential benefits to the user}. Note that the ``user'' is defined as the person wearing the headset, and may be either the attacker or the victim.

To build these demonstration applications, we first have to understand the impact of the snapback attack on the visual display.
In \Cref{fig:snapback_virtual}, we illustrate what happens to a virtual object on the user's display during a snapback attack.
After the user moves and during a snapback (when the sound starts or stops), the world coordinate system will re-initialize to (0,0) because of the false readings of IMU caused by the acoustic signal. 
Since the virtual cube is always at (3,2) in world coordinates, it will appear at that same position relative to the user, \ie higher in the sky after the user moved and snapback occurred.
In other words, the virtual cube will appear as head-locked content that is frozen in place on the display, violating XR UI design guidelines~\cite{lebeck2017securing}.

In addition to the head-locked virtual content, we also observe an additional effect that we previously did not find during the simulations: a \textbf{small drift} effect.
Namely, during the acoustic signal, sometimes the virtual objects will have random shifts of a few tens of centimeters, which is illustrated by the cube with dashed lines in \Cref{fig:snapback_virtual}. 
We hypothesize that this is due to slight noise in the pose estimate, but not severe enough to cause snap back (similar to the Misleading effect).
We leverage this small drift effect during several of our demonstration apps.

The four proof-of-concept attacks we demonstrate are summarized in \Cref{table:app_summary}, in terms of user positioning (user is mobile or stationary) and the effect on the virtual content.
In the second column, virtual content should be world-locked (fixed in the real world) but is instead head-locked (fixed on the user's display); for example, in the ``denial of user interaction'' attack, the snapback effect causes an opaque virtual wall to be block the user's display.
In the third column, virtual content drifts unexpectedly in the user's display; for example, in the ``clickjacking attack'', a virtual keyboard that drifts even if the user is stationary.
Details of the attacks follow in the next four subsections.

\begin{table}[]
\begin{tabular}{l|ll|}
\cline{2-3}
 &
  \multicolumn{2}{c|}{\textbf{Undesired effect}} \\ \cline{2-3} 
 &
  \multicolumn{1}{l|}{\textbf{\begin{tabular}[c]{@{}l@{}}Virtual content\\is head-locked\end{tabular}}} &
  \textbf{\begin{tabular}[c]{@{}l@{}}Virtual content\\ drifts\end{tabular}} \\ \hline
\multicolumn{1}{|l|}{\textbf{\begin{tabular}[c]{@{}l@{}}User\\ mobile\end{tabular}}} &
  \multicolumn{1}{l|}{\begin{tabular}[c]{@{}l@{}}Denial of user\\ interaction (\S\ref{sec:app_blocking_wall})\\ Secure zone\\ invasion (\S\ref{sec:app_secure_zone})\end{tabular}} &
  \begin{tabular}[c]{@{}l@{}}Manipulating\\ user input (\S\ref{sec:app_vr_gaming})\end{tabular} \\ \hline
\multicolumn{1}{|l|}{\textbf{\begin{tabular}[c]{@{}l@{}}User\\ stationary\end{tabular}}} &
  \multicolumn{1}{l|}{N/A} &
  Clickjacking (\S\ref{sec:app_clickjacking}) \\ \hline
\end{tabular}
\caption{Summary of proof-of-concept end-to-end attacks}
\label{table:app_summary}
\end{table}

\begin{figure*}[h]
    \begin{minipage}{0.8\textwidth}
        \centering
        \begin{subfigure}{0.3\textwidth}
            \includegraphics[width=\linewidth]{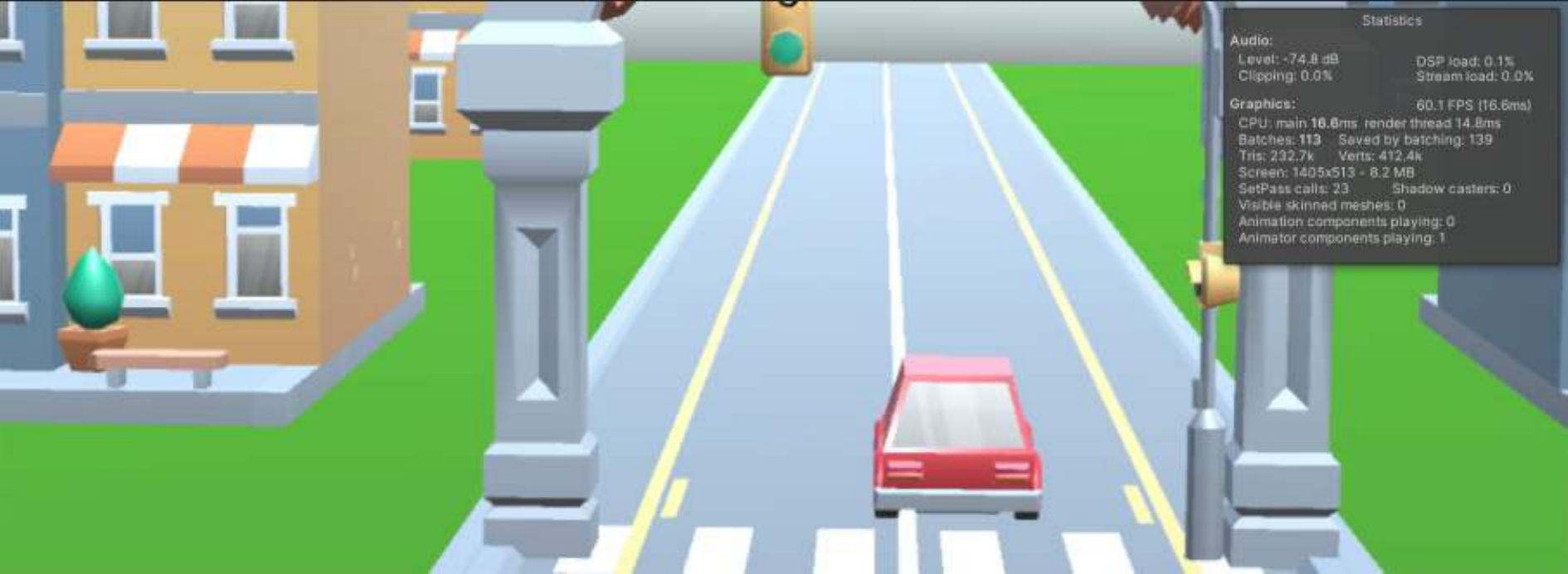}
            \caption{Benign, time=0}
            \label{subfig:car_nosound_frame1}
        \end{subfigure}
        \hfill
        \begin{subfigure}{0.3\textwidth}
            \includegraphics[width=\linewidth]{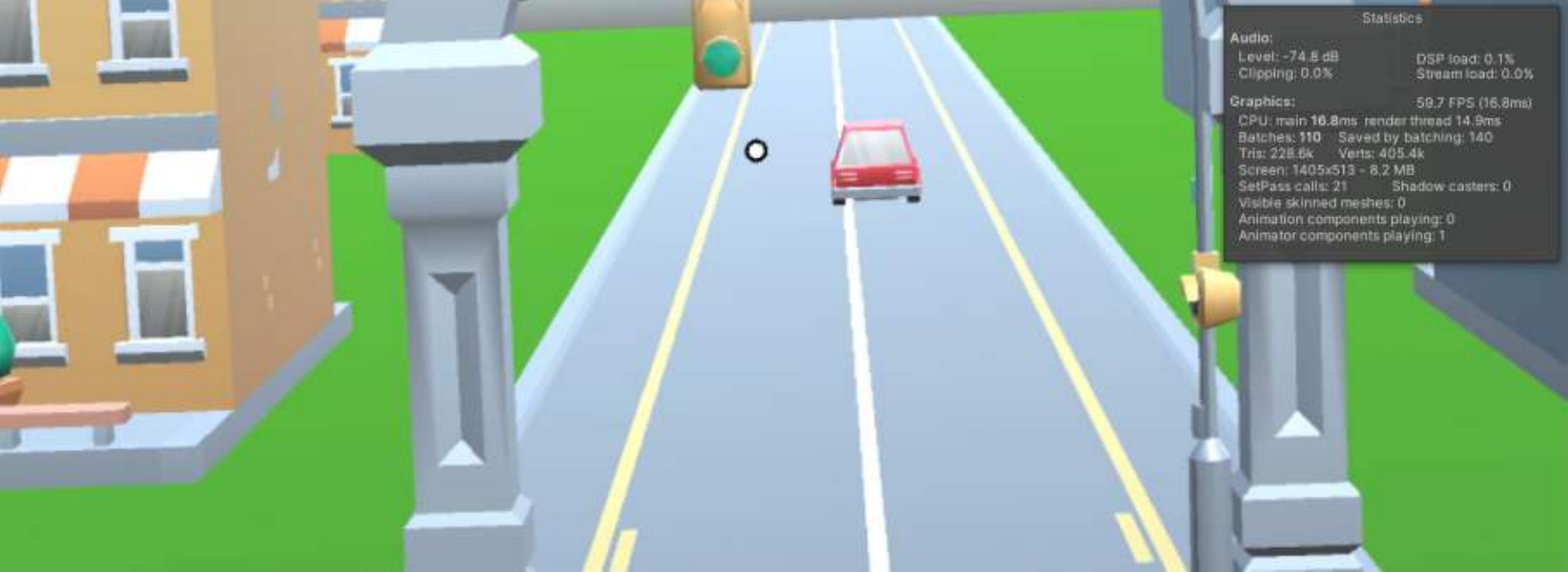}
            \caption{Benign, time=1}
            \label{subfig:car_nosound_frame2}
        \end{subfigure}
        \hfill
        \begin{subfigure}{0.3\textwidth}
            \includegraphics[width=\linewidth]{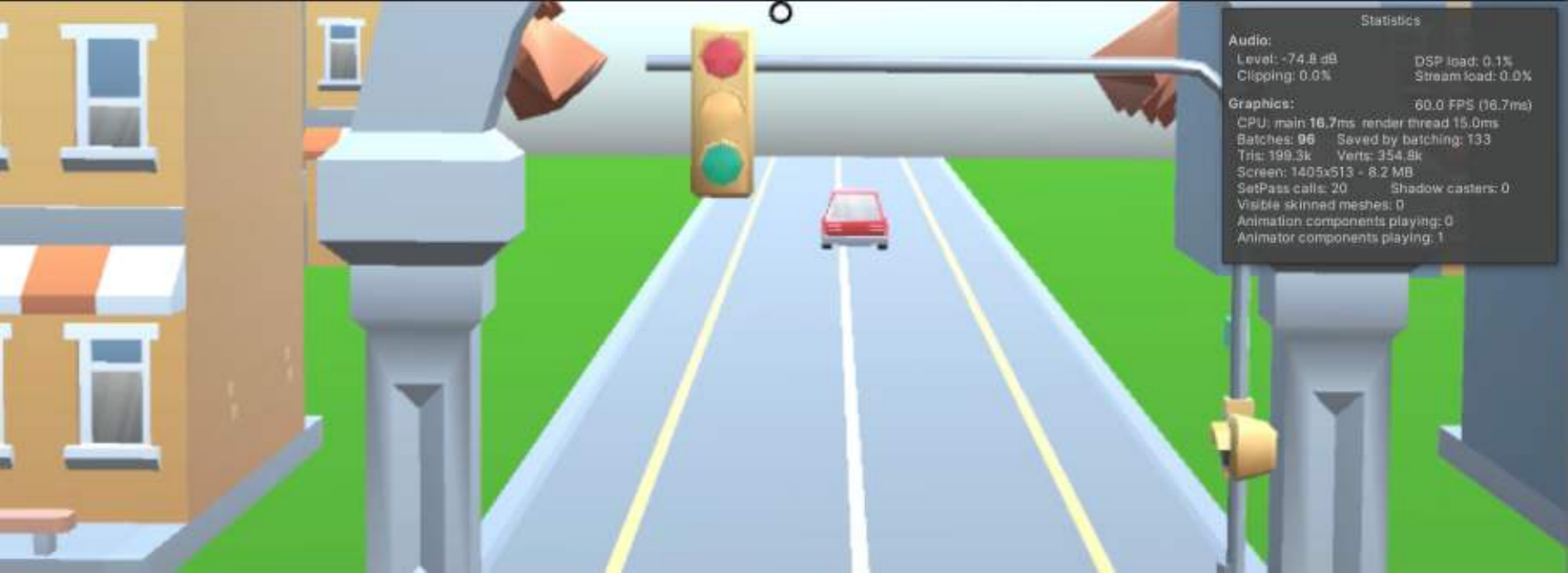}
            \caption{Benign, time=2}
            \label{subfig:car_nosound_frame3}
        \end{subfigure}
        
        \vspace{0.5em} 

        \begin{subfigure}{0.3\textwidth}
            \includegraphics[width=\linewidth]{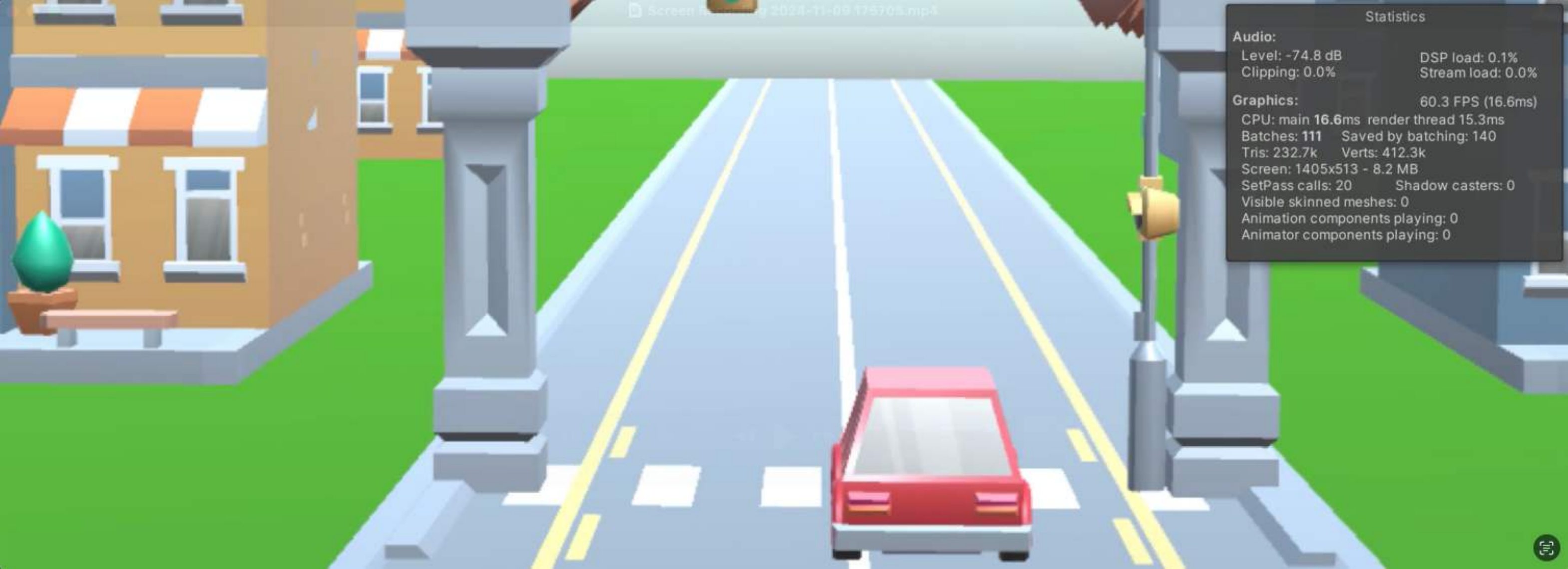}
            \caption{With sound, time=0}
            \label{subfig:car_withsound_frame1}
        \end{subfigure}
        \hfill
        \begin{subfigure}{0.3\textwidth}
            \includegraphics[width=\linewidth]{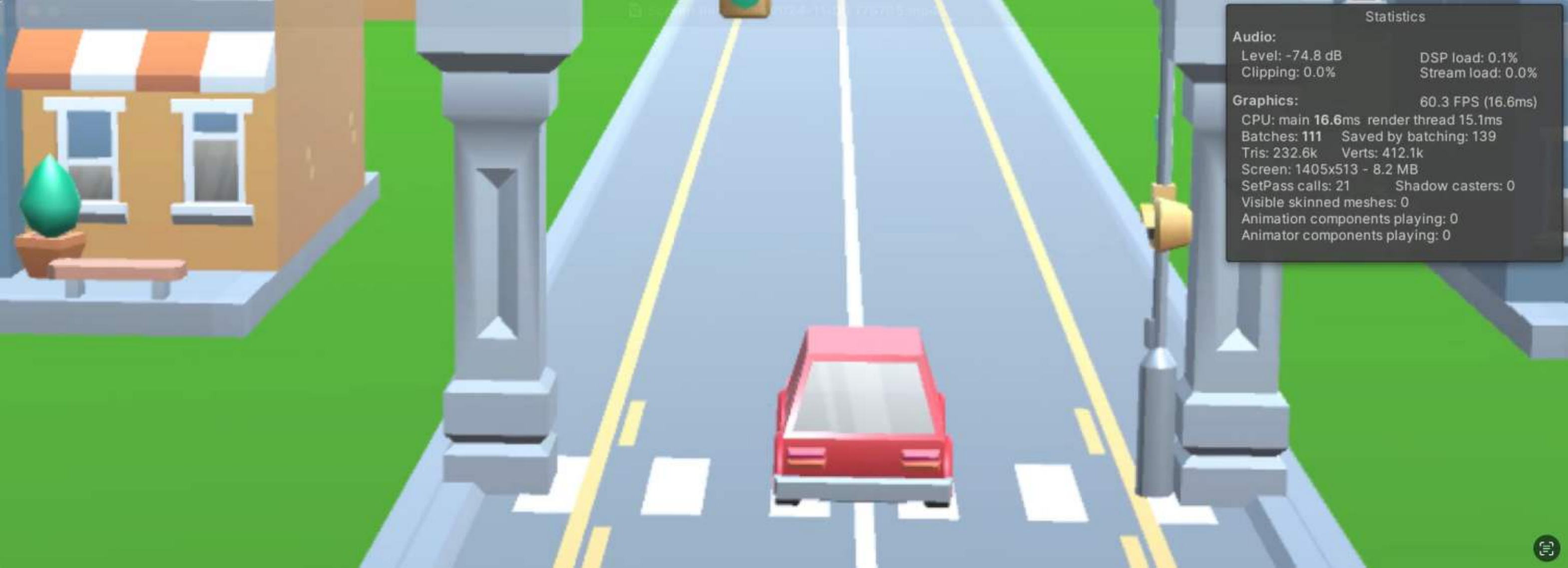}
            \caption{With sound, time=1}
            \label{subfig:car_withsound_frame2}
        \end{subfigure}
        \hfill
        \begin{subfigure}{0.3\textwidth}
            \includegraphics[width=\linewidth]{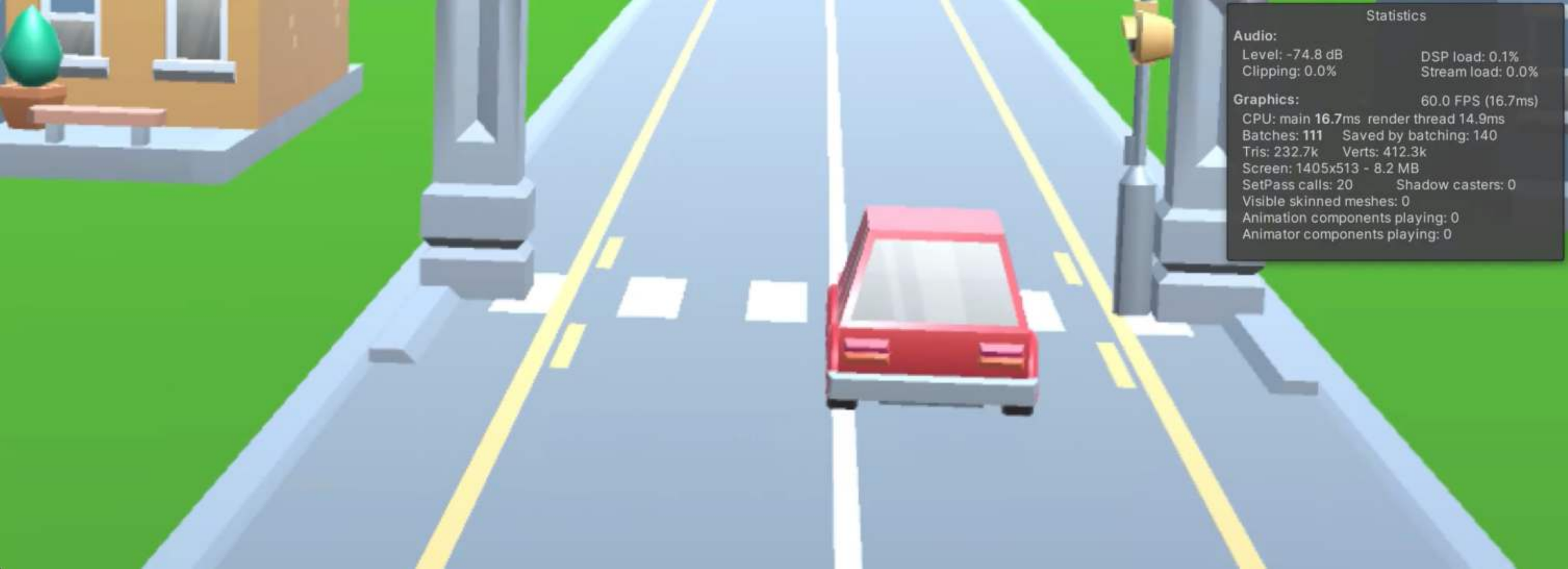}
            \caption{With sound, time=2}
            \label{subfig:car_withsound_frame3}
        \end{subfigure}
    \end{minipage}
    \hfill
    \begin{minipage}{0.15\textwidth}
        \centering
        \begin{subfigure}{\textwidth}
            \includegraphics[width=\linewidth]{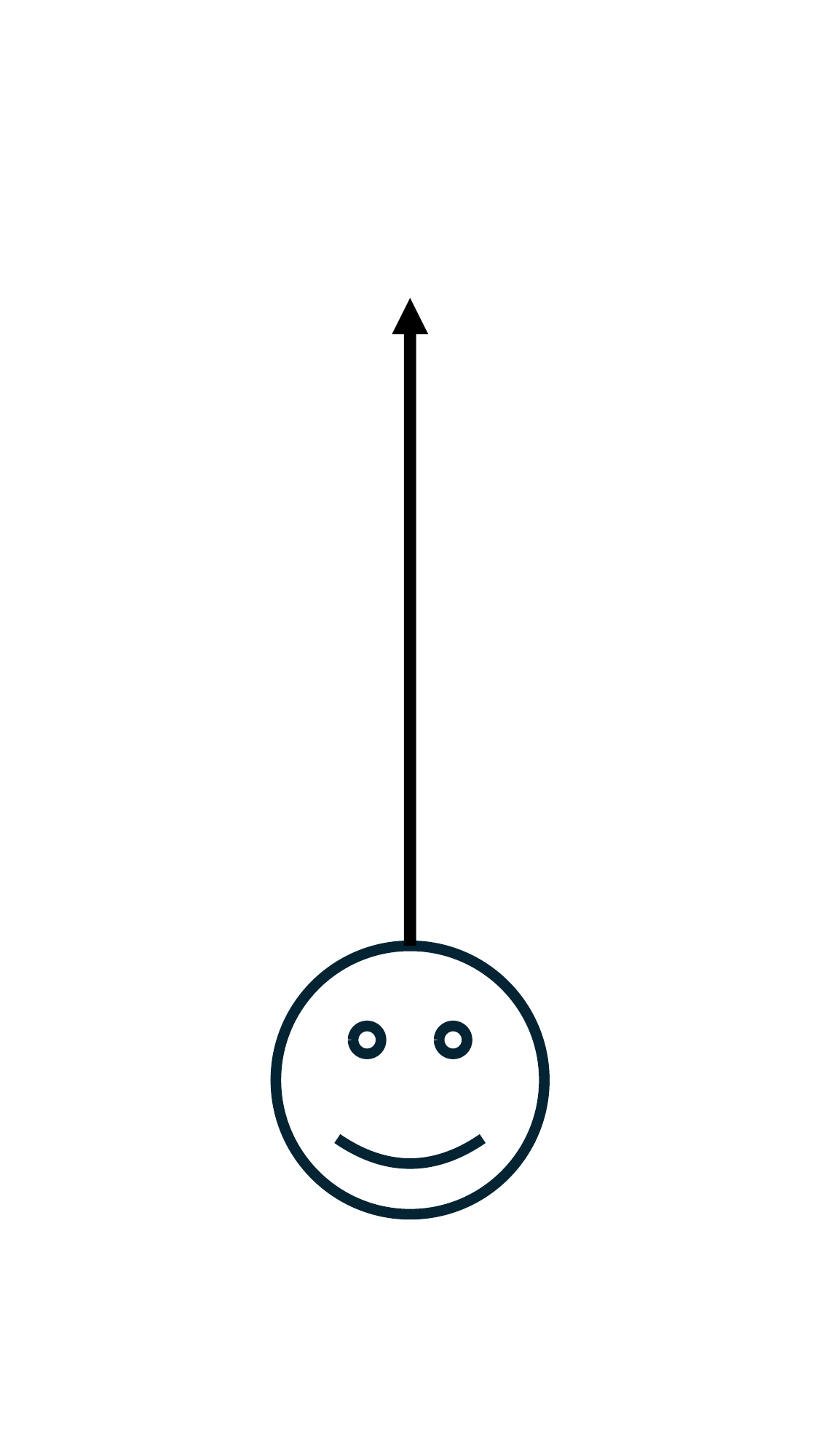}
            \vspace{-0.5cm}
            \caption{User's physical movements from time=0 to time=2.}
            \label{subfig:car_usertrace}
        \end{subfigure}
    \end{minipage}
    \caption{Manipulating user input attack. The top row shows the intended benign behavior of a virtual car driving forward, and the bottom row shows the effect of the acoustic attack: The car is unable to move from the starting point. 
    }
    \label{fig:car_game}
\end{figure*}

\subsubsection{Harm to User: Manipulating User Input}
\label{sec:app_vr_gaming}

\noindent \emph{\textbf{Attack motivation.}}
Gaming is a major driver of XR adoption, with 91 of the 100 most popular VR applications being games as of early 2023~\cite{nair2023unique}. In these XR games, players interact with virtual 3D objects (like cars or virtual avatars) in the virtual world using IMU sensors embedded in controllers or headsets. In this attack, we demonstrate how an attacker can manipulate a user’s input by injecting adversarial acoustic signals to cause unwanted effects in the game. 

\noindent \emph{\textbf{Attack design.}} We implement this attack within a car racing game where the player controls the car's direction by moving their headset. The four directions (forward, backward, left, right) align with the player’s head movements, allowing the car to follow the user’s orientation. This interaction method is common in many AR/VR applications (\eg Google Earth VR) and games (\eg BeatSaber), where avatars or virtual vehicles move in sync with the user’s head movements.

\begin{figure*}[h]
    \centering
    \begin{tabular}{ccccc}
        \begin{subfigure}[b]{0.18\textwidth}
            \centering
            \includegraphics[width=\textwidth]{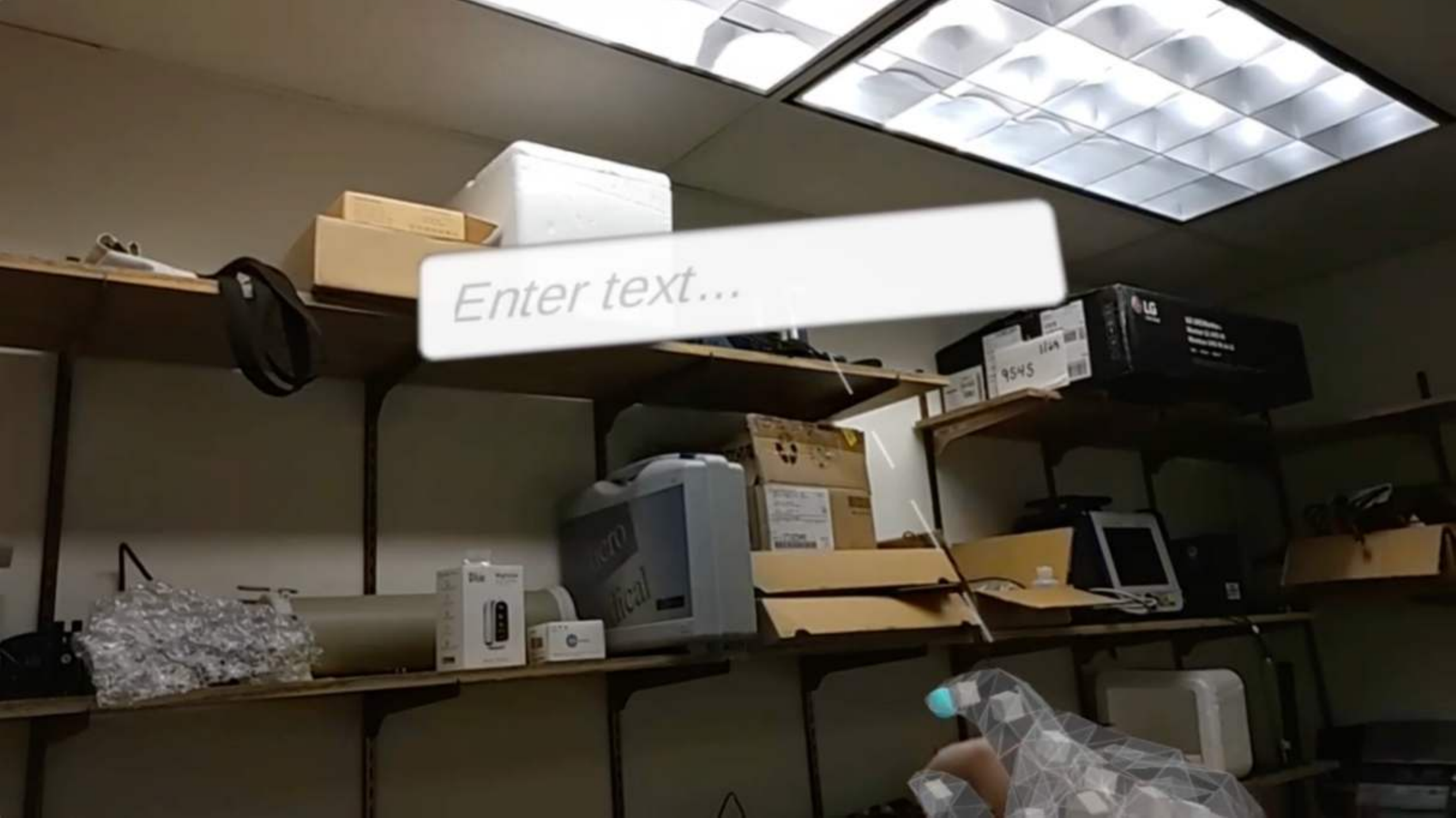}
            \caption{Benign, time=0}
            \label{subfig:keyboard_nosound_frame1}
        \end{subfigure} &
        \begin{subfigure}[b]{0.18\textwidth}
            \centering
            \includegraphics[width=\textwidth]{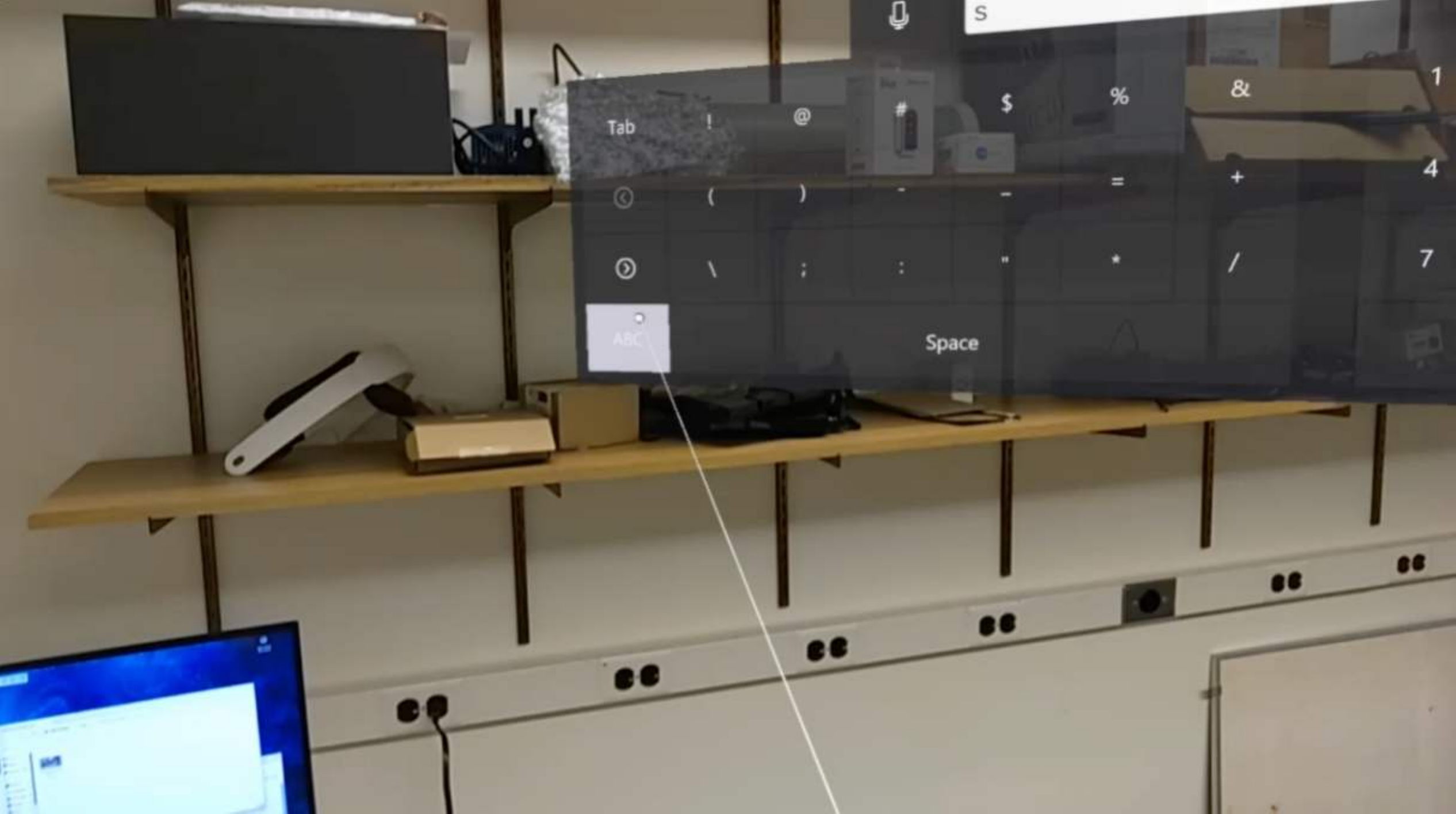}
            \caption{Benign, time=1}
            \label{subfig:keyboard_nosound_frame2}
        \end{subfigure} &
        \begin{subfigure}[b]{0.18\textwidth}
            \centering
            \includegraphics[width=\textwidth]{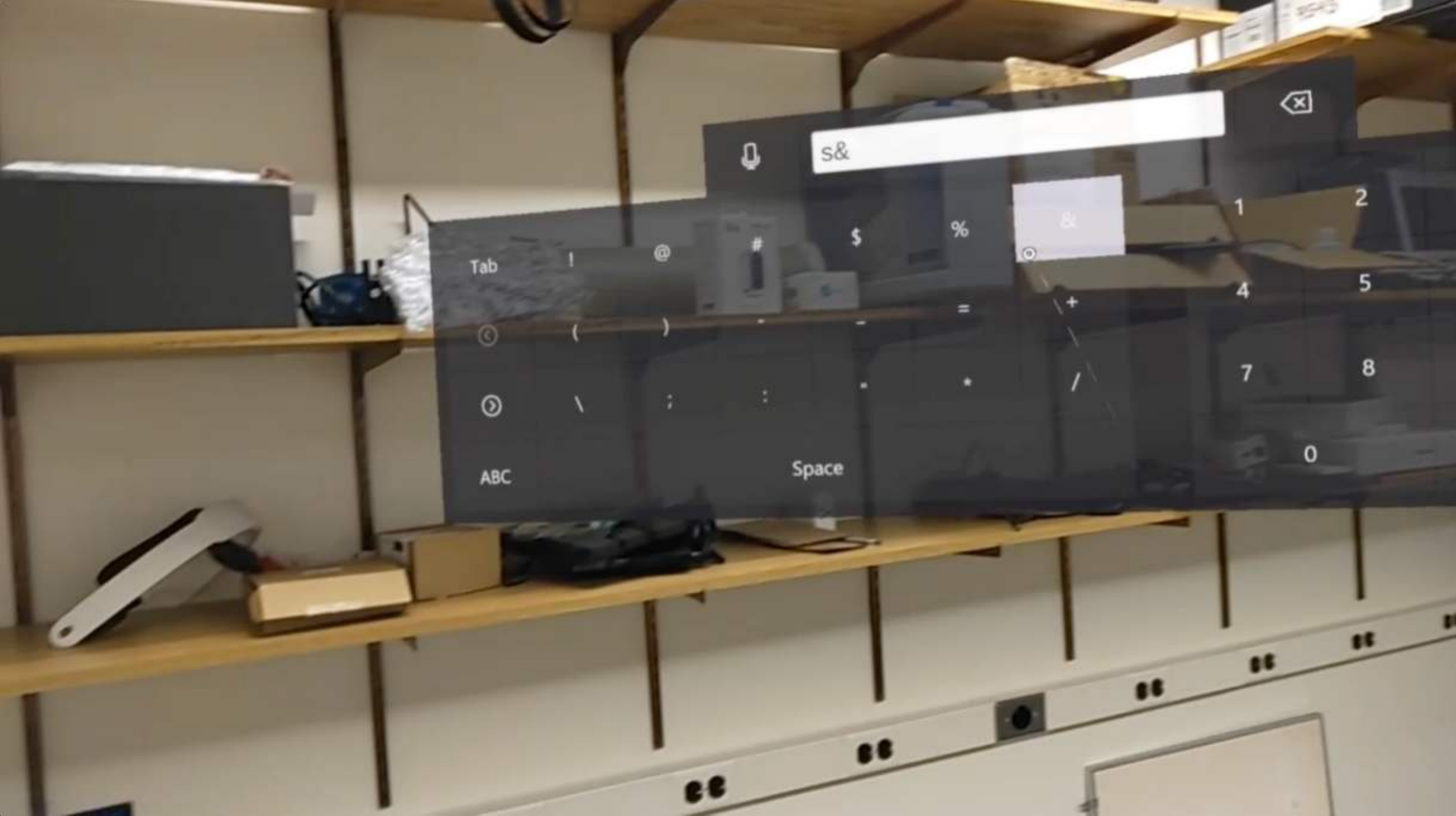}
            \caption{Benign, time=2}
            \label{subfig:keyboard_nosound_frame3}
        \end{subfigure} &
        \begin{subfigure}[b]{0.18\textwidth}
            \centering
            \includegraphics[width=\textwidth]{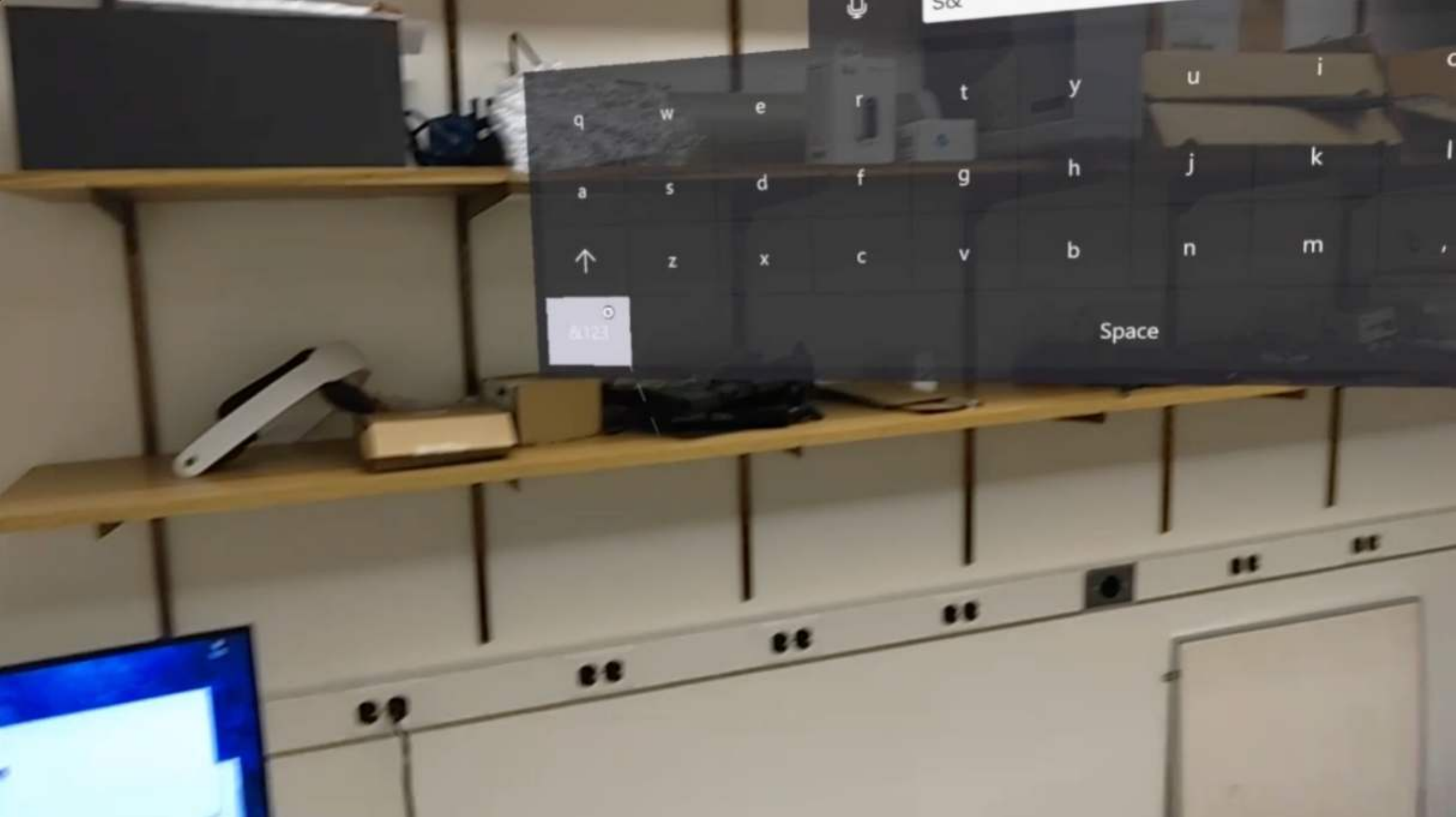}
            \caption{Benign, time=3}
            \label{subfig:keyboard_nosound_frame4}
        \end{subfigure} &
        \begin{subfigure}[b]{0.18\textwidth}
            \centering
            \includegraphics[width=\textwidth]{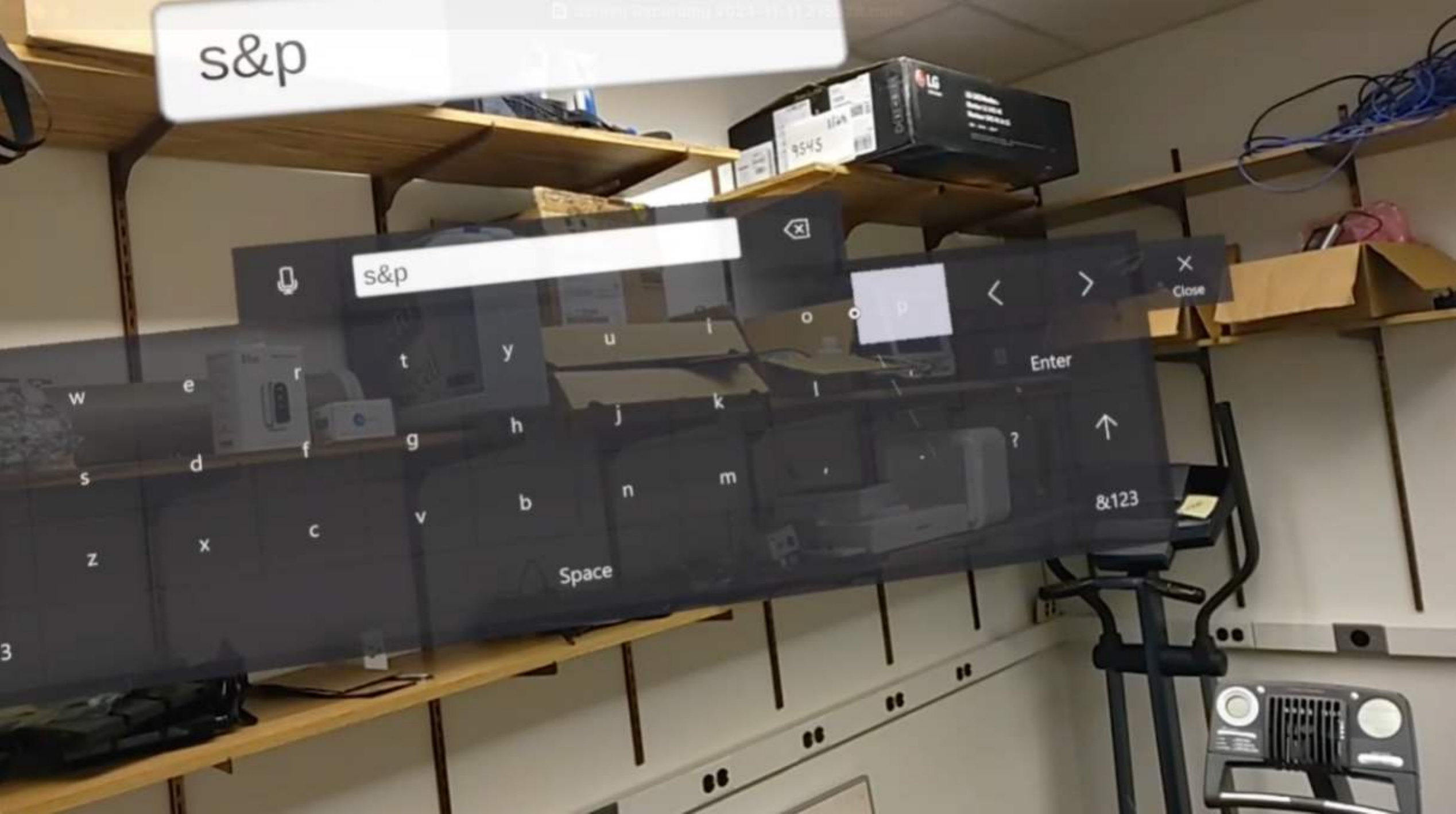}
            \caption{Benign, time=4}
            \label{subfig:keyboard_nosound_frame5}
        \end{subfigure} \\
        
        \begin{subfigure}[b]{0.18\textwidth}
            \centering
            \includegraphics[width=\textwidth]{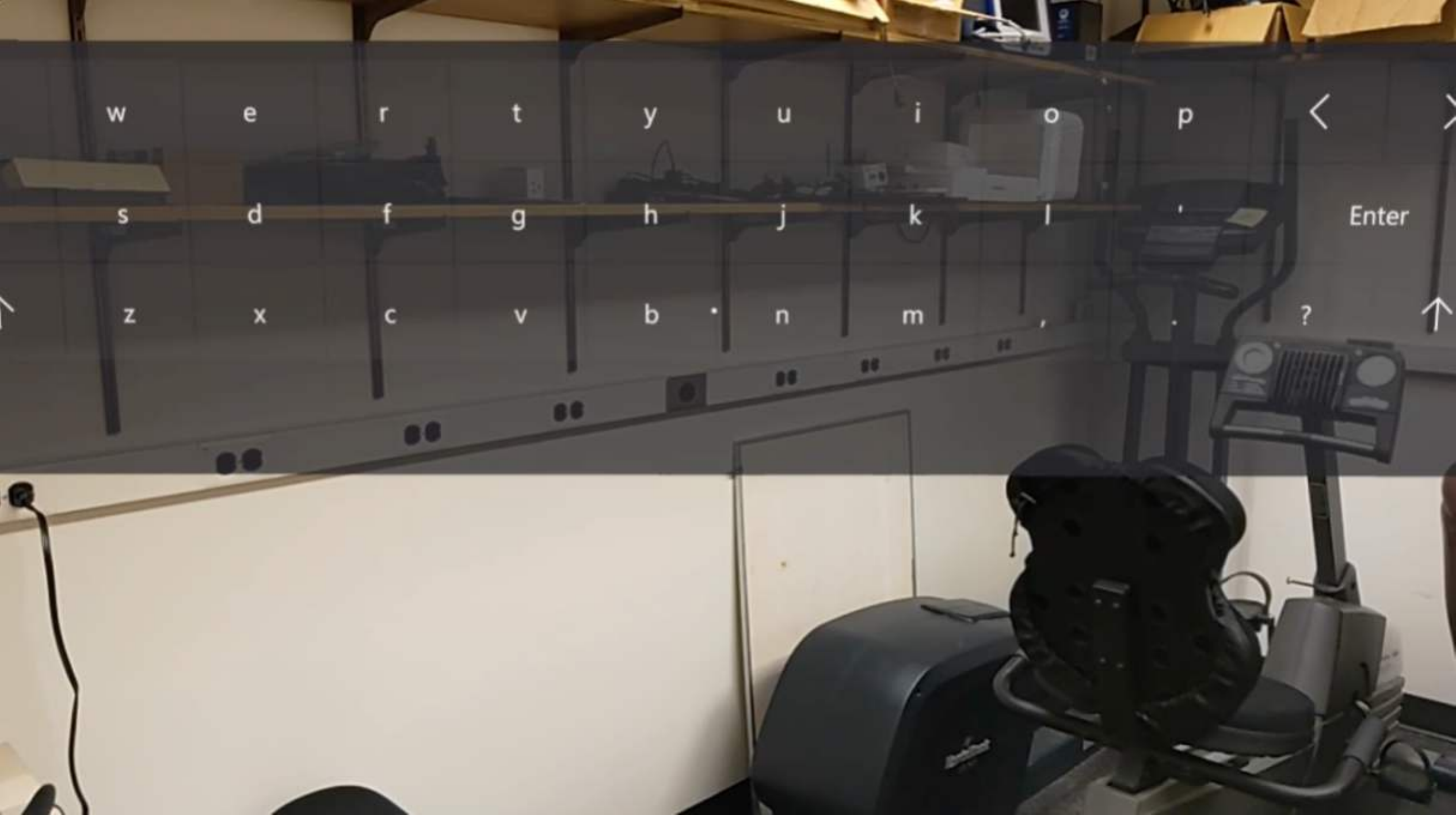}
            \caption{With sound, time=0}
            \label{subfig:keyboard_withsound_frame1}
        \end{subfigure} &
        \begin{subfigure}[b]{0.18\textwidth}
            \centering
            \includegraphics[width=\textwidth]{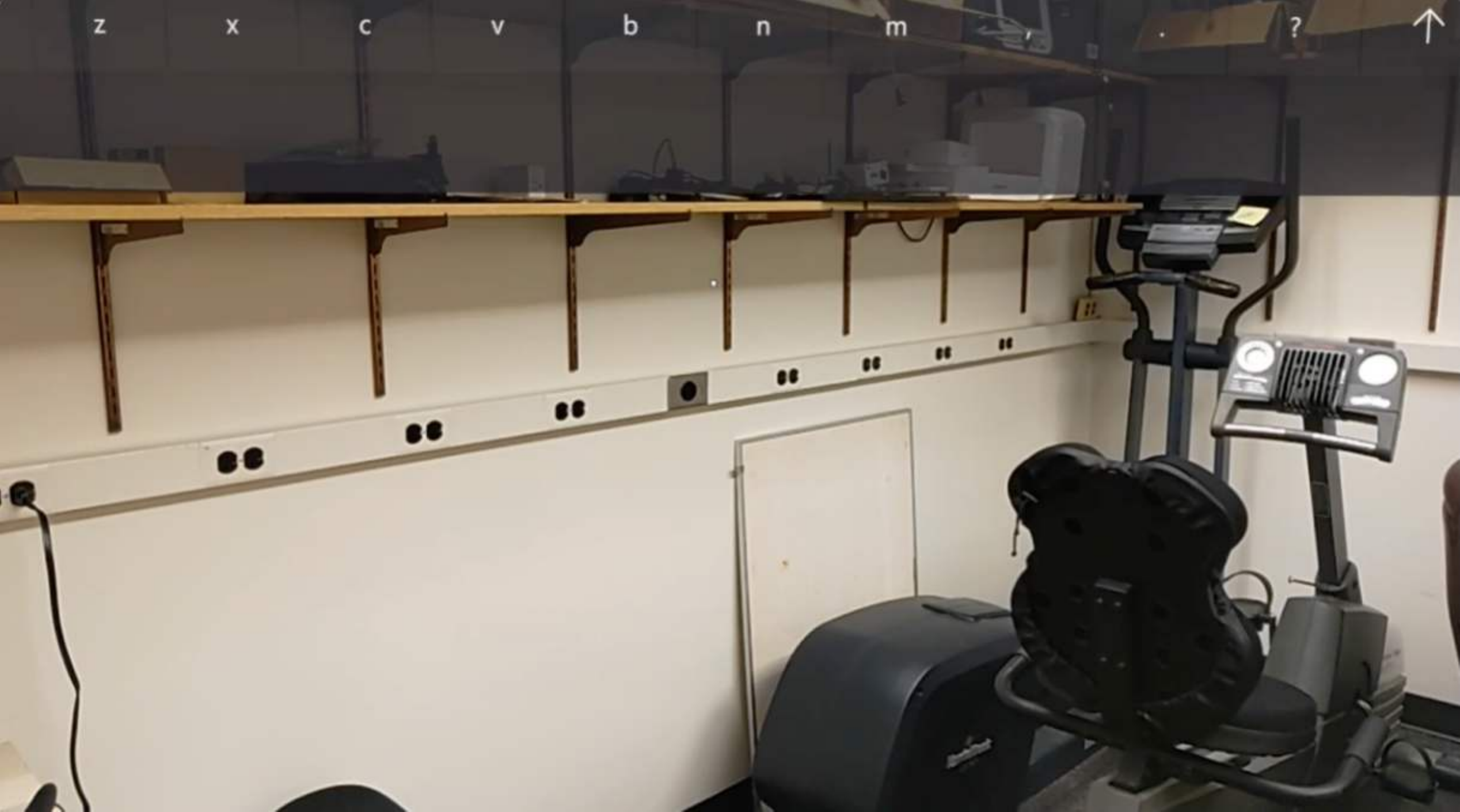}
            \caption{With sound, time=1}
            \label{subfig:keyboard_withsound_frame2}
        \end{subfigure} &
        \begin{subfigure}[b]{0.18\textwidth}
            \centering
            \includegraphics[width=\textwidth]{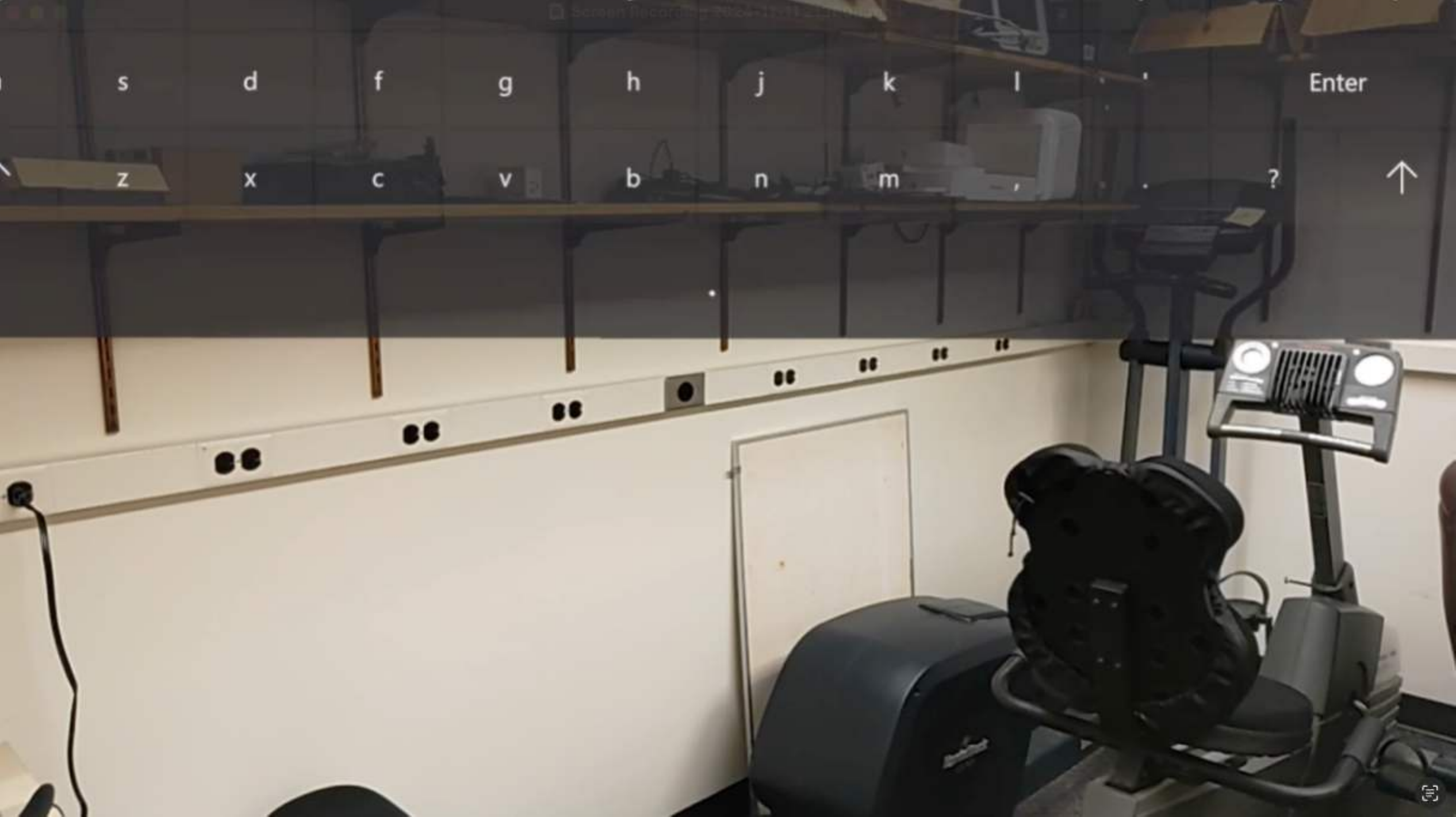}
            \caption{With sound, time=2}
            \label{subfig:keyboard_withsound_frame3}
        \end{subfigure} &
        \begin{subfigure}[b]{0.18\textwidth}
            \centering
            \includegraphics[width=\textwidth]{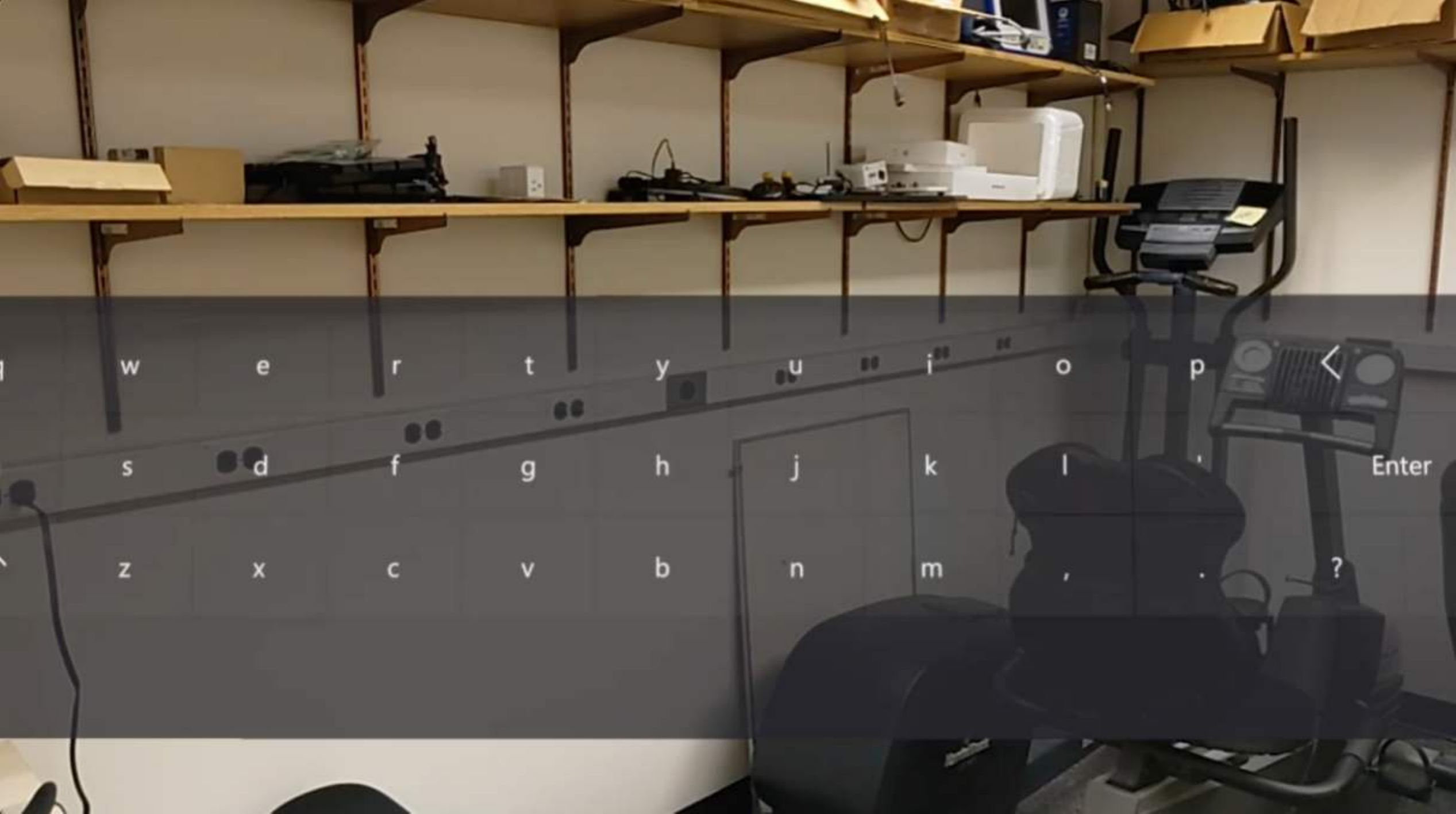}
            \caption{With sound, time=3}
            \label{subfig:keyboard_withsound_frame4}
        \end{subfigure} &
        \begin{subfigure}[b]{0.18\textwidth}
            \centering
            \includegraphics[width=\textwidth]{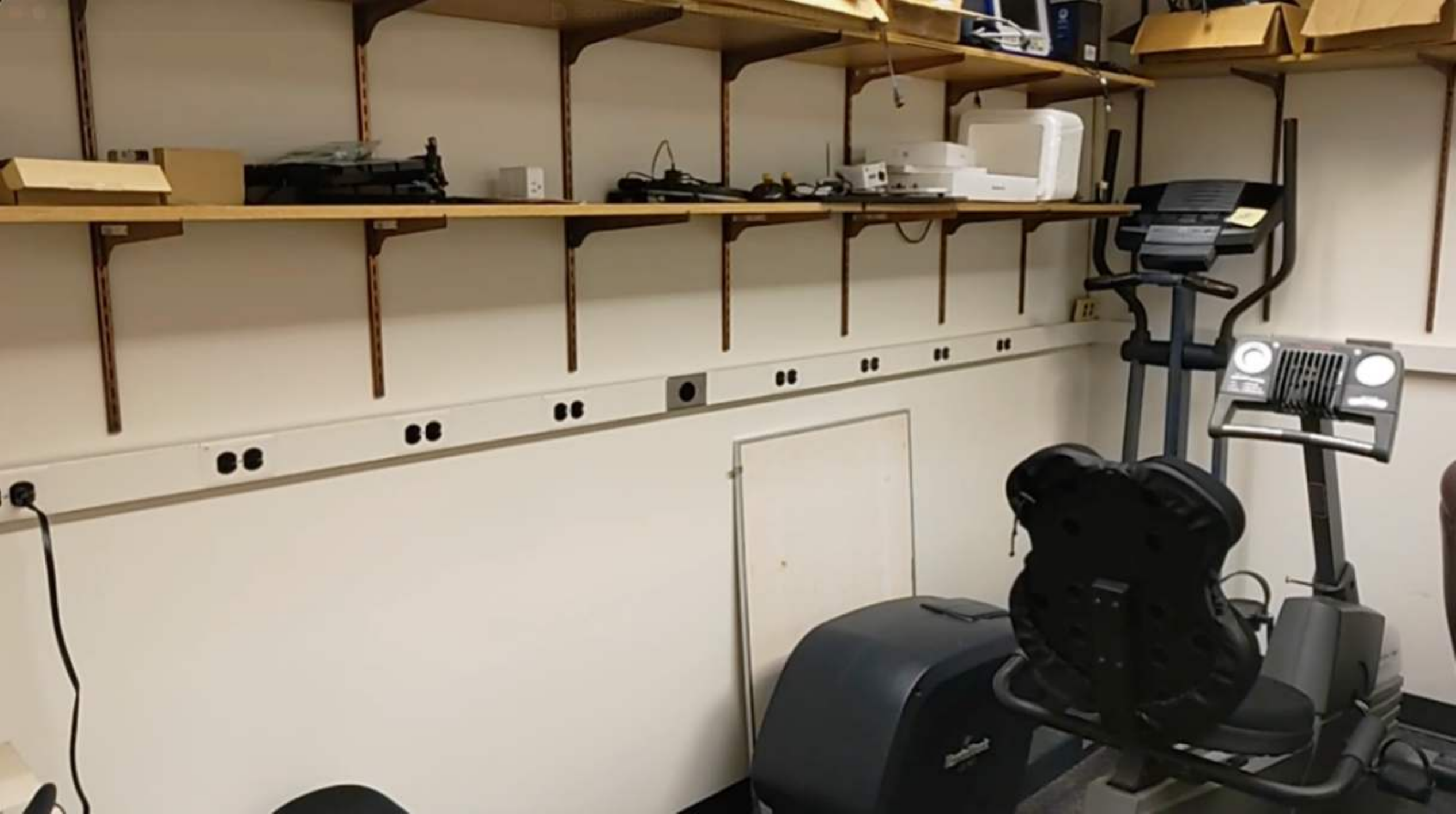}
            \caption{With sound, time=4}
            \label{subfig:keyboard_withsound_frame5}
        \end{subfigure}
    \end{tabular}
    \caption{
    Clickjacking attack. In the benign case (top row), the keyboard remains mostly fixed in front of the user.
    In the attack case (bottom row), the keyboard drifts up and down, inhibiting the user from interacting with the desired keys.
    }
    \label{fig:keyboard}
\end{figure*}

\noindent \emph{\textbf{Attack outcome.}} We place the headset on the remote-controlled car moving straight forward. According to the game design, the virtual car in the VR game should also move forward, aligned with the headset’s movement. 
The top row of \Cref{fig:car_game} shows the game's normal behavior as visualized on the HoloLens 2, where the virtual car moves straight ahead from the initial starting point near the crosswalk.

During the attack, we place the HoloLens on the same remote control car and repeat the game, but this time we inject acoustic signals targeting the HoloLens' IMU sensors. As shown in \Cref{subfig:car_withsound_frame1} and \Cref{subfig:car_withsound_frame2}, the car initially moves forward but then shifts slightly to the left (leftward shift caused by a small deviation in the car's movement). Due to the \textbf{snapback} effect (detailed in Section~\ref{sec:snapback_effect}) caused by the acoustic interference, the car’s position resets to zero, forcing it back to the starting point, as seen in  \Cref{subfig:car_withsound_frame3}.
In other words, we find that the car's position is fixed around the starting point when the acoustic signal is played, which destroys the functionality of the game for the victim user.

\subsubsection{Harm to User: Clickjacking} 
\label{sec:app_clickjacking}

\noindent \emph{\textbf{Attack motivation.}} The goal of clickjacking attacks is to deceive the victim into thinking they are clicking on one object, while in reality, they are interacting with a baited object. In an XR game, an attacker might use clickjacking to trick the user into clicking on XR advertisements that generate extra revenue. For example, prior research has shown that clickjacking attacks can bait or hijack user interactions in XR environments~\cite{cheng2024user, lee2021adcube}. In this attack, we demonstrate a clickjacking attack in AR using adversarial acoustic signals.

\noindent \emph{\textbf{Attack design.}} The AR app is designed for users to enter text via a virtual keyboard, with the victim user interacting through her hand gestures (air tap~\cite{hololens-gestures-intro}) on the HoloLens 2. The virtual keyboard is developed using Microsoft’s Mixed Reality Toolkit (MRTK). In a benign scenario, the keyboard remains anchored to fixed coordinates in front of the user to ensure accurate typing. However, in this proof-of-concept attack, the attacker directs sound at the headset to shift the position of the keyboard. 
The goal is to bait the victim into misclicking characters on the keyboard.

\noindent \emph{\textbf{Attack outcome.}} Without acoustic interference, the virtual keyboard remains stable in the same location across five consecutive frames, as shown in the top row of \Cref{fig:keyboard}. Regardless of the headset user’s movements, the virtual keyboard stays fixed.  However, after introducing acoustic sound directed at the headset, we observe that the keyboard’s position shifts. In the first two time instances (\Cref{subfig:keyboard_withsound_frame1} and \Cref{subfig:keyboard_withsound_frame2}), the keyboard drifts upward, with part of it moving out of the user’s field of view. However, because of the snapback effect, the keyboard then moves back down, returning to its original position in \Cref{subfig:keyboard_withsound_frame3} and \Cref{subfig:keyboard_withsound_frame4}. With continuous acoustic interference, the virtual keyboard shifts upward once more and disappears from the user’s field of view, as shown in \Cref{subfig:keyboard_withsound_frame5}. 

Thus with this continuous acoustic interference, the virtual keyboard oscillates up and down, complicating the user’s ability to accurately enter their intended input. In a more advanced attack scenario, the attacker could time the acoustic injection to coincide with the user entering sensitive information, such as passwords in a banking application. The exact timing of such injections could be precisely inferred through side channels~\cite{slocum2023going,zhang2023its}.

\begin{figure*}
    \centering
    \begin{tabular}{cccc}
        \begin{subfigure}[b]{0.22\textwidth}
            \centering
            \includegraphics[width=\textwidth]{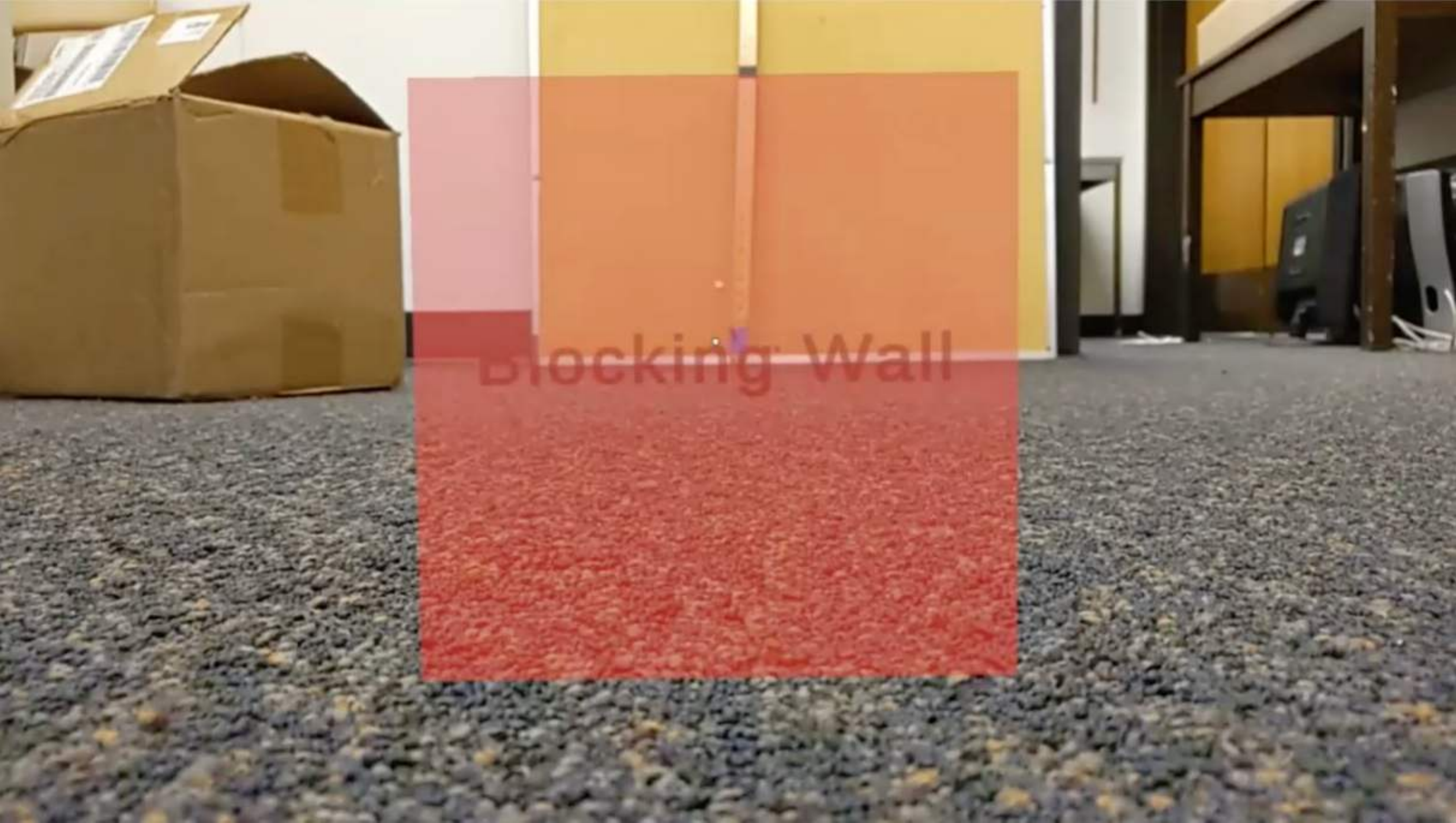}
            \caption{Benign, time=0}
            \label{subfig:wall_nosound_frame1}
        \end{subfigure} &
        \begin{subfigure}[b]{0.22\textwidth}
            \centering
            \includegraphics[width=\textwidth]{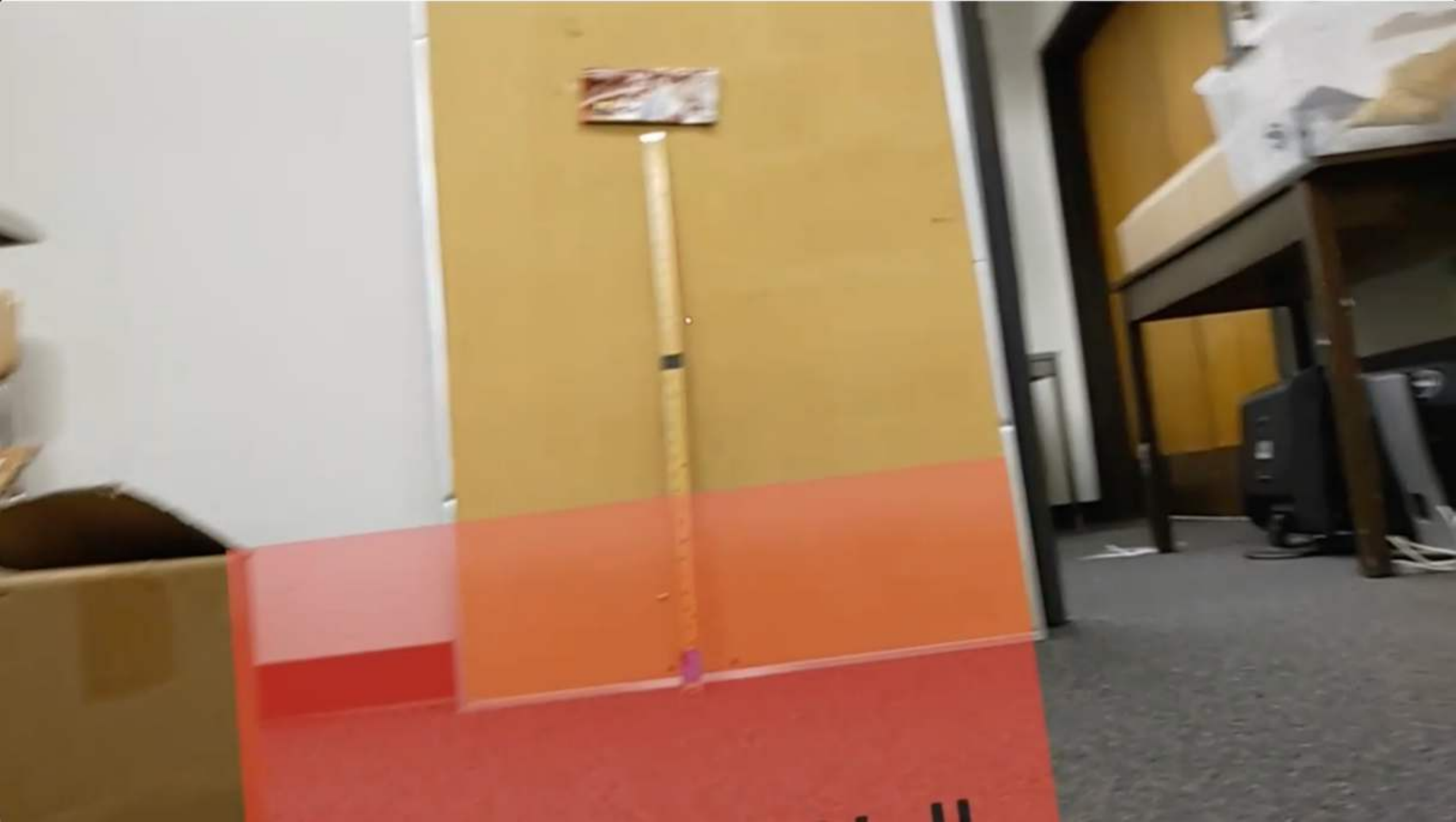}
            \caption{Benign, time=1}
            \label{subfig:wall_nosound_frame2}
        \end{subfigure} &
        \begin{subfigure}[b]{0.22\textwidth}
            \centering
            \includegraphics[width=\textwidth]{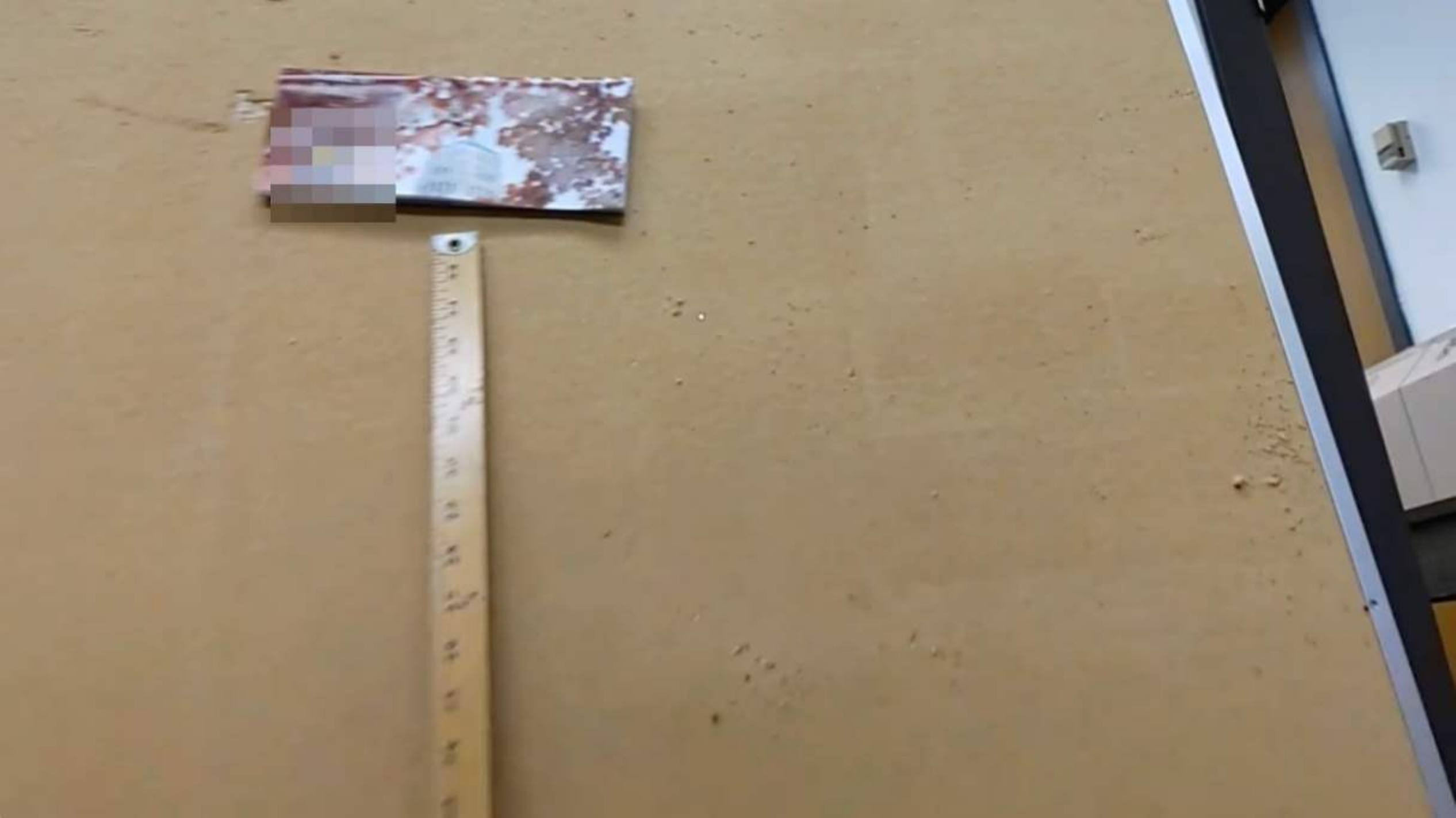}
            \caption{Benign, time=2}
            \label{subfig:wall_nosound_frame3}
        \end{subfigure} &
        \begin{subfigure}[b]{0.24\textwidth}
            \centering
            \includegraphics[width=\textwidth]{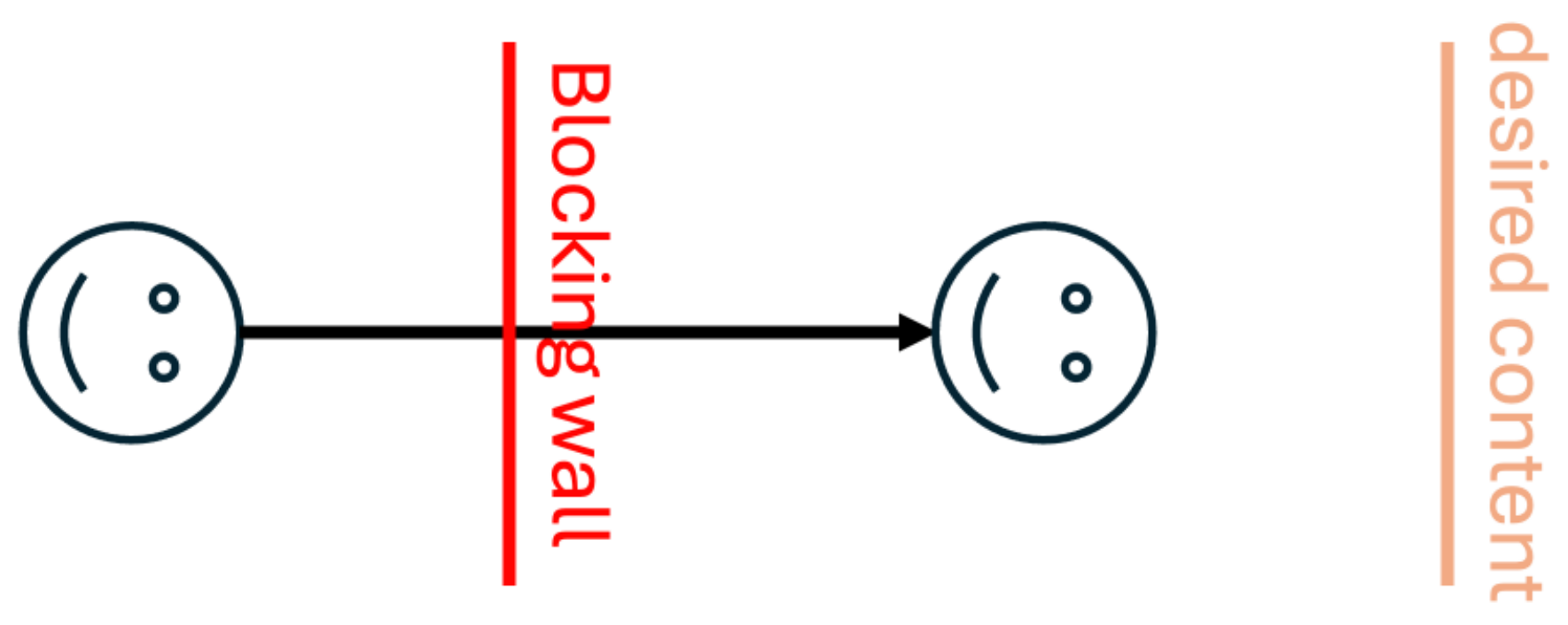}
            \caption{User's physical movements, benign case}
            \label{subfig:trace_wall_benign}
        \end{subfigure} \\
        
        \begin{subfigure}[b]{0.22\textwidth}
            \centering
            \includegraphics[width=\textwidth]{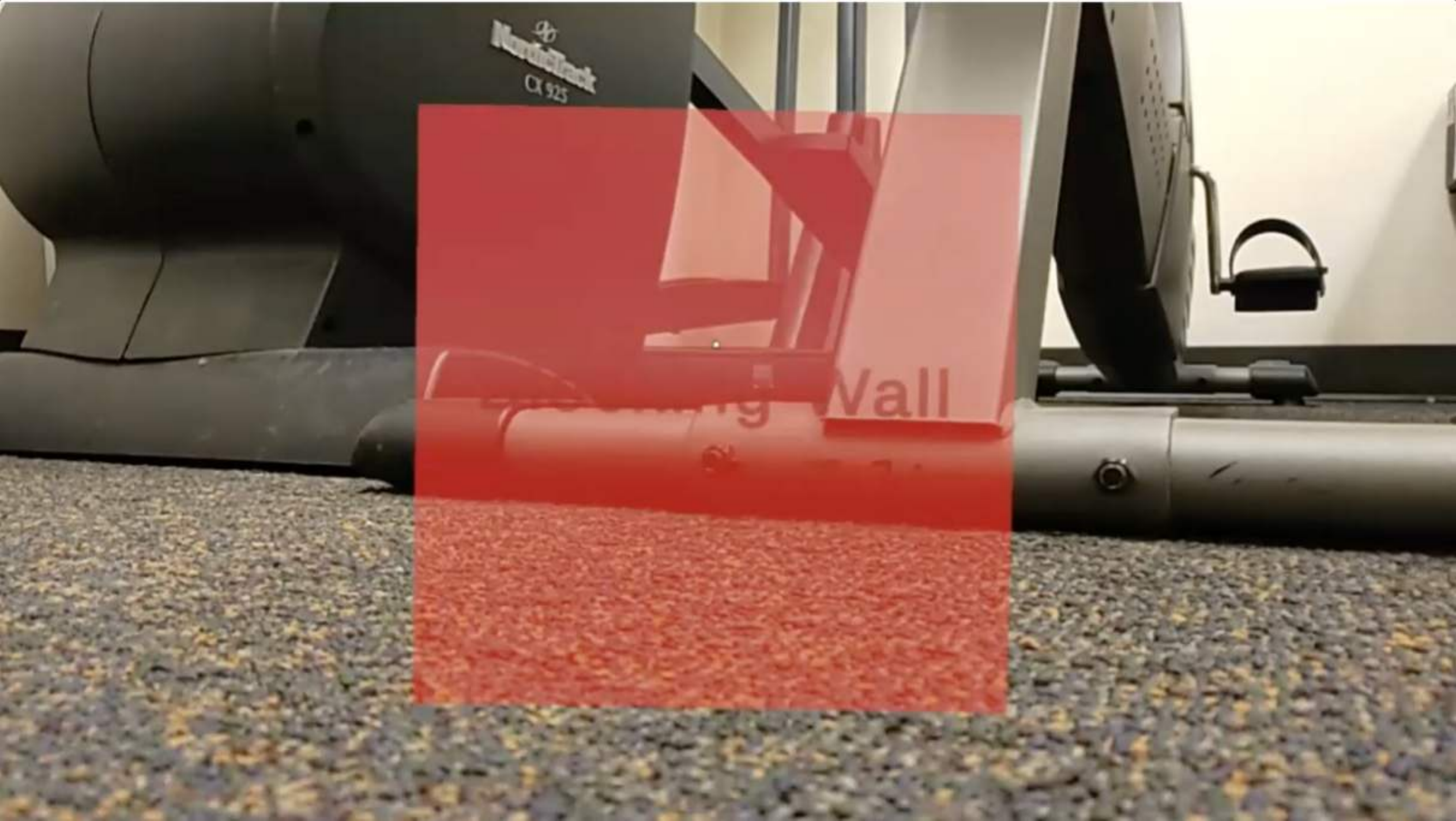}
            \caption{With sound, time=0}
            \label{subfig:wall_withsound_frame1}
        \end{subfigure} &
        \begin{subfigure}[b]{0.22\textwidth}
            \centering
            \includegraphics[width=\textwidth]{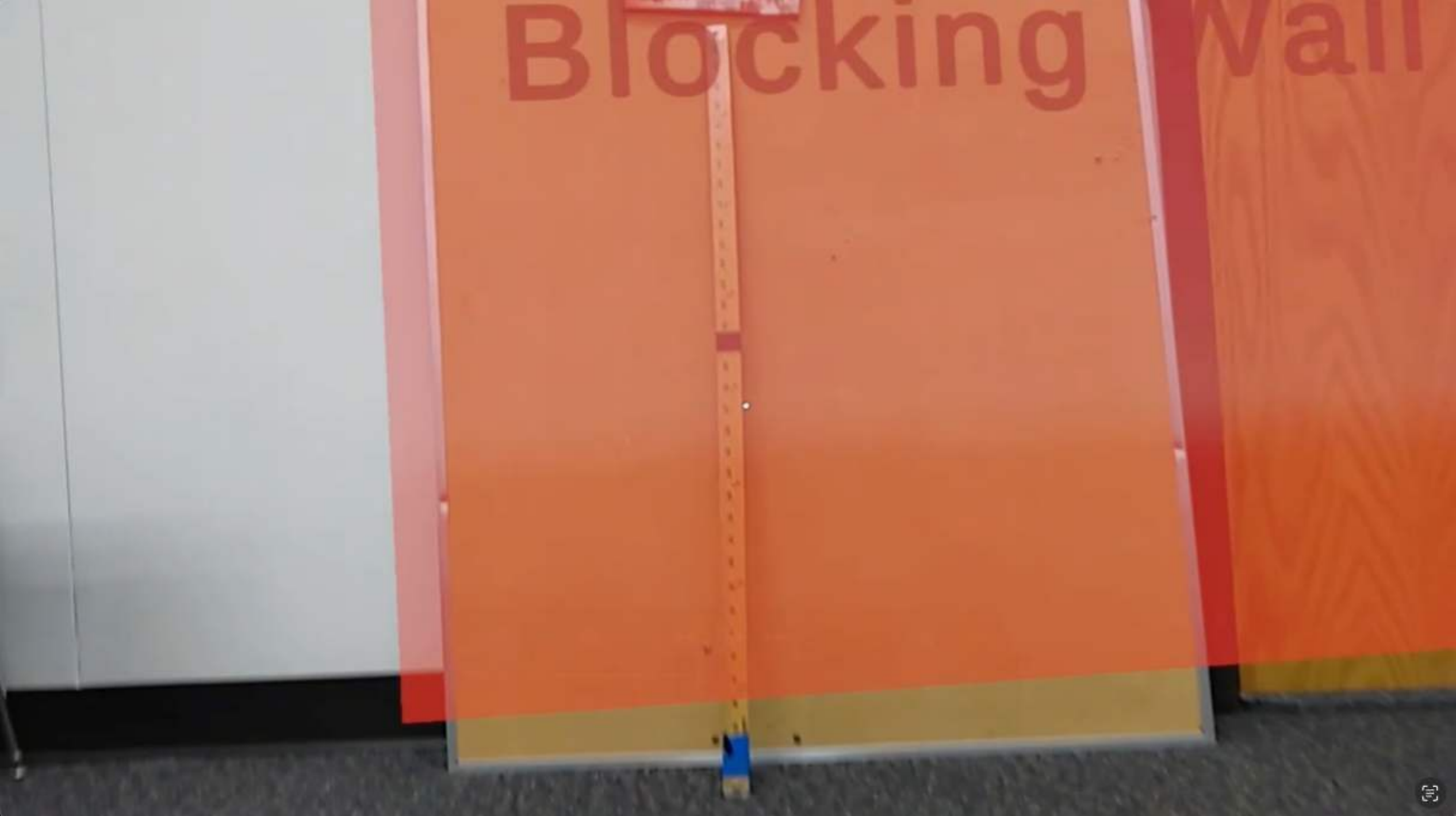}
            \caption{With sound, time=1}
            \label{subfig:wall_withsound_frame2}
        \end{subfigure} &
        \begin{subfigure}[b]{0.22\textwidth}
            \centering
            \includegraphics[width=\textwidth]{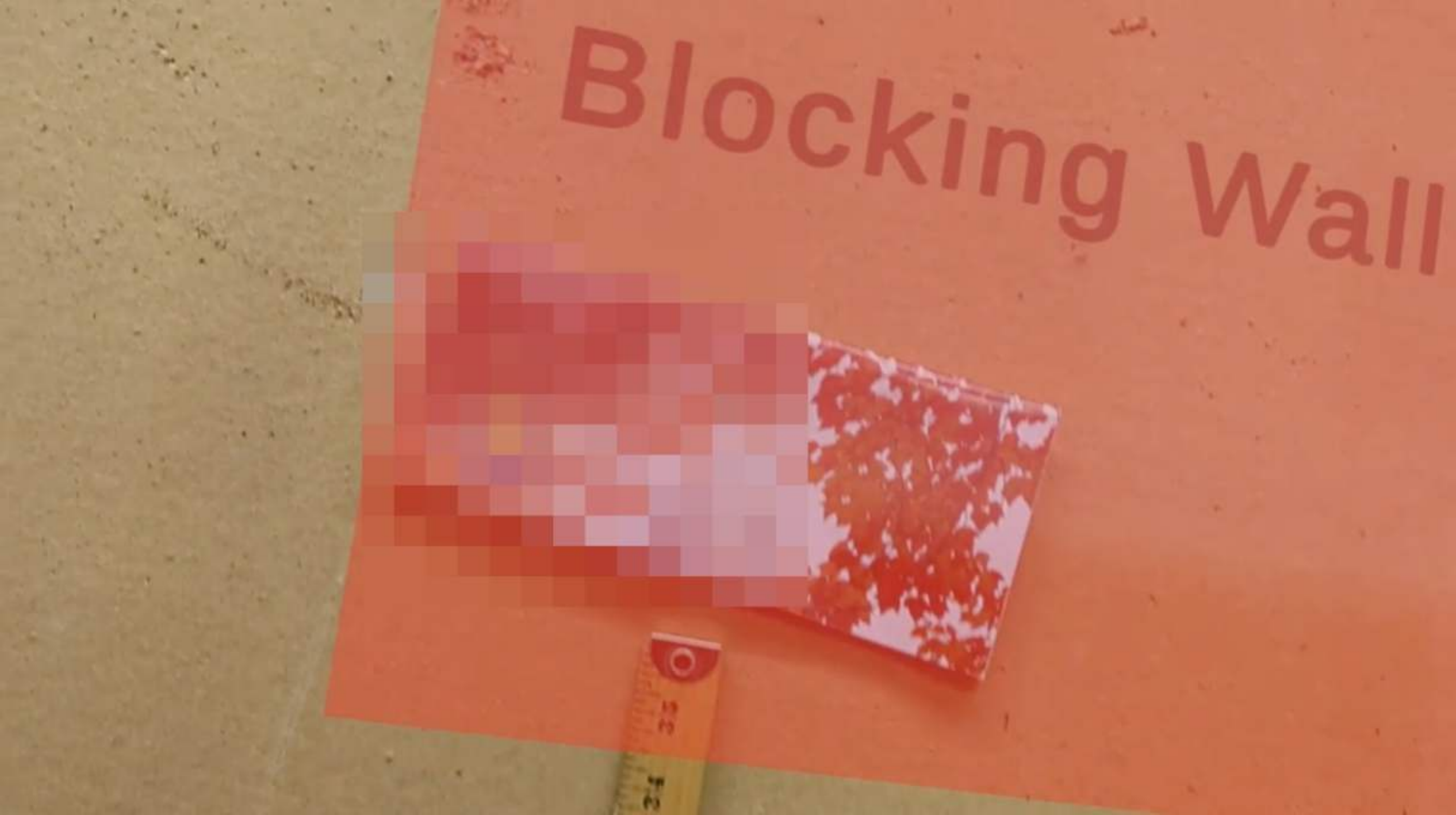}
            \caption{With sound, time=2}
            \label{subfig:wall_withsound_frame3}
        \end{subfigure} &
        \begin{subfigure}[b]{0.24\textwidth}
            \centering
            \includegraphics[width=\textwidth]{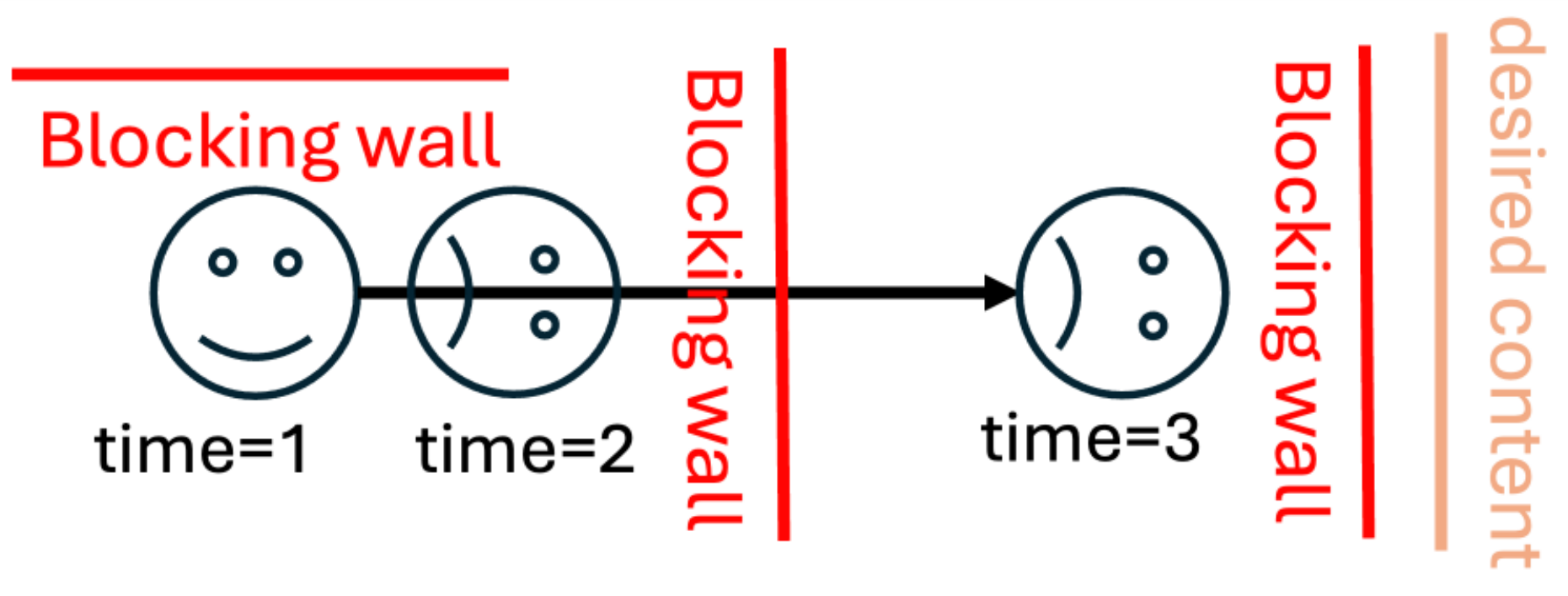}
            \caption{User's physical movements, With sound}
            \label{subfig:trace_wall_attack}
        \end{subfigure}
    \end{tabular}
    \caption{Denial of user interaction attack. 
    In the benign case (top row), the user is able to move pats the red blocking wall to interact with the brochure on the bulletin board.
    In the attack case (bottom row), the blocking wall prevents the user from clearly viewing or interacting with the brochure.
    In the user trajectory (d,h), the brown line is the content that the user wants to see, and the red line is the blocking wall.
   }
    \label{fig:wall}
\end{figure*}

\subsubsection{Harm to User: Denial of User Interaction Attack}
\label{sec:app_blocking_wall}

\noindent \emph{\textbf{Attack motivation.}} We introduce two variants of denial of user interaction attacks: \emph{denial-of-user-input} and \emph{visual blocking} attacks. \emph{Denial-of-user-input} attacks occur when an attacker blocks legitimate inputs, such as hand gestures or voice commands. Prior work~\cite{cheng2024user} demonstrates that an invisible ``cage'' can prevent a user from interacting with intended objects. \emph{Visual blocking} attacks arise when environmental factors—like poor lighting or occlusions—impair object visibility~\cite{huang2024ar,xiu2024lobstar,lebeck2017securing}. In this attack, the attacker’s virtual object can collide with the victim’s, causing visual obstruction. Consequently, the victim cannot detect or interact with the intended object, as it is concealed by the attacker’s interference.

\noindent \emph{\textbf{Attack design.}} 
Normally, in best practice UI design for XR~\cite{laviola20173d}, the virtual objects should remain static with respect to the real world.
In this attack, we leverage the \textbf{snapback} effect by continuously playing an acoustic signal to shift a virtual blocking object (represented by the ``Blocking Wall'' in \Cref{fig:wall}) and prevent the victim from viewing/interacting with the intended target.
The blocking wall object is rendered using Unity's shader~\cite{unity_shader}, which allows two rendering options: opaque and transparent, which we use to implement our two attack variants. For the denial-of-user-input attack, we render the blocking wall with transparency. Setting the transparency to 100\% makes the blocking wall completely invisible while still obstructing the user’s hand interactions, thereby implementing the virtual ``cage''\cite{cheng2024user}. For the visual blocking attack, we can set the blocking wall to opaque to obstruct the victim's view of important objects in the scene.

\noindent \emph{\textbf{Attack outcome.}} In a benign scenario, the ``Blocking Wall'' remains in a fixed position with respect to the real world, as shown in the top row of \Cref{fig:wall}.
As the headset moves forward and passes the wall, the user can view and interact with the object behind it, as illustrated in \Cref{subfig:wall_nosound_frame2} and \Cref{subfig:wall_nosound_frame3}. 
However, when the attacker injects acoustic sound, the virtual begins to move with the headset due to the snapback effect. This causes the blocking wall to stay continually in front of the victim.
The attack outcome depends on the rendering option: (1) If a transparent shader is chosen, the victim encounters an invisible wall that blocks all user interactions (\eg controller and hand gestures) with the items on the bulletin board; (2) If an opaque shader is used, an opaque wall obstructs the victim's visual perception of the real world.
A semi-transparent wall is shown in the second row of \Cref{fig:wall} to illustrate both scenarios.

\begin{figure*}[h]
    \begin{minipage}{0.8\textwidth}
        \centering
        \begin{subfigure}{0.24\textwidth}
            \includegraphics[width=\linewidth]{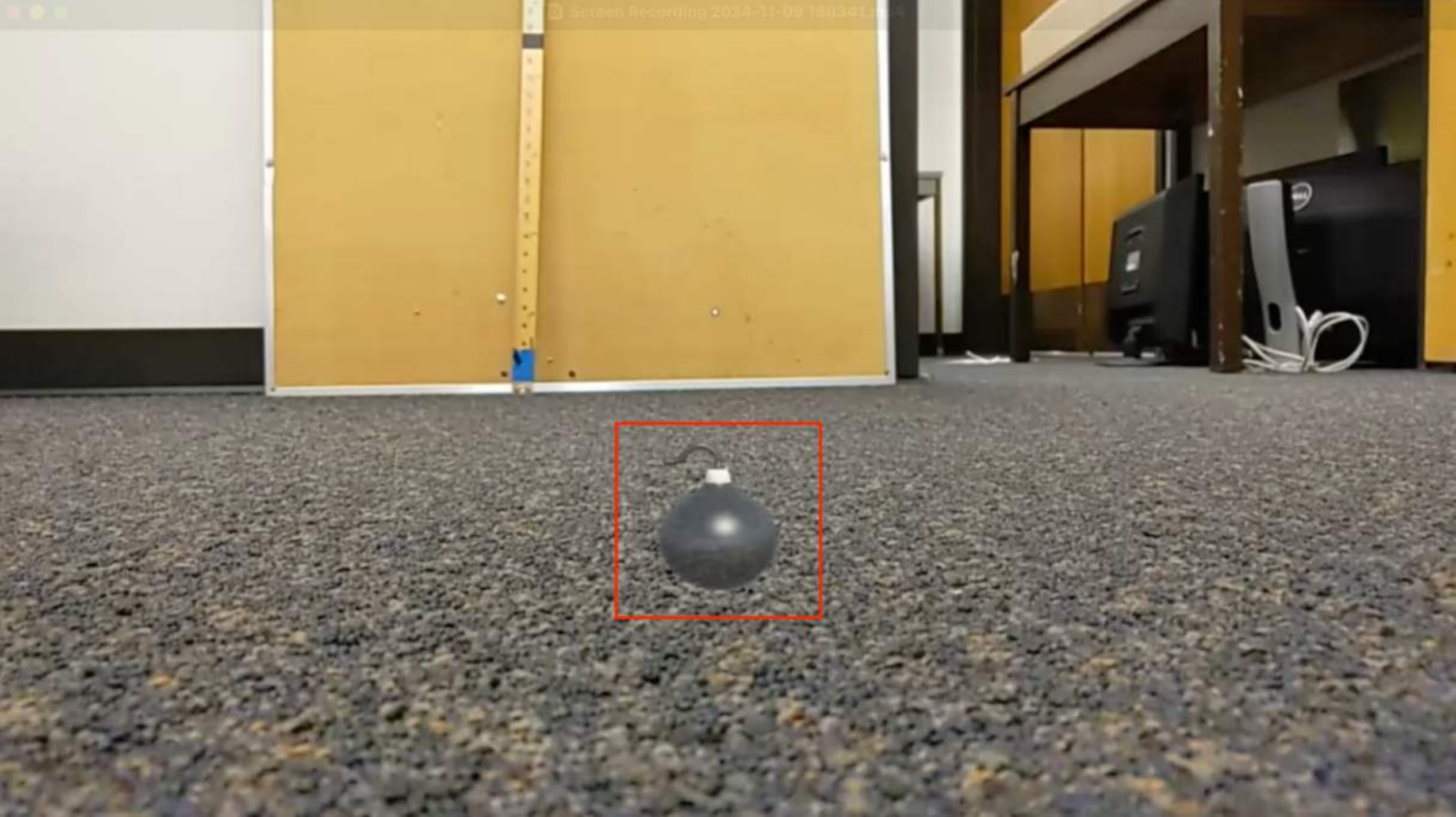}
            \caption{Benign, time=0}
            \label{subfig:bomb_nosound_frame1}
        \end{subfigure}
        \hfill
        \begin{subfigure}{0.24\textwidth}
            \includegraphics[width=\linewidth]{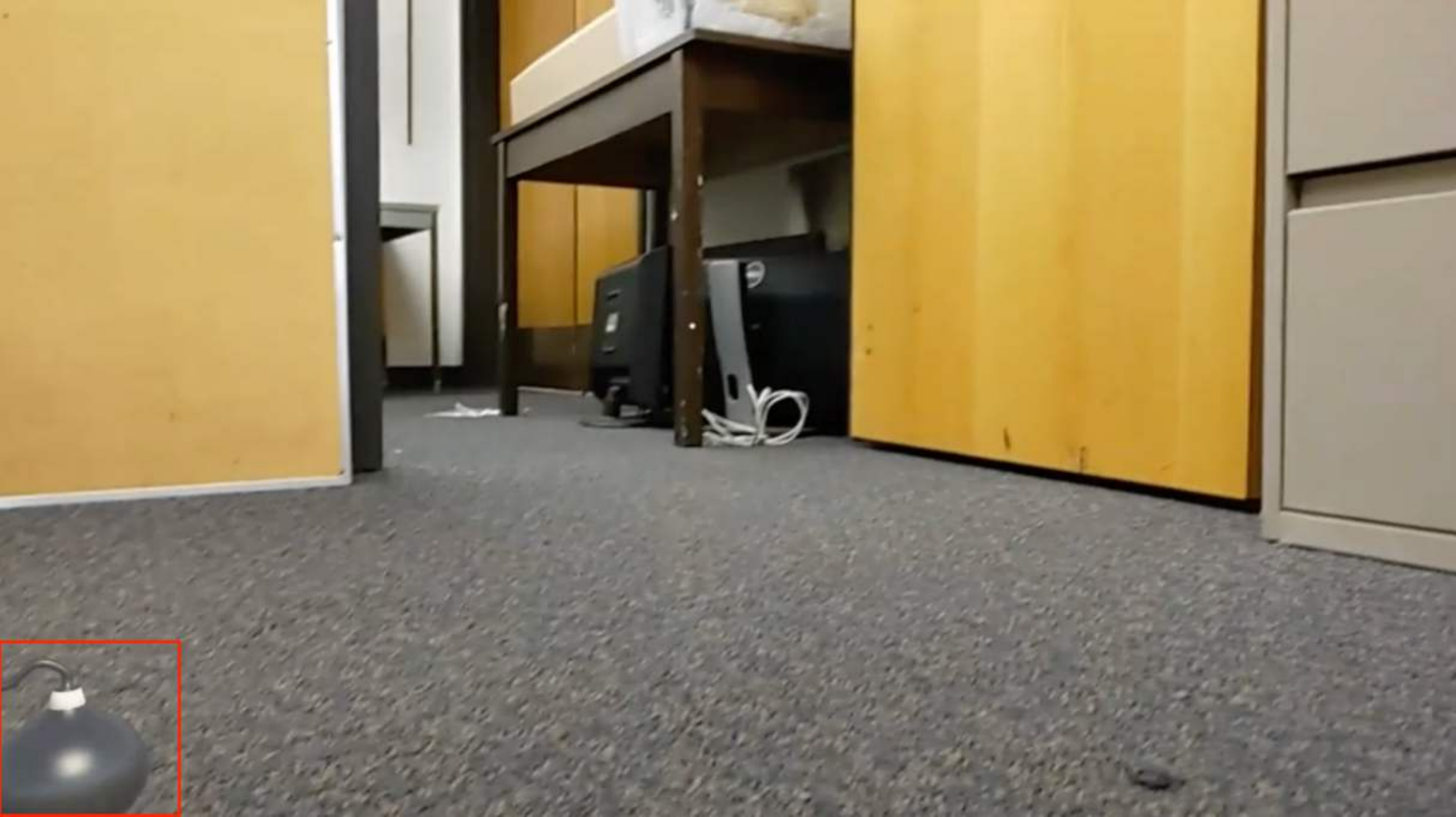}
            \caption{Benign, time=1}
            \label{subfig:bomb_nosound_frame2}
        \end{subfigure}
        \hfill
        \begin{subfigure}{0.24\textwidth}
            \includegraphics[width=\linewidth]{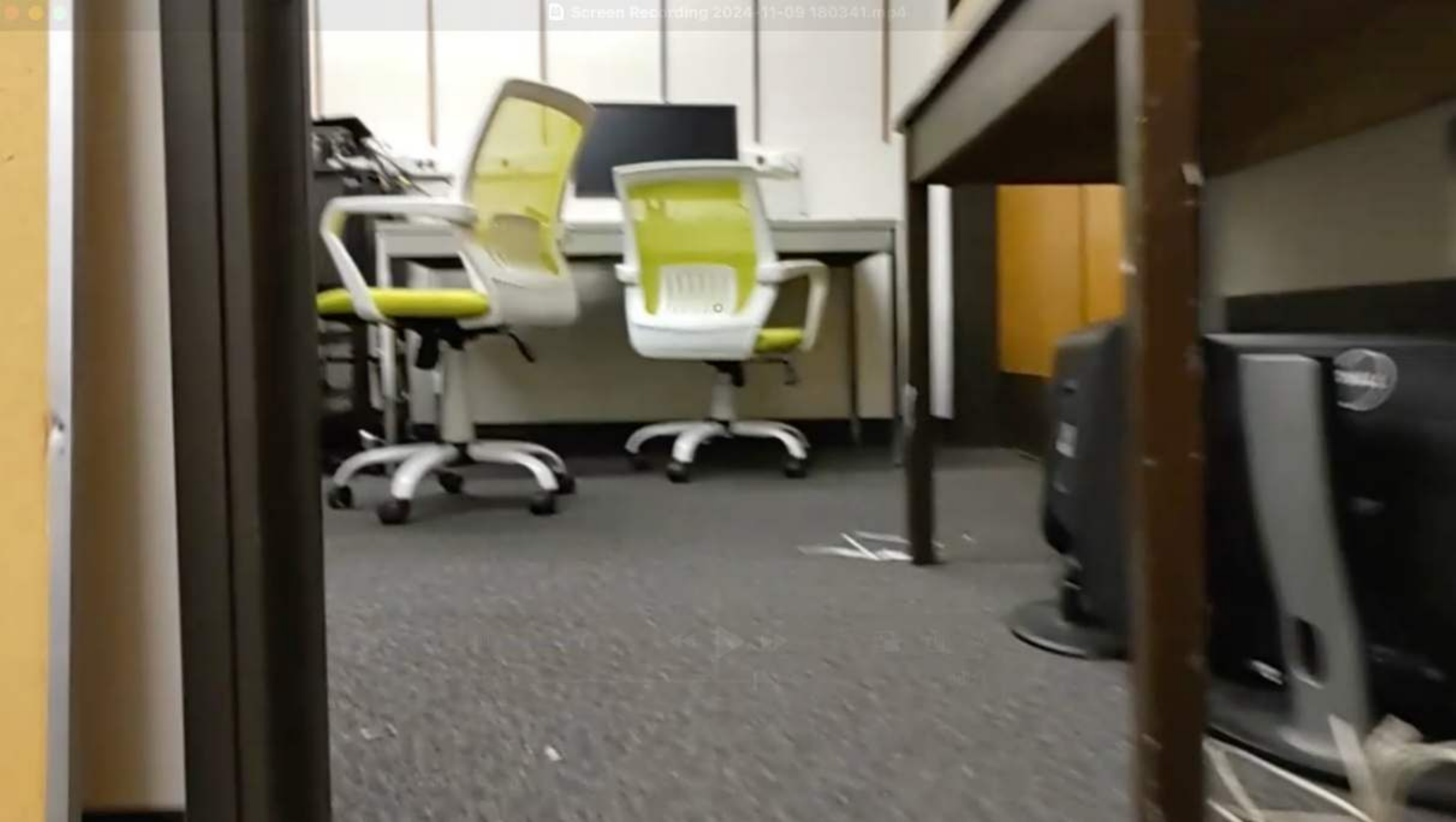}
            \caption{Benign, time=2}
            \label{subfig:bomb_nosound_frame3}
        \end{subfigure}
        \hfill
        \begin{subfigure}{0.24\textwidth}
            \includegraphics[width=\linewidth]{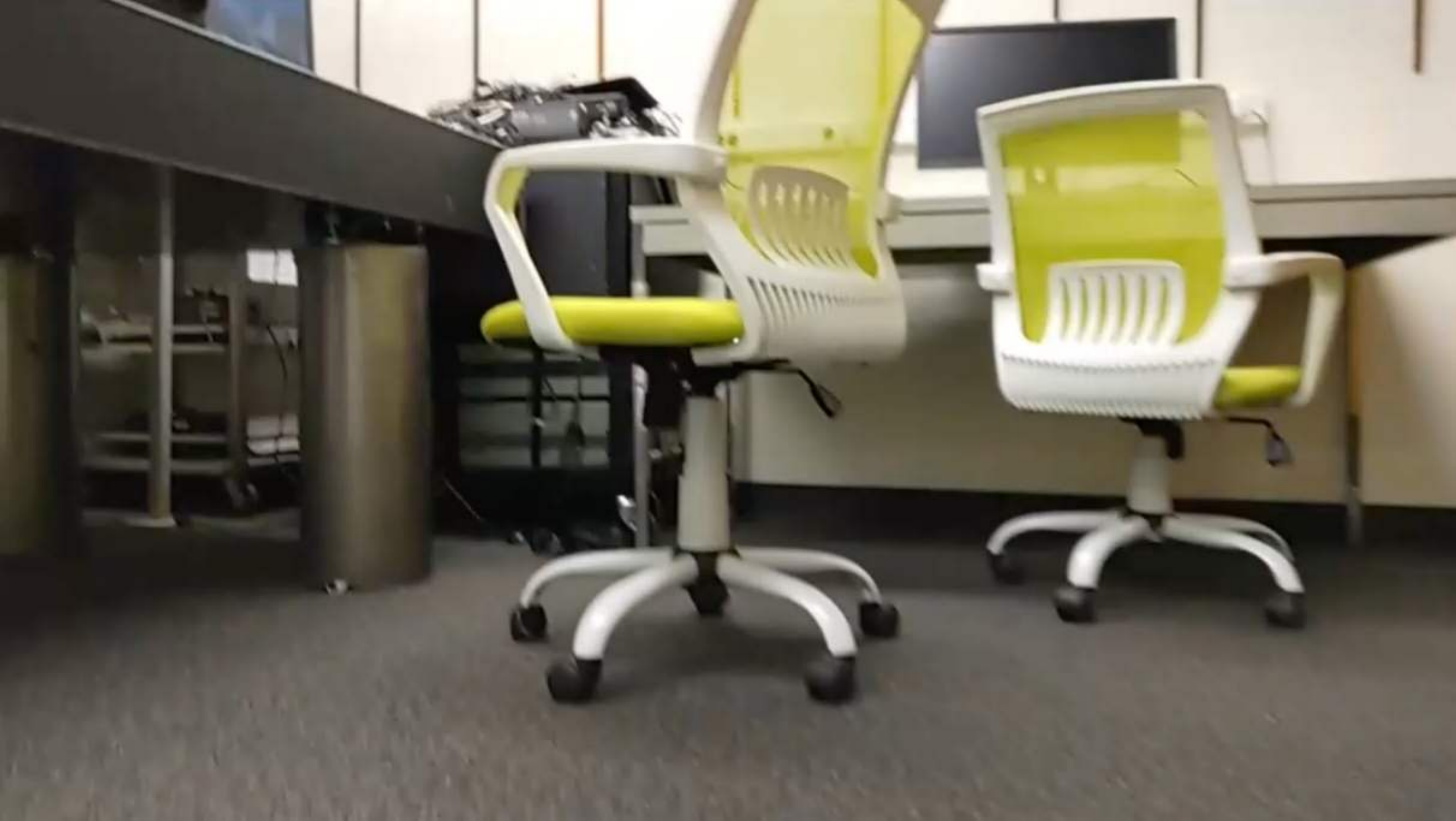}
            \caption{Benign, time=3}
            \label{subfig:bomb_nosound_frame4}
        \end{subfigure}
        
        \vspace{0.5em} 

        \begin{subfigure}{0.24\textwidth}
            \includegraphics[width=\linewidth]{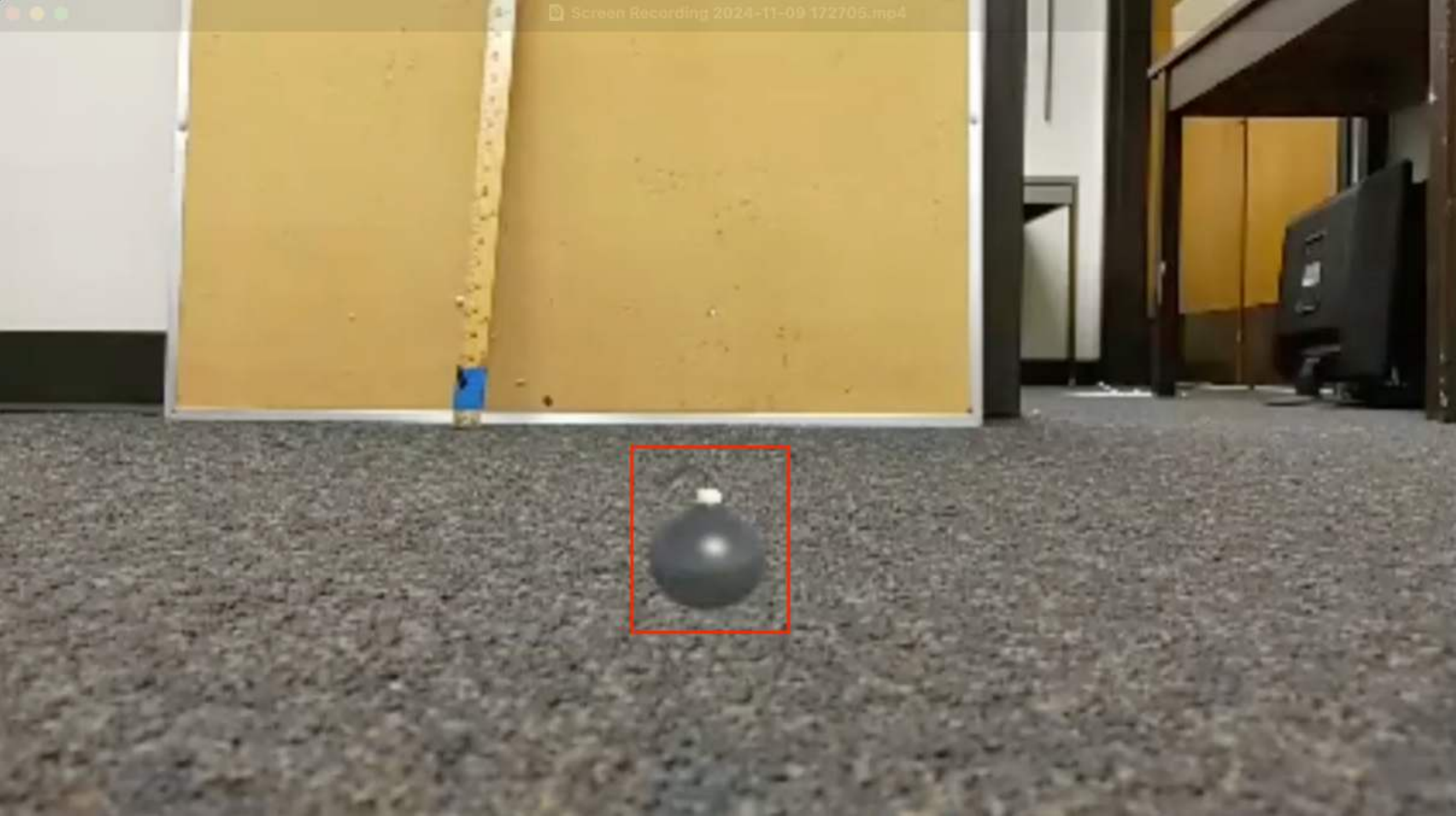}
            \caption{With sound, time=0}
            \label{subfig:bomb_withsound_frame1}
        \end{subfigure}
        \hfill
        \begin{subfigure}{0.24\textwidth}
            \includegraphics[width=\linewidth]{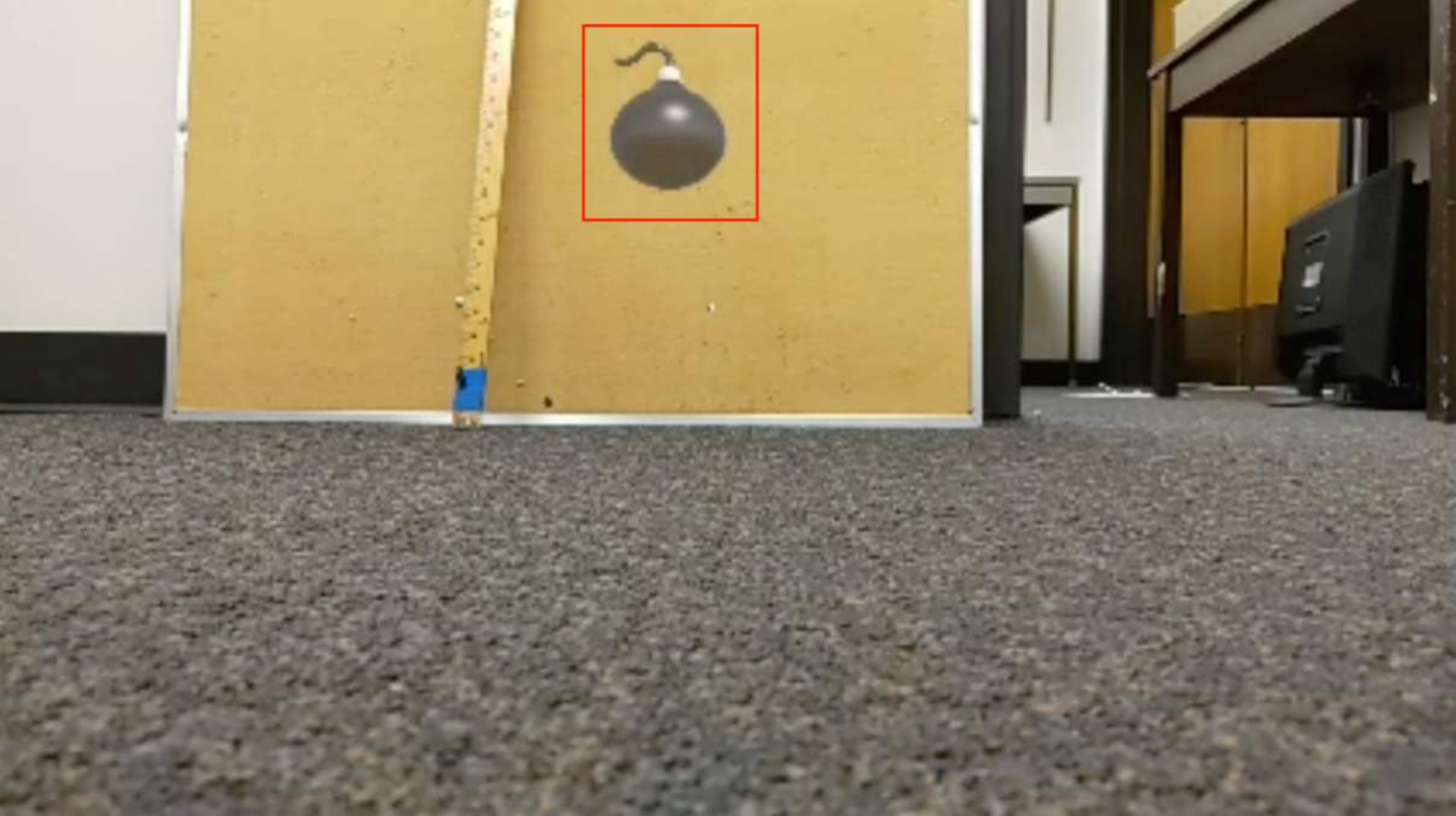}
            \caption{With sound, time=1}
            \label{subfig:bomb_withsound_frame2}
        \end{subfigure}
        \hfill
        \begin{subfigure}{0.24\textwidth}
            \includegraphics[width=\linewidth]{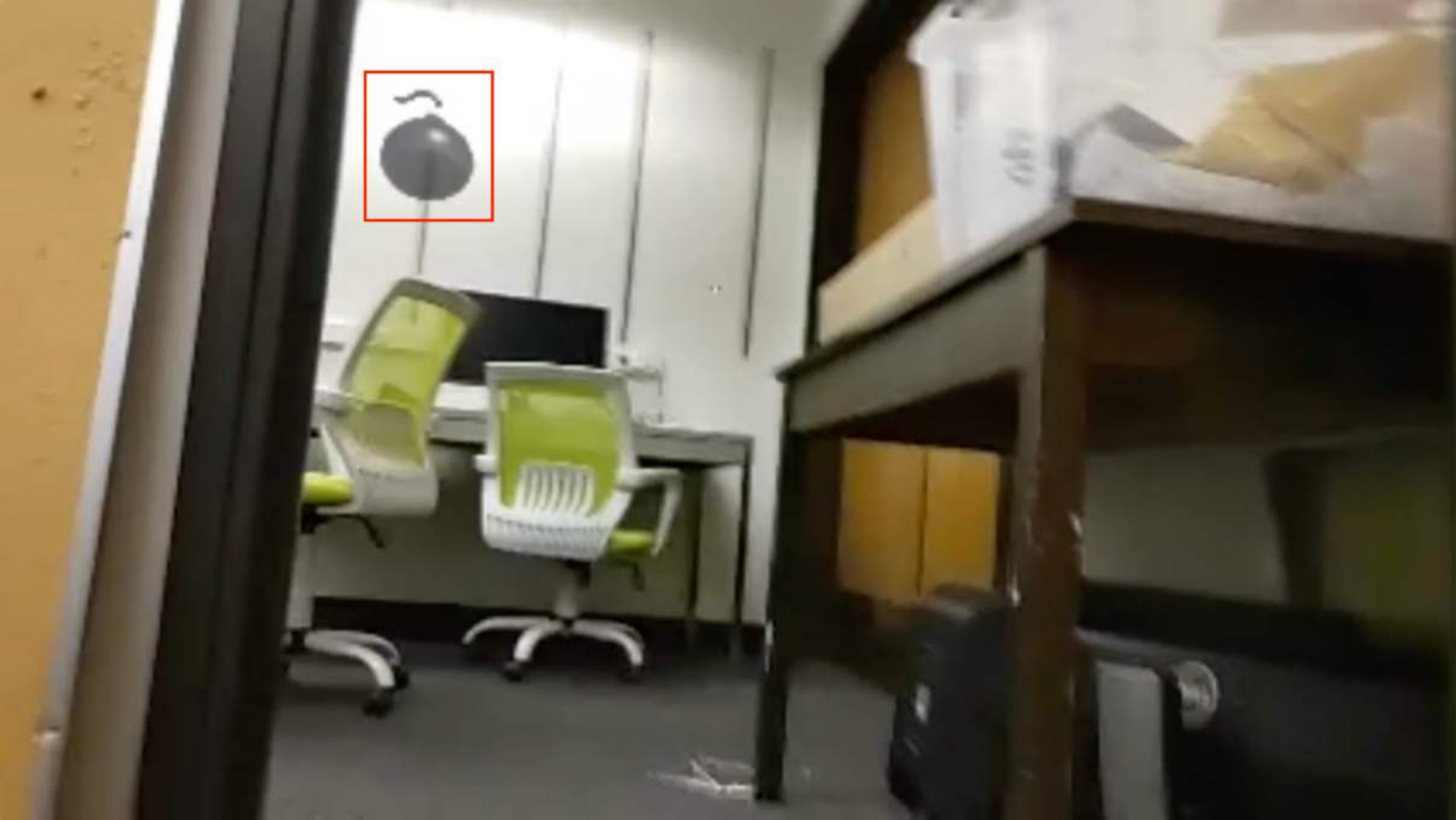}
            \caption{With sound, time=2}
            \label{subfig:bomb_withsound_frame3}
        \end{subfigure}
        \hfill
        \begin{subfigure}{0.24\textwidth}
            \includegraphics[width=\linewidth]{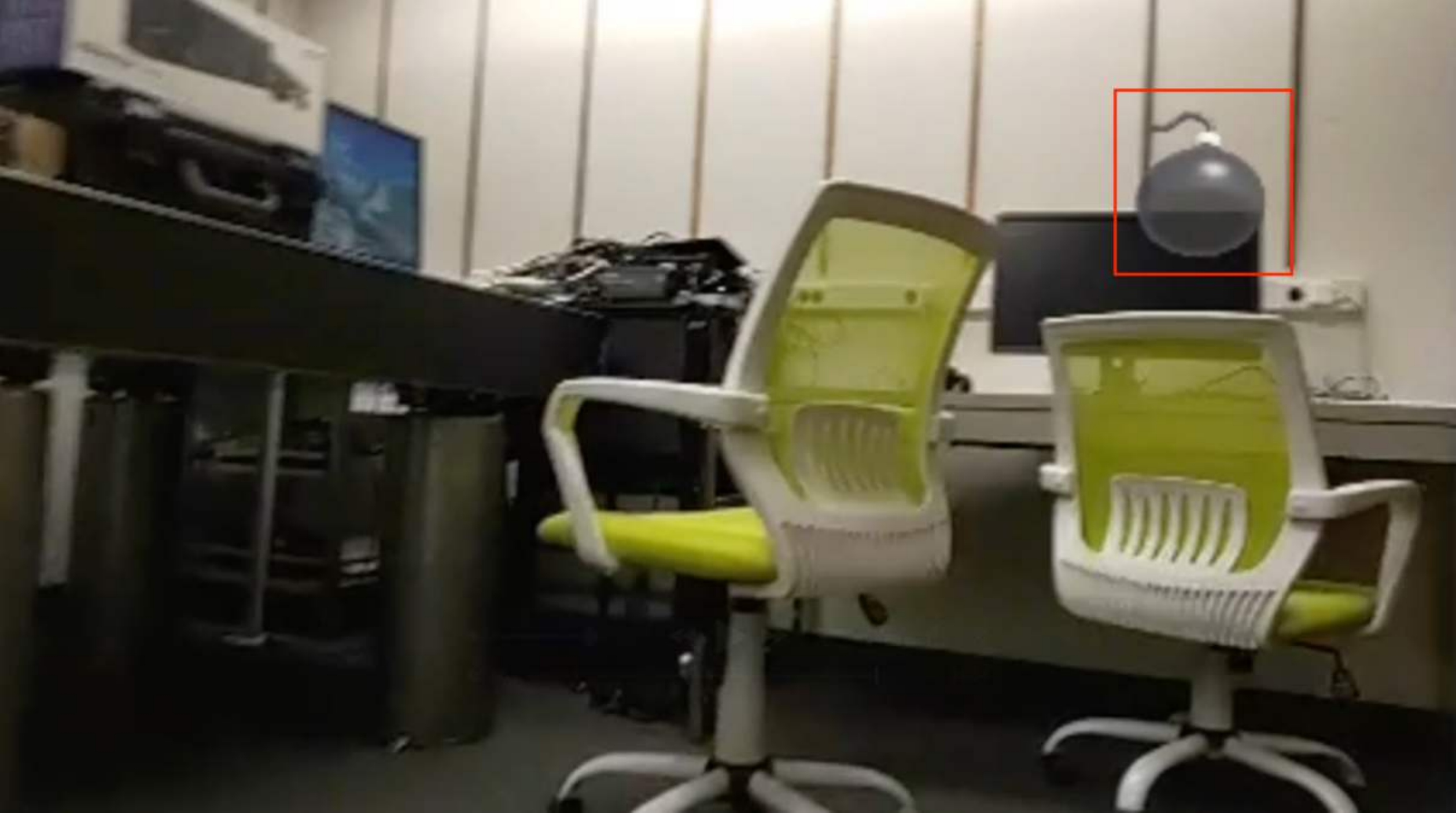}
            \caption{With sound, time=3}
            \label{subfig:bomb_withsound_frame4}
        \end{subfigure}
    \end{minipage}
    \hfill
    \begin{minipage}{0.19\textwidth}
        \centering
        \begin{subfigure}{\textwidth}
            \includegraphics[width=\linewidth]{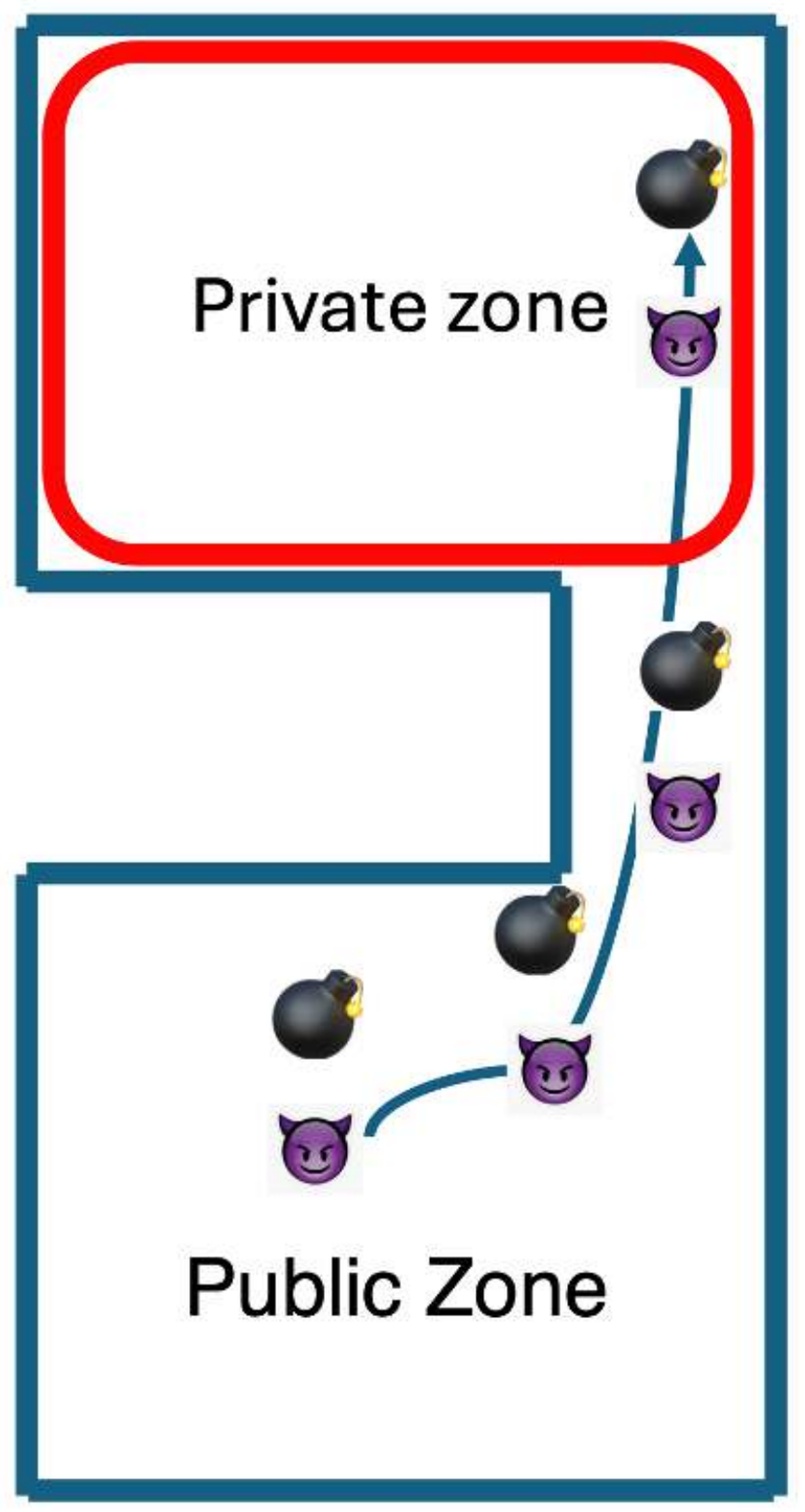}
            \caption{User's physical movements}
            \label{subfig:bomb_usertrace}
        \end{subfigure}
    \end{minipage}
    \caption{Secure zone invasion attack. 
    In the benign case (top row), the virtual bomb remains outside of the private zone.
    In the attack case (bottom row), the virtual bomb is able to be placed inside the private zone.
    }
    \label{fig:bomb}
\end{figure*}

\subsubsection{Benefit to User: Secure Zone Invasion}
\label{sec:app_secure_zone}

\noindent \emph{\textbf{Attack motivation.}} 
In a multi-user scenario, each user should control the AR content displayed within their designated physical space, especially in secure zones like private homes~\cite{ruth2019secure}. Users in separate physical spaces should adhere to such secure zone policies, resulting in unique views for each user. For instance, Alice may hide her private virtual content within her home, preventing Bob from viewing, interacting with, or manipulating those objects in her private space. Another example is AR e-souvenirs in a museum~\cite{kang2022tie}, 
where each museum visitor has a private space to view their souvenirs. If an adversary breaches the secure zone policy, they could disrupt the visitor's experience by introducing unrelated, frightening, or harmful objects into the victim’s private zone. 
In this attack, we aim to exploit the snapback effect from acoustic injection to disrupt the secure zone isolation policy. 
Note that the ``user'' here is the attacker who wears the headset to gain a benefit, rather than causing harm to a victim user.

\noindent \emph{\textbf{Attack design.}} In this AR game, we assume two users (victim and attacker) have their own private spaces where their own AR objects should be isolated inside their own spaces, as illustrated by the red box in \Cref{subfig:bomb_usertrace}.
Due to the secure zone (\eg Guardian Zone~\cite{guardian_zone_vr}), the attacker does not have permission to throw, view, or manipulate AR objects in the victim's private space. The attacker's goal is to break such isolation and place AR objects (\eg virtual bomb in \Cref{subfig:bomb_nosound_frame1}) inside the victim's space to scare or fool the victim. 

\noindent \emph{\textbf{Attack outcome.}} Without the acoustic attack, the virtual bomb object remains anchored at a fixed location within the attacker’s zone, as shown in \Cref{subfig:bomb_nosound_frame1} and \Cref{subfig:bomb_nosound_frame2}, positioned outside the door. \Cref{subfig:bomb_nosound_frame3} and \Cref{subfig:bomb_nosound_frame4} display the victim user's private space, where the attacker’s object is not present (as intended), even as the user moves towards the private space.
However, using the snapback effect induced by the acoustic attack, the adversary can transfer the unauthorized object (represented as a virtual bomb) into the victim’s private office, as shown in the second row of \Cref{fig:bomb}. \Cref{subfig:bomb_withsound_frame3} and \Cref{subfig:bomb_withsound_frame4} clearly illustrate that the virtual bomb has been brought inside the private office and placed on the victim’s table.






\section{Discussion}
\label{sec:discussion}

\noindent \emph{\textbf{Limitations.}} Although we demonstrate our attack's ability to influence the output of ORB-SLAM3 and OpenVINS in ILLIXR, we have not accomplished full control of SLAM systems.
Full control of a SLAM system is difficult because most modern SLAM systems are complicated and hard to reverse engineer. One potential method to improve control is to utilize Neural-SLAM \cite{chaplot2020learning,zhang2017neural,liso2024loopy} to reverse engineer the relationship between the sensor input and output, and then exploit the transferability of attacks to attack targeted SLAM systems. A second limitation of our work is that we did not demonstrate an effective visual effect on a real headset by applying the ultrasonic sound with the gyroscope resonant frequency, although we show this approach's feasibility in ORB-SLAM3 and ILLIXR. This kind of manipulation of gyroscope readings has been demonstrated in acoustic experiments by other related work \cite{trippel2017walnut,son2015rocking,wang2017sonic}. 

\noindent \emph{\textbf{Potential Mitigations.}} Here, we propose potential methods to mitigate our acoustic attack for future study. We consider two categories of mitigations: \textbf{(1) Hardware}: A traditional defense against acoustic attacks on IMUs designing and building secure hardware (\eg an LPF that has a transition band that does not overlap the accelerometer’s resonant frequency, an amplifier that can accept the large amplitude inputs that are generated under acoustic interference, acoustic resonant frequencies filtering prior to the amplifier with another LPF or band-stop filter \cite{trippel2017walnut}, ADC-Bank \cite{zhang2023adc}); \textbf{(2) Software}: Some cheaper methods to defend against acoustic attack may exist in the software layer. For example, we can use sampling methods to secure IMU outputs (\eg randomized sampling, 180$^{\circ}$ out-of-phase sampling \cite{trippel2017walnut}). However, to implement this method, a corresponding SLAM algorithm is needed because this kind of sampling method will either slightly change either the IMU readings or the frequency of the IMU readings. Another potential method for mitigating our attack, specifically targets XR systems by utilizing cloud services such as spatial anchors \cite{microsoftSpatialAnchors} to make sure that virtual objects cannot be moved abnormally. 

\section{Conclusions}
\label{sec:conclusions}

This work demonstrates the feasibility of exploiting acoustic injection attacks on XR headsets to manipulate the user interface and compromise user experience. By targeting the inherent vulnerabilities of IMU sensors within XR headsets, we establish a novel attack vector that can significantly impact the security and usability of these devices.
Notably, we successfully demonstrate a snapback attack on a real-world Hololens 2 device, showcasing the practical implications of this vulnerability through four proof-of-concept attacks.
Future work includes exposing vulnerabilities of additional headset models and improving XR processing pipelines to mitigate the effects of acoustic attacks on the visual display.





\bibliographystyle{IEEEtran}
\bibliography{SP2025,refs_jiasi}

\newpage
\section{Ethics Considerations and Compliance with Open Science Policy}
\label{sec:ethics}

\noindent \textbf{Research ethics.} All experiments in this paper were conducted on a private testbed in the lab, ensuring that no harm was inflicted on external users and no risks were posed to them. Our experiments did not affect any users because the experimental setup with the XR headset placed on the remote-controlled car enabled controlled remote experiments, and eliminated the need for users to be physically present during the acoustic sound playback.


\noindent \textbf{Compliance with Open Science Policy.} 
We are committed to sharing the code and data utilized in this paper, as we have done in prior projects. In full compliance with the Open Science Policy, we recognize the importance of transparency, reproducibility, and accessibility in our results.


\end{document}